\documentclass[structabstract]{aa}
\usepackage{amsmath,amsfonts,amssymb,amsxtra} 
\usepackage{txfonts}
\usepackage{graphicx}
\usepackage{natbib}
\usepackage{float}
\usepackage{tabularx}
\usepackage{ltxtable}
\usepackage[figuresright]{rotating}

\usepackage{color}

\newcommand{\Msun}{${M_\odot}$}
\newcommand{\Rsun}{${R_\odot}$}
\newcommand{\Xnmax}{ $X^\mathrm{max}(^{14}\mathrm{N})$}

\newcommand{\lsim}{\mathrel{\hbox{\rlap{\lower.55ex \hbox {$\sim$}}
 \kern-.3em \raise.4ex \hbox{$<$}}}}
\newcommand{\gsim}{\mathrel{\hbox{\rlap{\lower.55ex \hbox {$\sim$}}
 \kern-.3em \raise.4ex \hbox{$>$}}}}

\begin{document}

\title{Evolution of massive population III stars with rotation and magnetic fields}
\author{ S.-C. Yoon  \and A. Dierks  \and  N. Langer}

\institute {
Argelander-Institut f\"ur Astronomie der Universit\"at Bonn, Auf dem H\"ugel 71, 53121 Bonn, Germany
}

\date{Received:  / Accepted: }

\abstract
{
}
{
We present a new grid of massive population III (Pop III) star models including the
effects of rotation on the stellar structure and chemical mixing, and magnetic
torques for the transport of angular momentum. This grid covers the range of
 mass from 10 to 1000~\Msun{}, and  rotational velocity from zero to
100\% of the critical rotation on the zero-age main sequence. Based on the
grid, we also present a phase diagram for the expected final fates (i.e, 
core-collapse supernova, gamma-ray bursts and pair-instability supernovae of diverse
types) of rotating massive Pop III stars.}
{The model grid has been calculated with a stellar evolution code.  We adopted
the recent calibration made with the VLT-FLAMES data for the overshooting
parameter and the chemical mixing efficiency due to rotation.  The
Spruit-Tayler dynamo  was assumed for magnetic torques.}
{Our non-rotating models become redder than the previous models in the
literature because of the larger overshooting parameter adopted in this study.  In
particular, convective dredge-up of the helium core material into the hydrogen
envelope is observed in our non-rotating very massive star models ($\gsim
200$~\Msun), which is potentially important for the chemical yields.  On the
other hand, the stars become bluer and more luminous with a higher rotational
velocity. With the Spruit-Tayler dynamo, our models with a sufficiently high
initial rotational velocity can reach the critical rotation earlier and lose
more mass as a result, compared to the previous models without magnetic fields.
The most dramatic effect of rotation is found with the so-called chemically
homogeneous evolution (CHE),  which is observed for a limited mass and
rotational velocity range.  CHE has several important consequences: 1) Both
primary nitrogen and ionizing photons are abundantly produced.  2) Conditions
for gamma-ray burst progenitors are fulfilled for an initial mass range of 13 -
84~\Msun{}.  3) Pair instability supernovae of type Ibc are expected for 84 -
190~\Msun{}. 4) Both a pulsational pair instability supernova and a  gamma-ray burst (GRB) may
occur from the same progenitor of $\sim$ 56 - 84~\Msun{}, which might
significantly influence the consequent GRB afterglow.  We find that CHE
does not occur for very massive stars ($ > 190$~\Msun), in which case  the
hydrogen envelope expands to the red-supergiant phase and the final angular
momentum is too low to generate any explosive event powered by rotation.}
{}


\maketitle


\section{Introduction}

The first generations of stars (or population III stars) in the early Universe
are supposed to play a key role in shaping the early Universe.  It is believed
that the first stars were intrinsically massive, mainly because of the lack of
efficient coolants in the primordial gas \citep[e.g.][]{Bromm99,
Abel00,Nakamura01}.  This means that they should  be one of the most important
sources of ultra-violet photons for reionization at high redshift.  The
chemical evolution of the early Universe is critically determined by the first
stars because most of the elements heavier than helium are uniquely provided by
them via supernova explosions and possibly stellar winds. The chemical
abundance patterns observed in extremely metal-poor stars or in damped
Lyman-$\alpha$ systems at high redshift might imprint the nucleosyntheses in the
first stars \citep[e.g.][]{Depagne02, Christlieb02, Cayrel04,  Erni06,
Frebel11, Cooke11, Kobayashi11}.   Some of the first stars may produce very
bright events such as supernovae (SNe) or gamma-ray bursts (GRBs), which are
potentially observable.  The recent discoveries of some GRBs at very high
redshift \citep{Kawai06, Greiner09, Salvaterra09, Tanvir09, Cucchiara11} indeed
imply that such luminous events from the first stars may serve as indispensable
tools for the probe of the early Universe with the next generation of
telescopes. 

Many authors already addressed these questions using stellar evolution models
\citep{Woosley95, Marigo01, Heger02, Marigo03, Chieffi04, Umeda03,  Umeda05,
Ekstroem08, Heger10}. One of the most intriguing questions here is how rotation
affects the evolution and the final fates of the first stars.  Previous studies
on massive star evolution  indicate that both the rotationally induced chemical
mixing and the centrifugal force can significantly alter the mass loss rate,
the stellar structure, and the consequent evolution and nucleosynthesis
\citep{Maeder00b, Heger00}.  This effect should become even more important for
metal-poor massive stars \citep[e.g.][]{Meynet02, Yoon06, Hirschi07},  for
which angular momentum loss through radiation-driven stellar winds is generally
less efficient \citep[e.g.,][]{Vink01, Kudritzki02}.  In addition, a
recent numerical study on  star formation in the early Universe by
\citet{Stacy11} shows that the rotational velocity of the first stars can be
close to the break-up limit, suggesting that the effects of rotation on the evolution 
of Pop III stars should not be ignored.

There exist only a few studies on the role of rotation for the evolution of the
first stars.  \citet{Marigo03} calculated rigidly rotating massive metal free
stars (120 -- 1000~\Msun), finding that the centrifugally driven mass loss may
be significant when the rotational velocity at the equatorial surface reaches
the critical limit.  \citet{Ekstroem08} considered the internal transport of
angular momentum due to hydrodynamic processes and rotationally induced
chemical mixing in their evolutionary models for masses of 9 -- 200~\Msun{} at
zero metallicity. They  showed the importance of the centrifugally driven
mass loss, as well as the interesting consequences of chemical mixing in
the nucleosynthetic yields.  

In the present paper, we present a new grid of evolutionary sequences of
metal-free massive stars including the transport of angular momentum and
chemical species, for initial masses ranging from 10~\Msun{} to 1000~\Msun.
While only a fixed value for the rotational velocity was assumed in Marigo et al.
and Ekstr\"om et al. we considered several different  rotational velocities for
each mass. As shown below,  this enables predicting the final fates
of the first stars (core-collapse, GRB, pair-instability SN, etc.)  in the
parameter space spanned by the initial mass and the initial velocity
(Sect.~\ref{sect:fate}).  The present study also differs from the previous work
because we included magnetic torques according to the Spruit-Tayler dynamo
\citep{Spruit02}.  The adopted efficiency of the angular
momentum transport in our work is therefore in-between that of Marigo et al.,
who enforced rigid rotation, and that of Ekstr\"om et al., who considered only
hydrodynamic transport processes (see Sect.~\ref{sect:method} for a more detailed
discussion on this). 

The paper is organized as follows. We describe the numerical method, physical
assumptions,  and the newly constructed model grid in Sect.~\ref{sect:method}.
The evolution of basic physical properties for different initial masses and
rotational velocities and its consequence on the ionizing fluxes from rotating
Pop III stars are discussed in Sect.~\ref{sect:evolution}.  The effect of
chemical mixing due to convection and rotationally induced hydrodynamic
instabilities on the nucleosynthesis is investigated in
Sect.~\ref{sect:mixing}.  The conditions for the so-called chemically
homogeneous evolution are discussed in Sect.~\ref{sect:che}.  In
Sect.~\ref{sect:fate}, we explain how the final structure of the first stars
would be affected by rotation, and discuss the possible diversity of the
explosions of the first stars.  Finally we summarize our results and
discussions in Sect.~\ref{sect:conclusion}.

\section{Numerical methods}\label{sect:method}

\subsection{Basics on the stellar evolution code}

We used the stellar evolution code that is described in \citet{Yoon06} and
references therein. It solves a set of stellar structure equations using the
Heney-type implicit method.  The  acceleration term in the momentum
equation was ignored to suppress pulsations that often occur because of
partial ionization of hydrogen when the surface temperature becomes lower than
about 5000~K \citep[see, for example,][]{Yoon10}. 
The opacities are taken from \citet{Iglesias96} for $T > 10^4$~K, and from
\citet{Alexander94} for lower temperatures.    
The nuclear network includes
63 nuclear reactions, with more than 30 isotopes \citep{Heger00}.  
The NeNa and
MgAl chains are not considered in this study.  We adopted the rate of
\citet{Caughlan85} for the $^{12}\mathrm{C}(\alpha, \gamma)^{16}\mathrm{O}$
reaction, multiplied by a factor of 1.6. 
The considered reactions for carbon burning
include
$2\mathrm{^{12}C} + \gamma \rightarrow \mathrm{^{20}Ne}+\alpha$, 
and $2\mathrm{^{12}C} + \gamma \rightarrow \mathrm{^{23}Na}+p$, 
with secondary reactions like  $^{23}\mathrm{Na}(p,\alpha)^{20}\mathrm{Ne}$
and $^{23}\mathrm{Na}(p,\gamma)^{24}\mathrm{Mg}$. 
Neon burning is mainly followed by 
the reactions $^{20}\mathrm{Ne}(\gamma, \alpha)^{16}\mathrm{O}$, 
and $^{20}\mathrm{Ne}(\alpha, \gamma)^{24}\mathrm{Mg}$. 
For oxygen burning, only $^{28}$Si and $^{32}$S were considered
as the main products.
The nuclear network is not coupled with the transport equation.
This may have consequences in the stellar models where protons are mixed into the 
convective layers associated with helium shell burning, as discussed in Sect.~\ref{sect:mixing}.

The effect of rotation on the stellar structure is considered following
\citet{Endal76}.  The transport of angular momentum and chemical species through
rotationally induced hydrodynamic instabilities are considered as  explained in
\citet{Heger00}. 
The effect of magnetic torques is also considered in our models. 
In the radiative layers, dynamo action can be achieved by the interplay between
amplification of toroidal fields ($B_\phi$) by differential rotation, and
continuous creation of radial fields ($B_\mathrm{r}$) by the Tayler
instability (the so-called Spruit-Tayler dynamo; \citealt{Spruit02}).  
To consider this effect, the steady-state solution for $B_\phi$
and $B_\mathrm{r}$ given by Spruit was used to calculate the effective
viscosity for the radial transport of angular momentum (see
\citealt{Petrovic05} for more details on the implementation of the Spruit-Tayler 
dynamo in our code). Even though the efficiency of the Spruit-Tayler dynamo is
currently much debated \citep[e.g.,][]{Braithwaite06, Zahn07, Gellert08,
Arlt11}, it still remains one of the most promising mechanisms that can
explain the observed spin rates of young neutron stars and isolated white
dwarfs, which should reflect the history of angular momentum redistribution in
the evolution of their progenitors \citep{Heger05, Suijs08}.  Note that very
weak magnetic fields suffice for the seed of the Spruit-Tayler dynamo
\citep{Spruit99, Spruit02}.  Since there exist various plausible mechanisms for
the generation of the primordial magnetic fields in the early Universe
\citep[see][for a recent review]{Kandus11}, our assumption of the presence of
weak seed magnetic fields in the first stars can be justified. 

For convective overshooting and rotationally induced chemical mixing, we adopted
the parameters given by \citet{Brott11}, who have recently calibrated the
mixing efficiency using the data of the VLT-FLAMES Survey of massive stars.  We
used 0.1 and 0.028 for the $f_\mu$ and $f_c$ parameters, respectively
\citep[see][for the definition of these parameters]{Heger00}. We applied
overshooting for the hydrogen-burning core only, for which a parameter of
$0.335 H_\mathrm{P}$ was adopted.  We ignored overshooting for the stars beyond
the core hydrogen exhaustion, where the efficiency of overshooting is expected
to be small.  This is because the  convective cores during the advanced stages
tend to grow in mass, which creates strong compositional stratification at the
upper boundary of the convective core 
\citep{Langer91}.

\begin{figure*}
\centering
\includegraphics[width=\columnwidth]{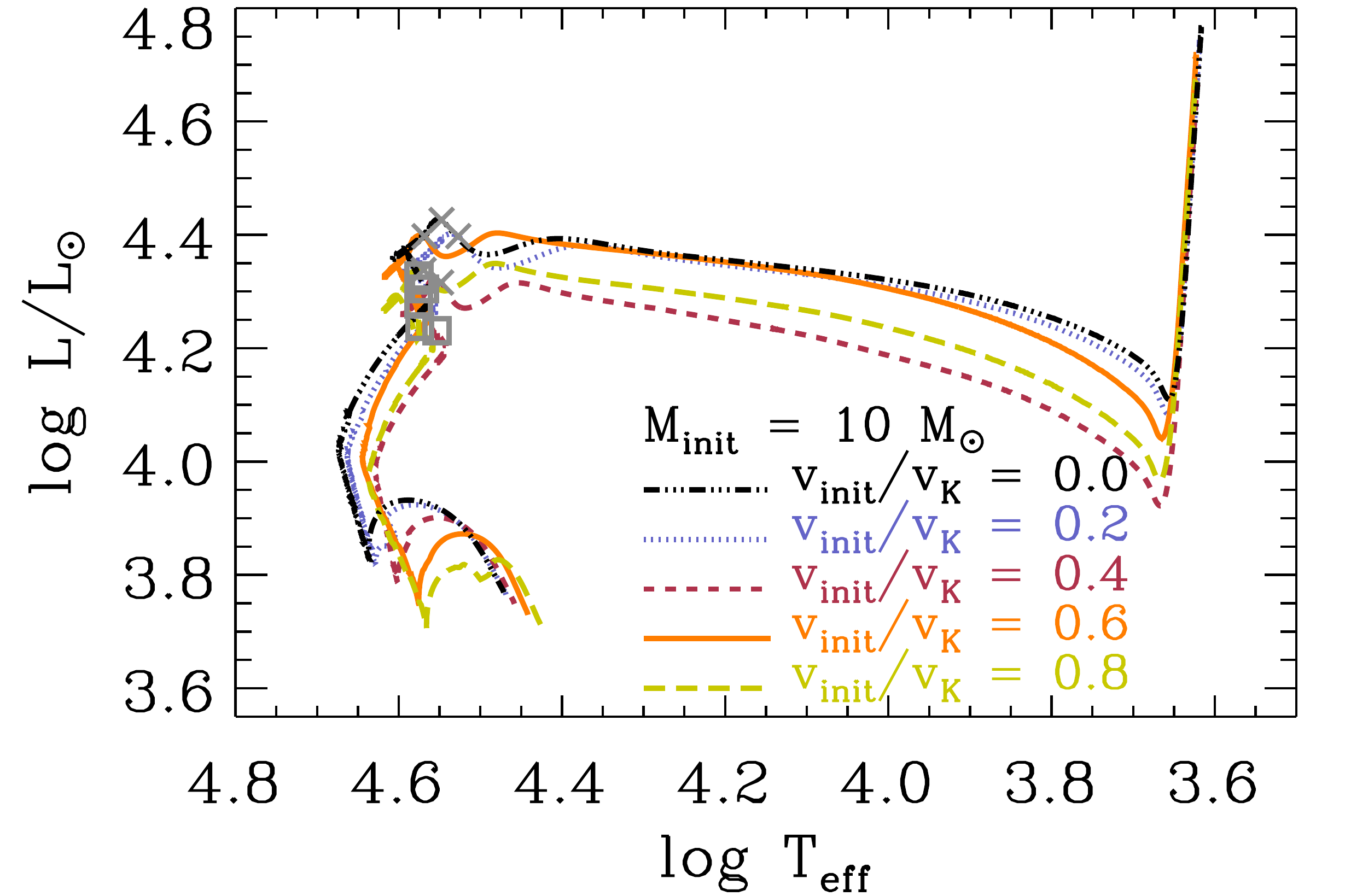}
\includegraphics[width=\columnwidth]{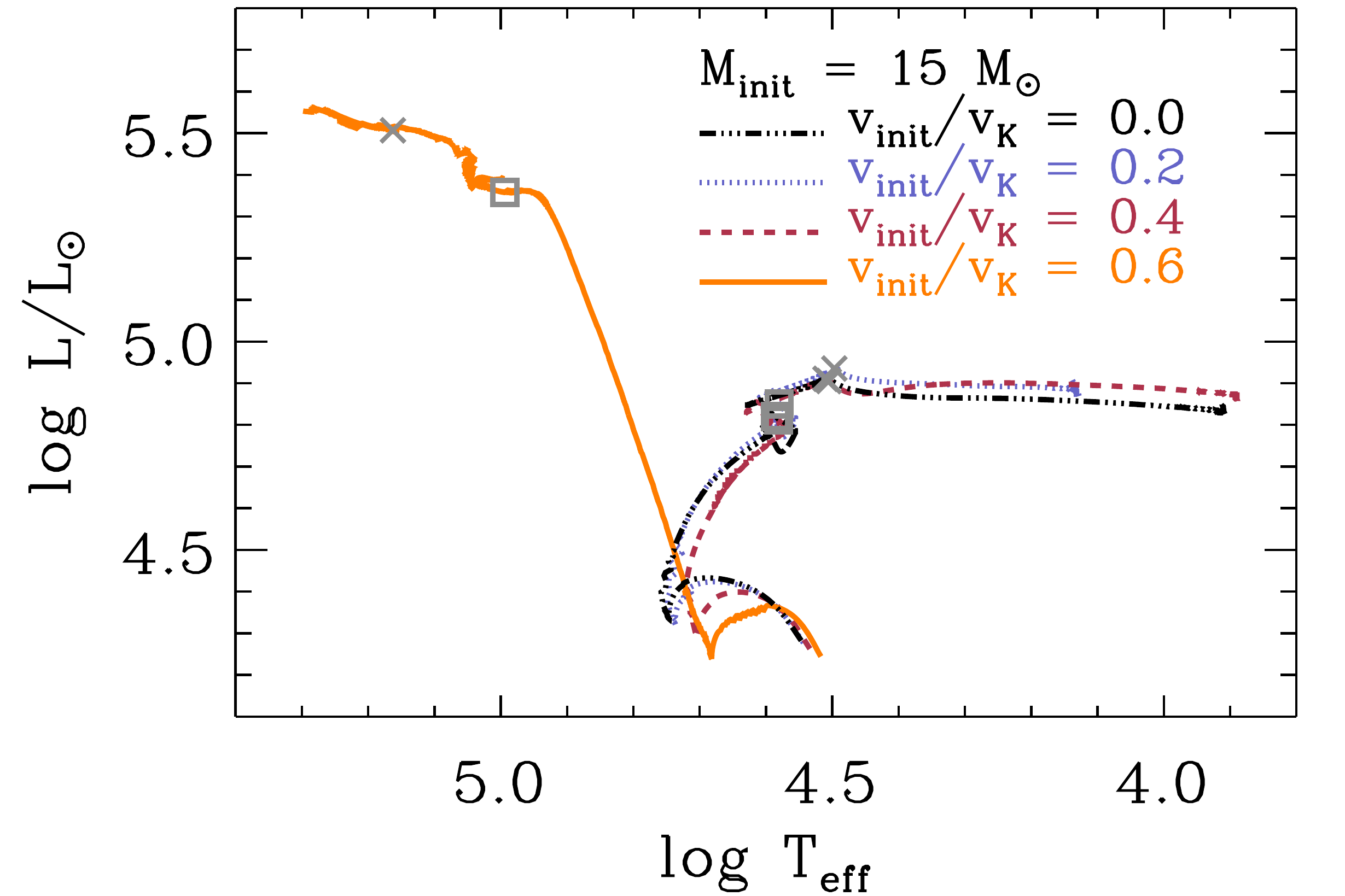}
\includegraphics[width=\columnwidth]{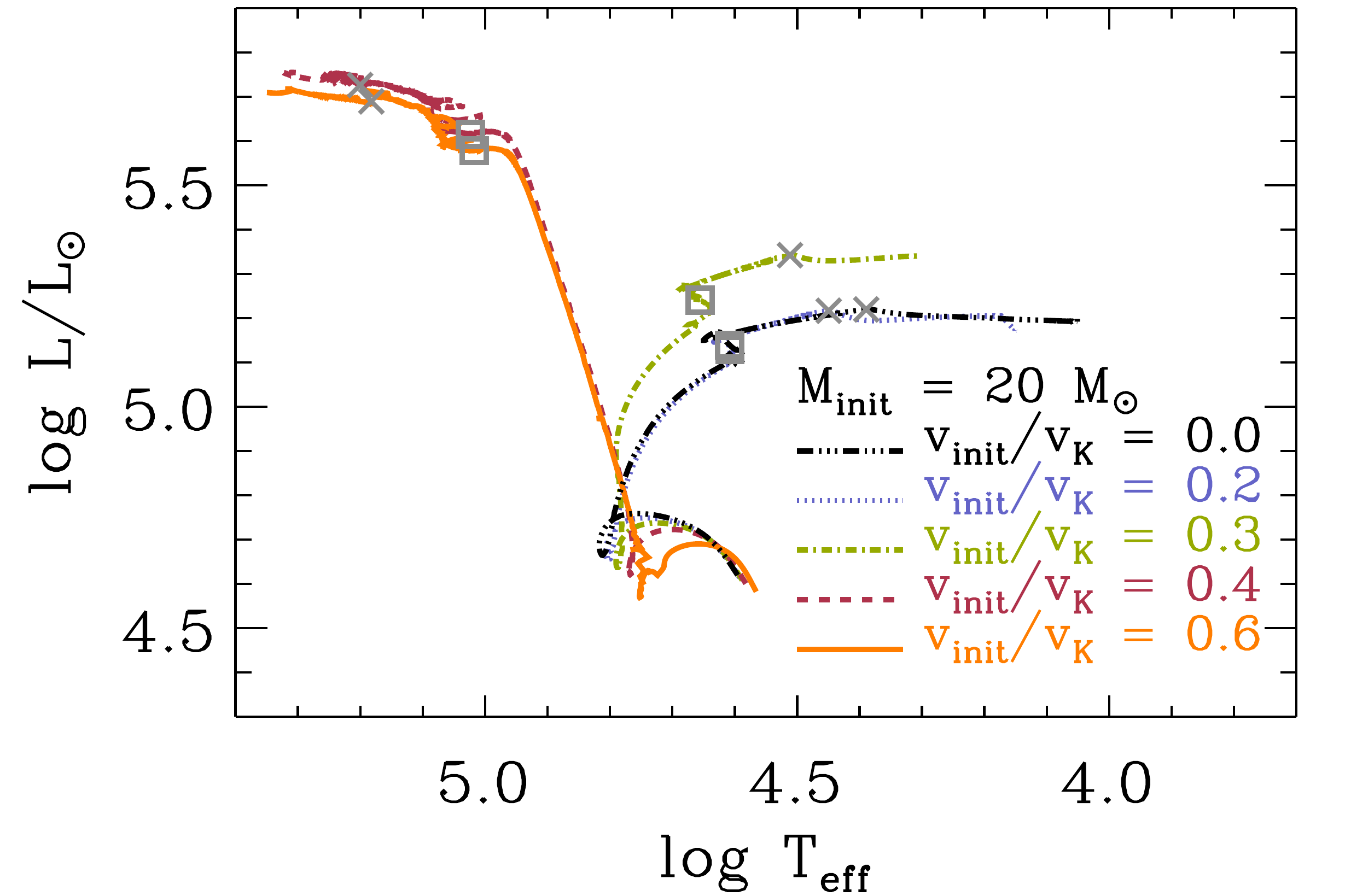}
\includegraphics[width=\columnwidth]{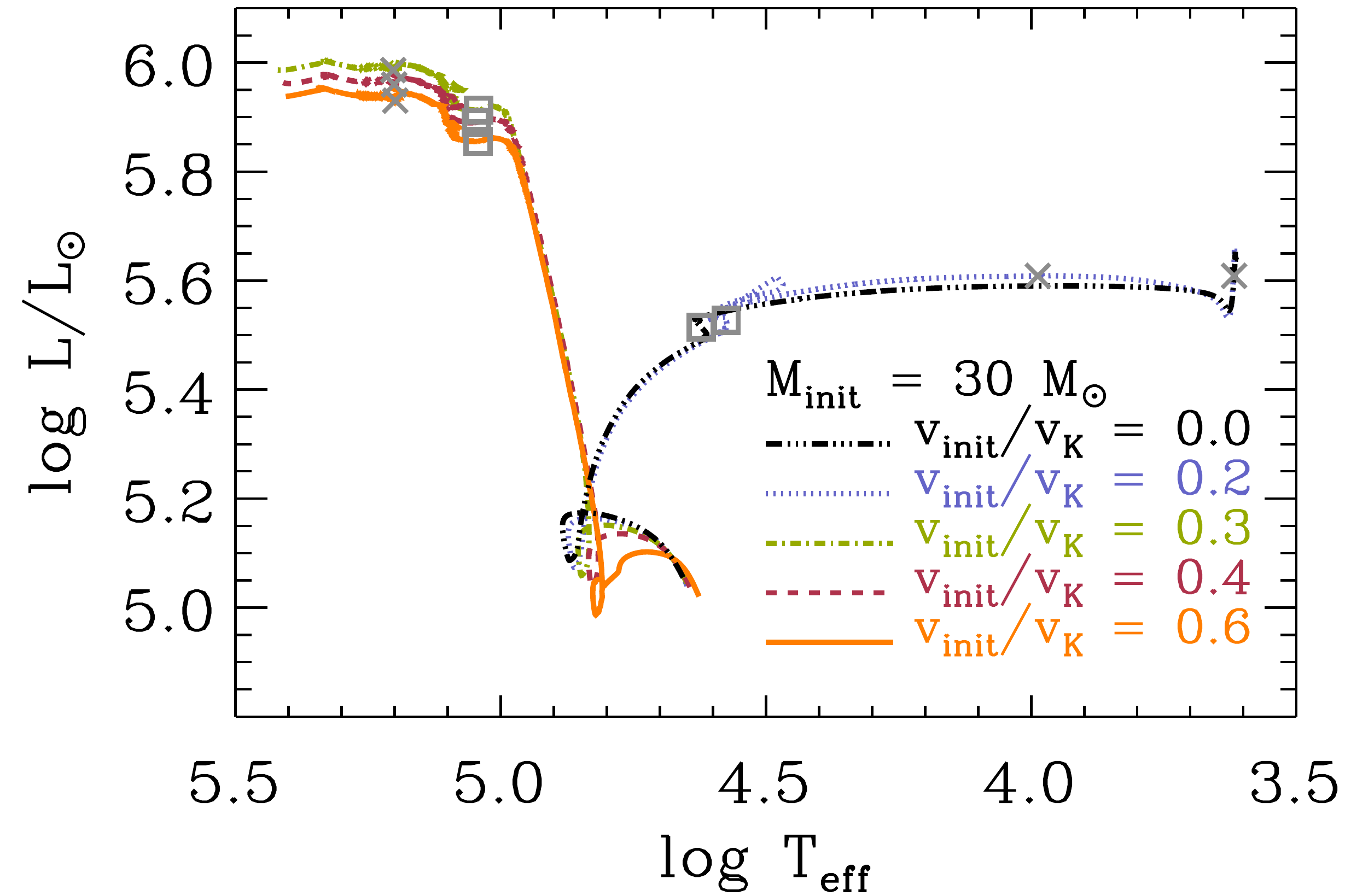}
\includegraphics[width=\columnwidth]{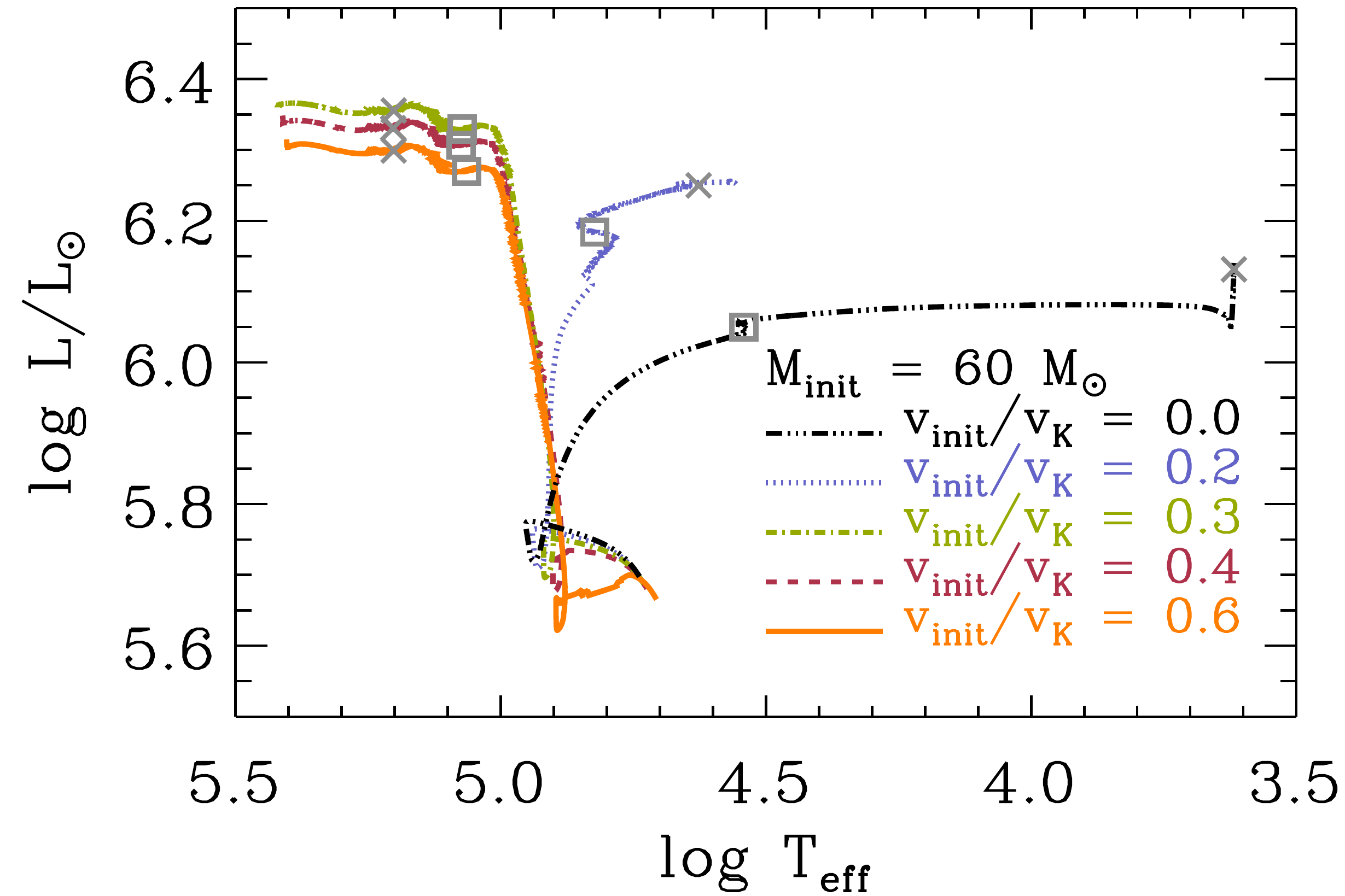}
\includegraphics[width=\columnwidth]{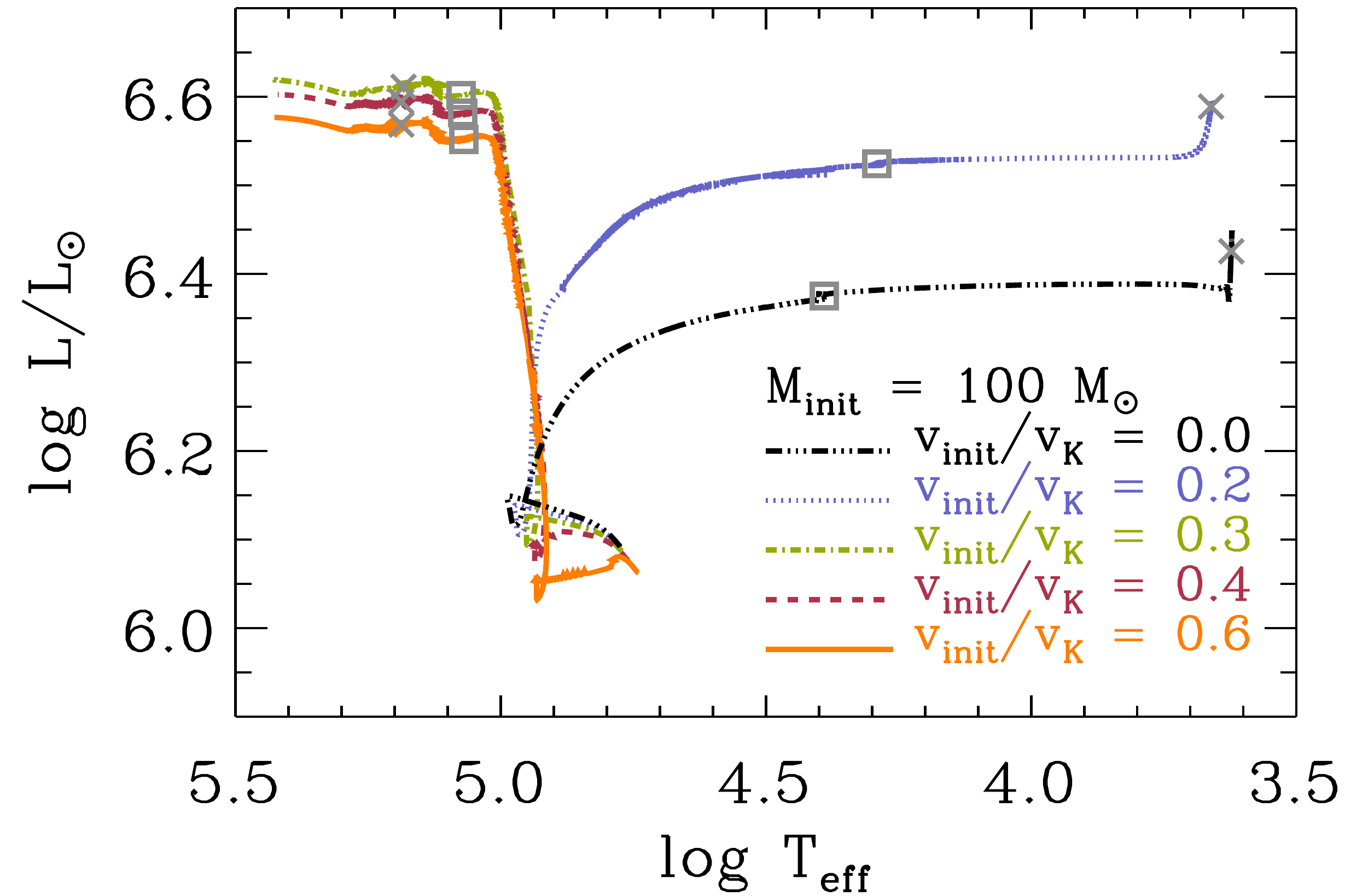}
\caption{
Evolution of the models in the HR diagram
for 10, 15, 20, 30, 60 and 100~\Msun. 
Evolutionary tracks for different initial 
rotational velocities at a given initial mass are given 
in each panel as indicated by the labels. 
The end points of core hydrogen burning and core helium burning
are marked by an square and a cross,
respectively, on each evolutionary track.}
\label{fig:hr1}
\end{figure*}

\begin{figure*}
\centering
\includegraphics[width=\columnwidth]{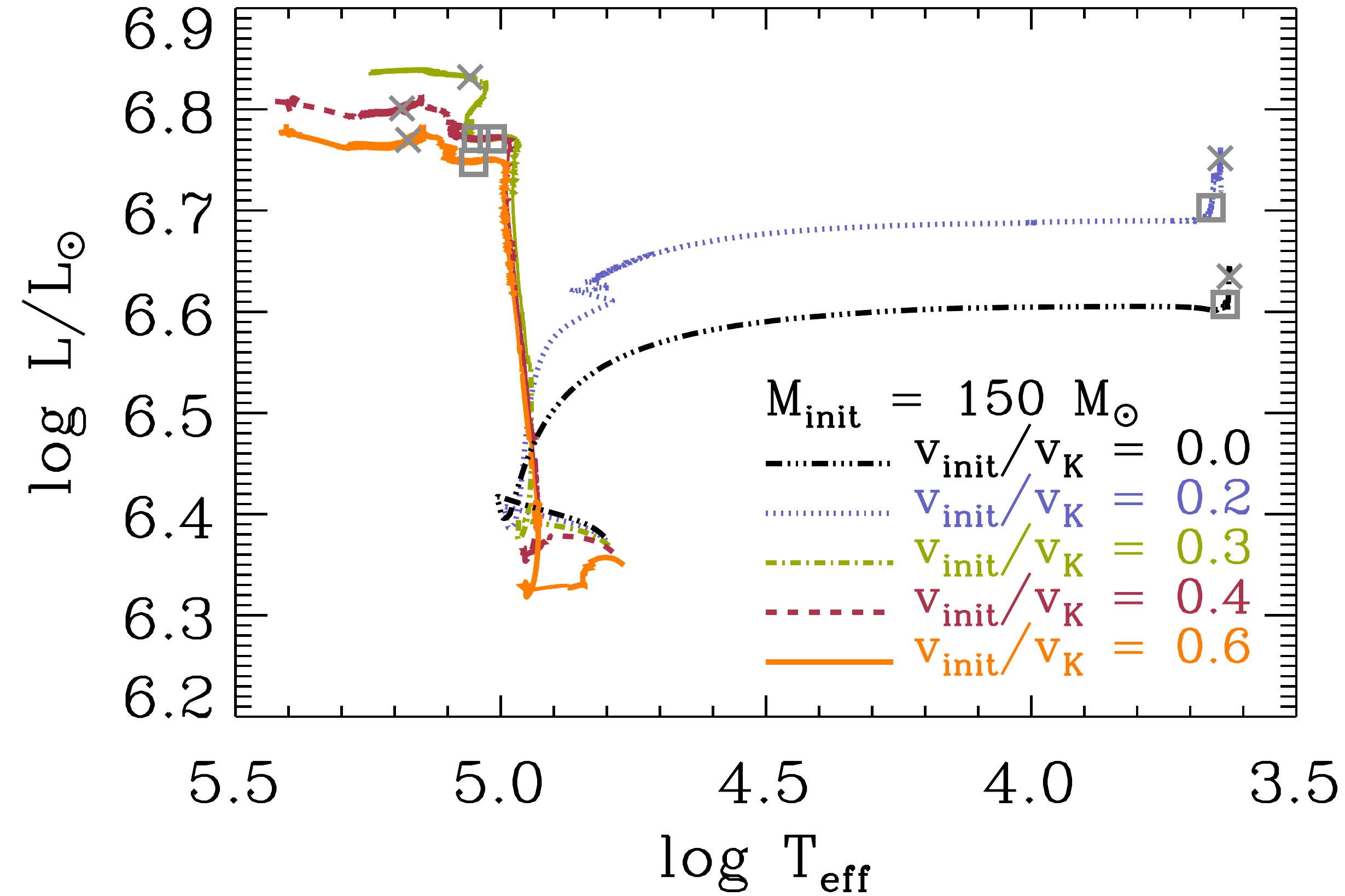}
\includegraphics[width=\columnwidth]{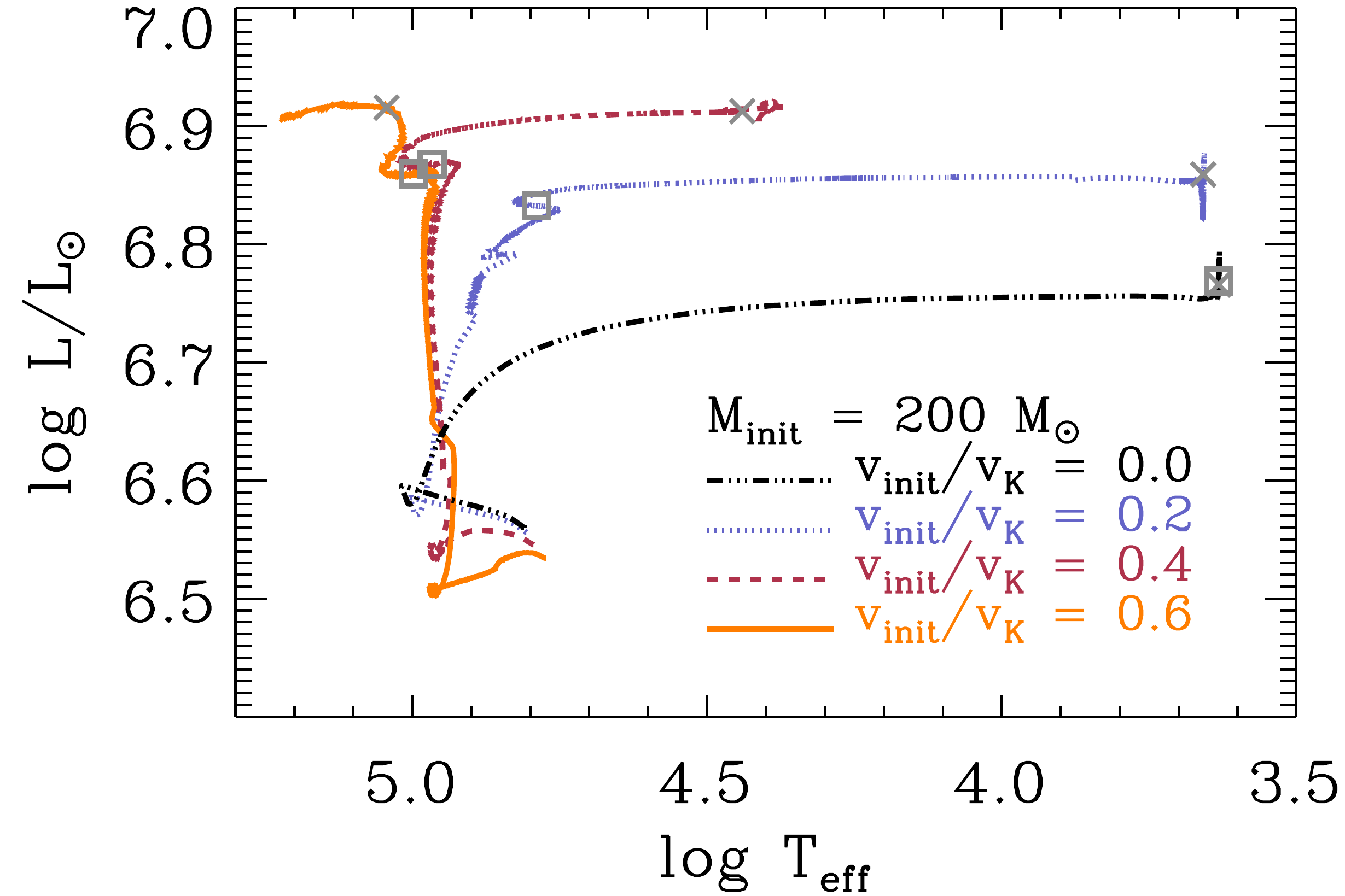}
\includegraphics[width=\columnwidth]{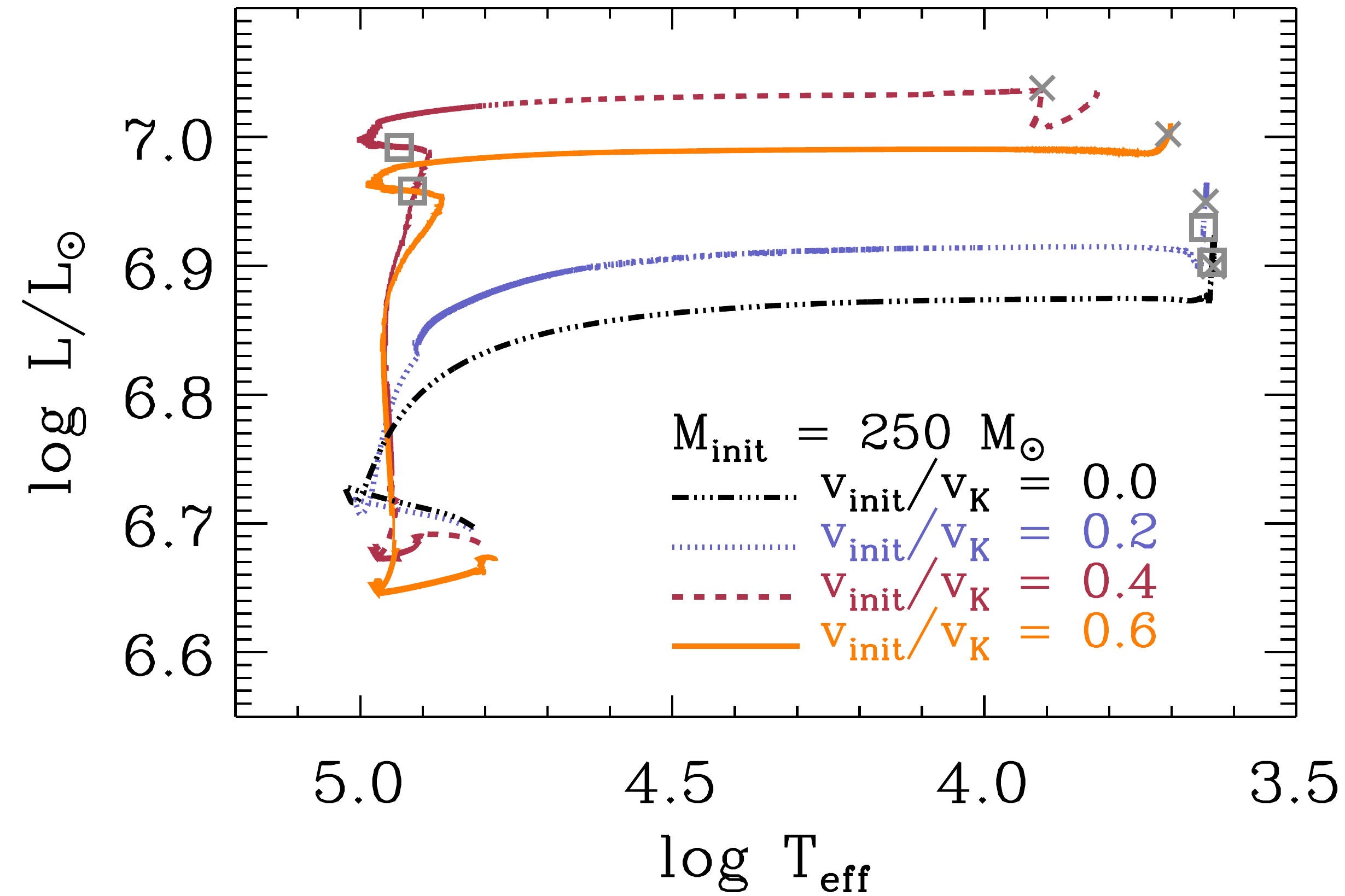}
\includegraphics[width=\columnwidth]{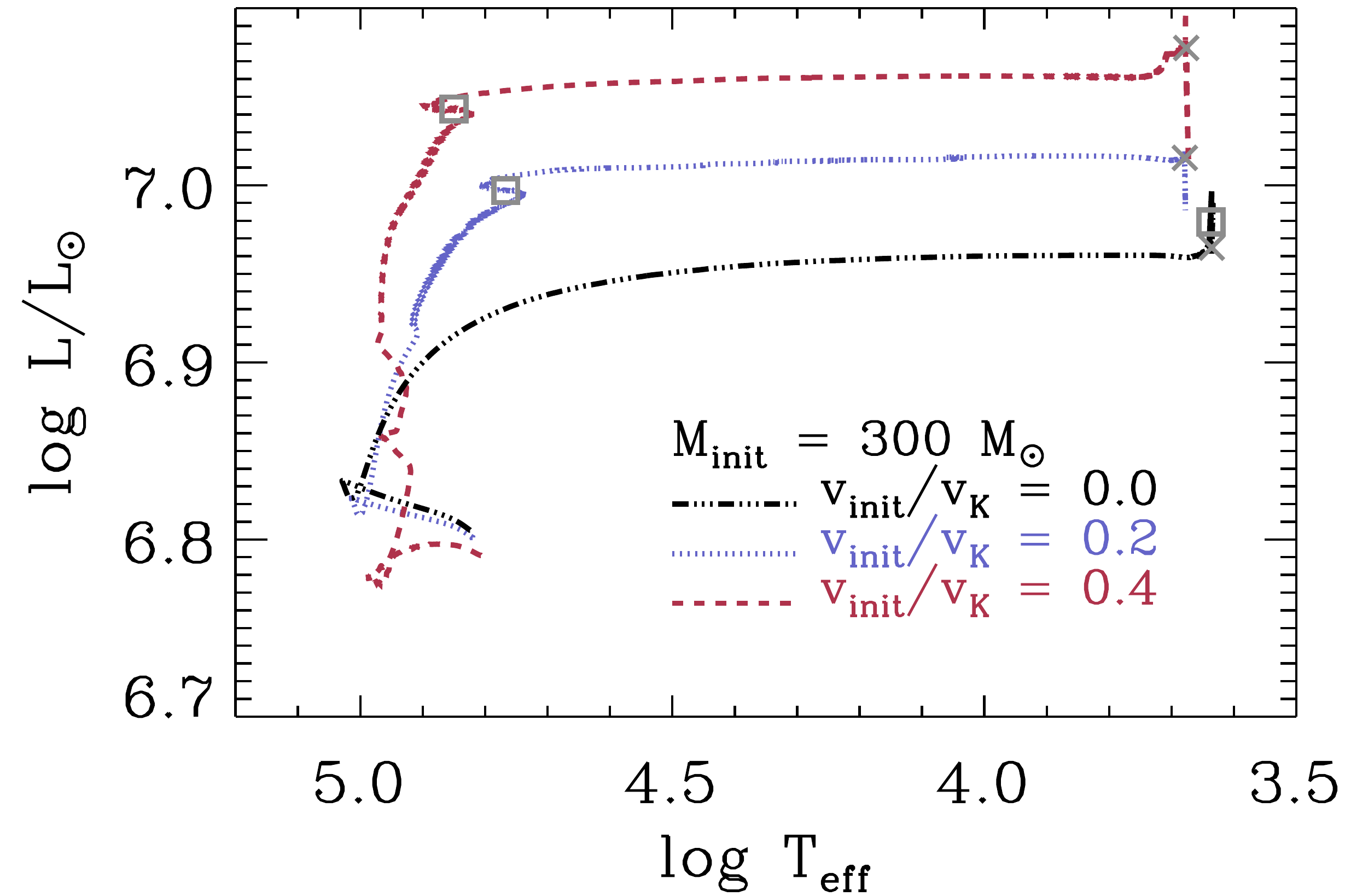}
\includegraphics[width=\columnwidth]{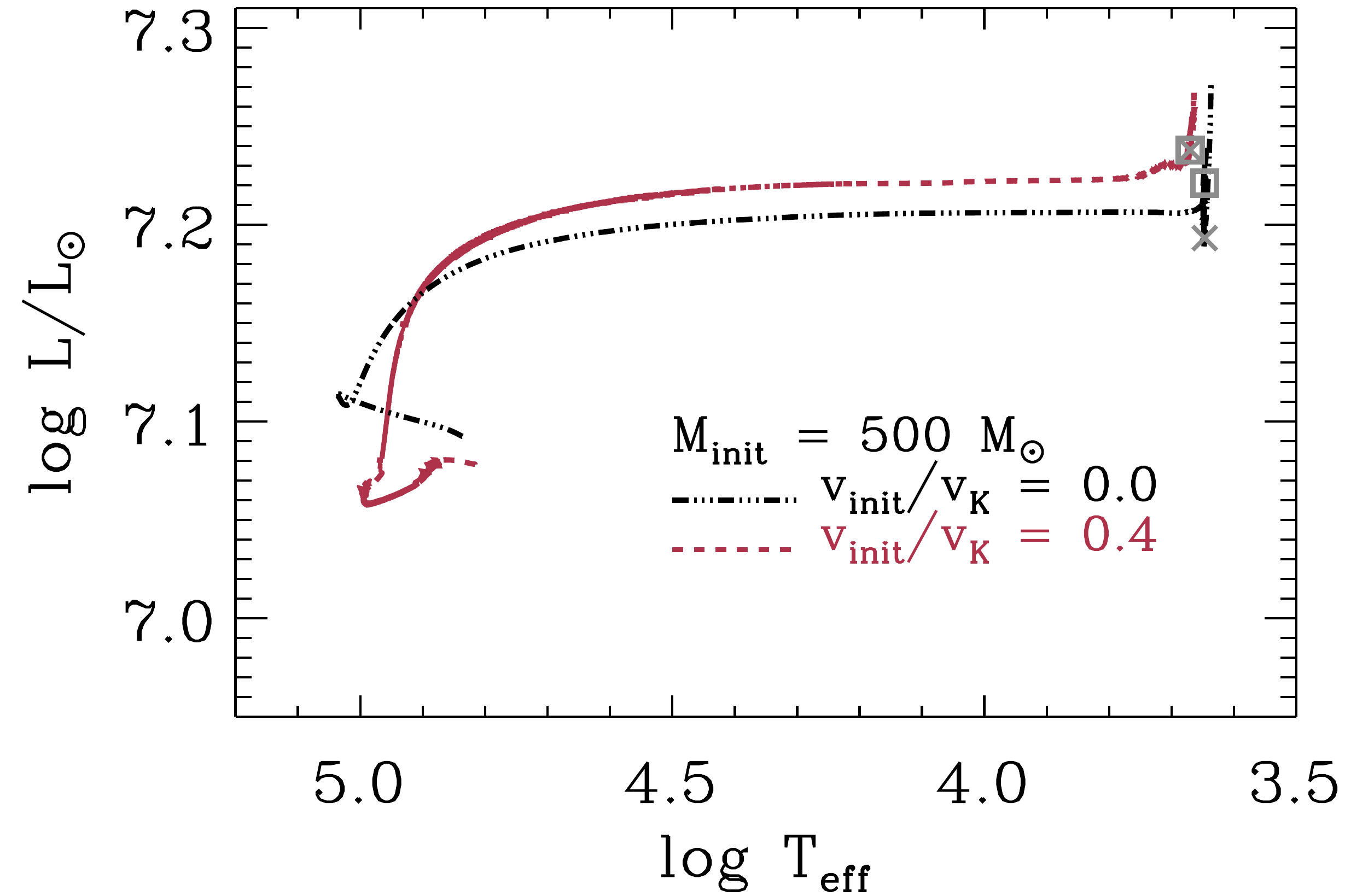}
\includegraphics[width=\columnwidth]{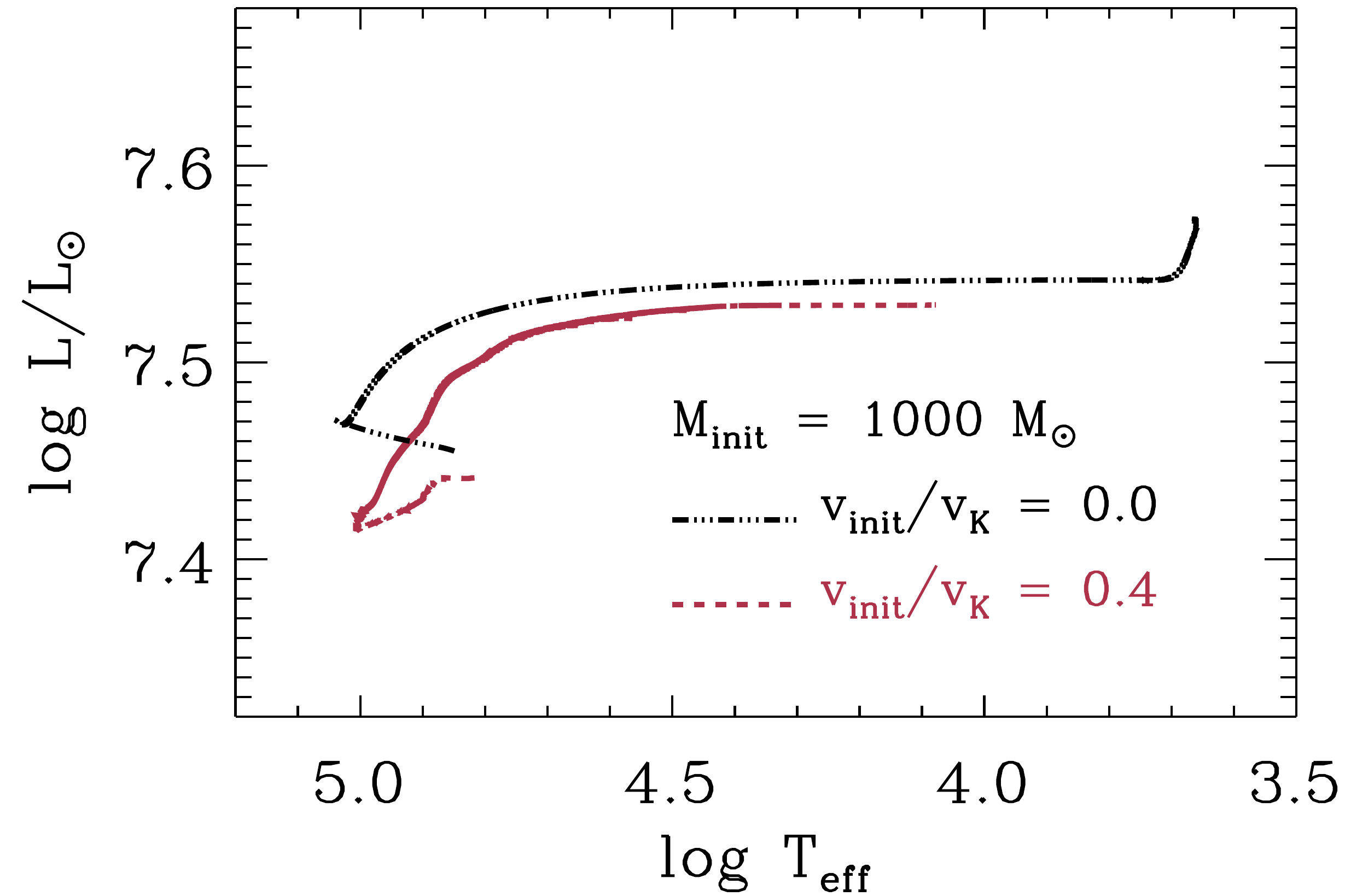}
\caption{
Same as in figure~\ref{fig:hr1}, but 
for 150, 200, 250, 300, 500, and 1000~\Msun.}
\label{fig:hr2}
\end{figure*}

\subsection{Mass loss}\label{sect:wind}

\citet{Krticka06} found that mass loss due to stellar winds from metal-free hot
massive stars is practically negligible.  A very low mass loss rate of
$10^{-14}~M_\odot~\mathrm{yr^{-1}}$ is predicted only when the star reaches the
Eddington limit.  Therefore, we considered stellar wind mass loss only if there
is the surface enrichment in CNO elements, as the following.  For hydrogen-rich
stars, for which the surface hydrogen mass fraction  ($X_\mathrm{s}(H)$) is greater 
than 0.3, we used the mass loss rates of \citet{Kudritzki89} if $T_\mathrm{s} >
10^4$~K and \citet{Nieuwenhuijzen90} otherwise, with a metallicity dependence
of $Z^{0.69}$.  On the other hand, \citet{Vink05} found that the CNO lines play
the dominant role for driving winds from helium-rich stars  (i.e.,
$X_\mathrm{s}(H) < 0.3$)  for $Z  < 10^{-3} Z_\odot$, while the iron lines become
important for higher metallicities. In our models,  therefore,  the mass loss
enhancement caused by the surface enrichment of CNO elements for helium-rich stars
is considered only up to $Z = 10^{-5}$, following the prescription by
\citet{Nugis00}, which has a metallicity dependence of $ Z^{0.5}$.  

Some cautionary remarks should be made here.  The recent studies by
\citet{Krticka09} and \citet{Muijres11} show that the effect of the  surface
enrichment of CNO elements  on the mass loss enhancement from very metal-poor
stars is only moderate.  This means that the above prescription on the
radiation-driven wind may \emph{overestimate} the mass loss rates in our
metal-free massive star models.  In particular, CNO elements are highly ionized
for very hot stars with surface temperatures higher than 50000~K, providing too
few efficient lines for driving winds. As we will see below,  the helium-rich
stars resulting from the chemically homogeneous evolution in our model grid
have surface temperatures significantly higher than 50000 K. 

However, this uncertainty should not have any significant consequence for the
conclusions of the present study, because the mass loss history in all our
rotating model sequences is dominated by the mechanical winds that occur at the
so-called $\omega\Gamma$ limit,  which is the critical limit
where the effective gravity at the stellar surface becomes zero because of  the
radiation pressure \emph{and} the centrifugal force ~\citep{Langer97, Maeder00a}.  
This effect is treated
as in the following equation: 
\begin{equation}\label{eq1}  
\dot{M}(v) =
\mathrm{max}\left[10^{-14}~M_\odot~\mathrm{yr^{-1}},
\dot{M}(v=0)\right] \left(\frac{1}{1-\Omega}\right)^{0.43}~,
\end{equation}
where
\begin{equation}\label{eq2}  
\Omega :=
\frac{v}{v_\mathrm{crit}}~~\mathrm{and}~~v_\mathrm{crit} =
\sqrt{\frac{GM}{R}(1-\Gamma)}~.  
\end{equation}
Here, $v$ denotes the rotational velocity
at the equatorial surface,  $\dot{M}(v=0)$ the mass loss rate
given by our adopted prescription for non-rotating stars that is described above, 
and $\Gamma$ the Eddington factor.  
 
The enhancement of the mass loss rate when the star approaches the  $\omega\Gamma$ limit is
reflected in the term $(1/(1-\Omega))^{0.43}$. 
We evaluate $\Gamma$ using the radiative opacity for the outermost layers
where the optical depth is lower than 100. 
Because a singularity can occur at $\Omega =
1$ according to Eq.~\ref{eq1}, we limit the mass loss rate such that the
mass loss time scale is longer than the thermal time scale of the star
as in \citet{Yoon10b}: 
\begin{equation}\label{eq3} 
\dot{M} = \mathrm{min}\left[ \dot{M}(v),
0.3\frac{M}{\tau_\mathrm{KH}} \right]~.  
\end{equation}
where $\tau_\mathrm{KH}$ is the Kelvin-Helmoltz time scale, and
$\dot{M}(v)$ is given by Eq.~\ref{eq1}. 

The wind material that is mechanically ejected at the $\omega \Gamma$ limit may
form a decretion disk.  Recently, \citet{Krticka11} argued that viscous
coupling between the star and the decretion disk may significantly reduce the
mass loss rate. This might potentially alter the final masses of the rotating
models, but its detailed investigation is beyond the scope of the present
study.

\subsection{Initial conditions and model grid}\label{sect:grid}

The initial conditions and some important properties of the evolutionary models
are summarized in Tables~\ref{tab1} and ~\ref{tab:properties}. We consider
twelve different initial masses ($M_\mathrm{init} =$ 10, 15, 20, 30, 60, 100,
150, 200, 250, 300, 500, and 1000~\Msun).  Several different initial rotational
velocities were considered at each mass, from zero to 80\% of the Keplerian
value ($v_\mathrm{K}$) at the equatorial surface.  Solid-body rotation
was assumed for the initial velocity profile. The initial mass and the
initial rotational velocity are denoted by the model sequence number. For
example, the sequence ``m100vk04" indicates $M_\mathrm{init} = 100~M_\odot$ and
$v_\mathrm{init}/v_\mathrm{K} = 0.4$. 

In metal-free massive stars, the CNO cycle cannot be activated initially.  Because 
the energy generation due to the $pp$ chain is too weak to support a massive
star with $M \ge 20~M_\odot$  for a significant fraction
of the evolutionary time, the stellar core rapidly contracts until
enough carbon (i.e., $X(C) \sim 10^{-10}$) is produced by helium
burning at $T_\mathrm{c} \sim 10^{8}$ K.  Hydrogen burning with the CNO cycle
only begins thereafter.  This makes it difficult to make a relaxed model on the
zero-age main sequence for Pop III stars.  For this reason, we included a small
amount of $^{3}$He for the initial chemical composition,  which may result
from deuterium burning,  such that
$X(\mathrm{H}) = 0.76$, $X(^{4}\mathrm{He}) = 0.23999$ and $X(^{3}\mathrm{He})
= 0.00001$.  In this way, all our initially relaxed models are supported  by
the energy generation due to $^{3}$He burning, and  the radii of our
initial models are  much larger than those obtained when the main sequence
begins.  Although this initial configuration looks artificial, it is not
unrealistic given that the first stars should undergo rapid mass accretion
phase that leads to significant expansion of the envelope before they reach
the zero-age main sequence (ZAMS) ~\citep{Omukai03, Ohkubo09}. 

In Table~\ref{tab1}, we provide the information for the  mass and rotational
velocity when the stars arrive on the ZAMS, as well as their initial values.
Given the efficient transport of angular momentum
by the Spruit-Tayler dynamo, near solid-body rotation is maintained 
during this initial contraction phase.  
It should be noted that \citet{Ekstroem08} determined rotational velocities
only when the stars are close to the ZAMS, and thus our ZAMS values should be
compared with theirs, and not with our initial values.


\begin{sidewaystable*}
\begin{minipage}[t][ 0mm]{\textwidth}
\caption{Initial conditions for the rotating model sequences, and the corresponding values
on the zero-age main sequence (ZAMS).  
}\label{tab1}
\centering
\begin{tabularx}{\linewidth}
{c | >{\centering\arraybackslash}X >{\centering\arraybackslash}X >{\centering\arraybackslash}X >{\centering\arraybackslash}X >{\centering\arraybackslash}X %
     >{\centering\arraybackslash}X | >{\centering\arraybackslash}X >{\centering\arraybackslash}X >{\centering\arraybackslash}X >{\centering\arraybackslash}X %
      >{\centering\arraybackslash}X  >{\centering\arraybackslash}X >{\centering\arraybackslash}X}
\hline
        & $M_\mathrm{i}$    & $v_\mathrm{init}/v_\mathrm{K}$ & $v_\mathrm{init}/v_\mathrm{crit}$ & $v_\mathrm{init}$ & $R_\mathrm{init}$ & $J_\mathrm{init}$  %
        & $M_\mathrm{ZAMS}$ & $v_\mathrm{ZAMS}/v_\mathrm{K}$ & $v_\mathrm{ZAMS}/v_\mathrm{crit}$ & $v_\mathrm{ZAMS}$ &$R_\mathrm{ZAMS}$  & $J_\mathrm{ZAMS}$  \\
        & [$M_\odot$]       &                &                 & [$\mathrm{km~s^{-1}}$] &  [$R_\odot$]  & [$\mathrm{erg~s^{-1}}$] %
        & [$M_\odot$]       &                &                 & [$\mathrm{km~s^{-1}}$] &  [$R_\odot$]  & [$\mathrm{erg~s^{-1}}$] & Mode \\
\hline
m10vk02 &  10.0          & 0.20                     &  0.21                               &   159.8  &    2.94        & $6.03\times10^{51}$   %
        &  10.0          & 0.29                     &  0.31                               &   325.5  &    1.49        & $6.03\times10^{51}$ & NE \\
m10vk04 &  10.0          & 0.40                     &  0.42                               &   314.9  &    3.03        & $1.16\times10^{52}$  %
        &  10.0          & 0.61                     &  0.65                               &   662.7  &    1.63        & $1.16\times10^{52}$ & NE \\
m10vk06 &  10.0          & 0.60                     &  0.63                               &   459.1  &    3.20        & $1.61\times10^{52}$  %
        &  10.0          & 0.90                     &  0.95                               &   933.6  &    1.77        & $1.56\times10^{52}$ & NE \\
m10vk08 &  10.0          & 0.80                     &  0.84                               &   596.8  &    3.36        & $2.00\times10^{52}$  %
        &   9.6          & 0.90                     &  0.94                               &   918.2  &    1.77        & $1.48\times10^{52}$  & NE \\
\hline
m15vk02 &  15.0          & 0.20                     &  0.22                               &   173.1  &    3.75        & $1.29\times10^{52}$  %
        &  15.0          & 0.32                     &  0.35                               &   422.5  &    1.61        & $1.29\times10^{52}$  & NE\\
m15vk04 &  15.0          & 0.40                     &  0.43                               &   341.1  &    3.86        & $2.49\times10^{52}$  %
        &  15.0          & 0.69                     &  0.75                               &   869.9  &    1.81        & $2.49\times10^{52}$  & NE \\
m15vk06 &  15.0          & 0.60                     &  0.64                               &   496.8  &    4.08        & $3.45\times10^{52}$  %
        &  14.6          & 0.88                     &  0.95                               &  1064.9  &    1.90        & $2.89\times10^{52}$  & CHE \\
\hline
m20vk02 &  20.0          & 0.20                     &  0.22                               &   184.0  &    4.42        & $2.21\times10^{52}$  %
        &  20.0          & 0.34                     &  0.37                               &   497.8  &    1.74        & $2.21\times10^{52}$  &NE \\
m20vk03 &  20.0          & 0.30                     &  0.33                               &   274.3  &    4.48        & $3.26\times10^{52}$  %
        &  20.0          & 0.53                     &  0.59                               &   764.4  &    1.84        & $3.26\times10^{52}$ & NE \\
m20vk04 &  20.0          & 0.40                     &  0.44                               &   362.3  &    4.56        & $4.25\times10^{52}$  %
        &  20.0          & 0.75                     &  0.83                               &  1044.5  &    1.97        & $4.25\times10^{52}$ & CHE \\
m20vk06 &  20.0          & 0.60                     &  0.66                               &   527.3  &    4.82        & $5.88\times10^{52}$  %
        &  19.3          & 0.86                     &  0.94                               &  1160.1  &    2.00        & $4.47\times10^{52}$ & CHE \\
\hline
m30vk02 &  30.0          & 0.20                     &  0.24                               &   202.5  &    5.53        & $4.66\times10^{52}$  %
        &  30.0          & 0.34                     &  0.39                               &   554.6  &    2.15        & $4.66\times10^{52}$  & NE \\
m30vk03 &  30.0          & 0.30                     &  0.35                               &   300.5  &    5.59        & $6.87\times10^{52}$  %
        &  30.0          & 0.53                     &  0.61                               &   844.4  &    2.27        & $6.87\times10^{52}$  & CHE \\
m30vk04 &  30.0          & 0.40                     &  0.47                               &   396.7  &    5.70        & $8.93\times10^{52}$  %
        &  30.0          & 0.75                     &  0.86                               &   1158.0 &    2.42        & $8.93\times10^{52}$  & CHE\\
m30vk06 &  30.0          & 0.60                     &  0.69                               &   576.7  &    6.04        & $1.24\times10^{53}$  %
        &  28.5          & 0.78                     &  0.88                               &   1180.6 &    2.36        & $8.33\times10^{52}$  & CHE\\
\hline
m60vk02 &  60.0          & 0.20                     &  0.27                               &   237.6  &    7.95        & $1.62\times10^{53}$  %
        &  60.0          & 0.34                     &  0.42                               &   631.3  &    3.26        & $1.62\times10^{53}$  & NE \\
m60vk03 &  60.0          & 0.30                     &  0.39                               &   353.8  &    8.05        & $2.38\times10^{53}$  %
        &  60.0          & 0.53                     &  0.66                               &   964.6  &    3.43        & $2.38\times10^{53}$  & CHE\\
m60vk04 &  60.0          & 0.40                     &  0.52                               &   466.4  &    8.22        & $3.09\times10^{53}$  %
        &  59.4          & 0.70                     &  0.86                               &  1236.2  &    3.63        & $2.89\times10^{53}$  & CHE\\
m60vk06 &  60.0          & 0.60                     &  0.76                               &   675.3  &    8.78        & $4.25\times10^{53}$  %
        &  55.9          & 0.72                     &  0.88                               &  1260.3  &    3.51        & $2.67\times10^{53}$  & CHE\\
\hline
m100vk02 & 100.0          & 0.20                     &  0.29                              &   268.3  &   10.36        & $3.96\times10^{53}$  %
         & 100.0          & 0.34                     &  0.47                              &   704.0  &    4.37        & $3.96\times10^{53}$  & NE\\
m100vk03 & 100.0          & 0.30                     &  0.43                              &   390.1  &   10.52        & $5.82\times10^{53}$  %
         & 100.0          & 0.52                     &  0.70                              &  1061.7  &    4.66        & $5.82\times10^{53}$  & CHE\\
m100vk04 & 100.0          & 0.40                     &  0.57                              &   520.1  &   11.00        & $7.55\times10^{53}$  %
         &  98.4          & 0.67                     &  0.89                              &  1309.3  &    4.88        & $6.68\times10^{53}$  & CHE\\
m100vk06 & 100.0          & 0.60                     &  0.82                              &   750.4  &   11.79        & $1.03\times10^{54}$  %
         &  93.2          & 0.68                     &  0.90                              &  1319.4  &    4.73        & $6.13\times10^{53}$  & CHE\\
\hline
m150vk02 & 150.0          & 0.20                     &   0.31                             &   295.1  &   12.83        & $7.97\times10^{53}$  %
         & 150.0          & 0.36                     &   0.54                             &   820.5  &    5.48        & $7.97\times10^{53}$  & NE \\
m150vk03 & 150.0          & 0.30                     &   0.47                             &   438.5  &   13.05        & $1.17\times10^{54}$  %
         & 150.0          & 0.53                     &   0.77                             &  1169.0  &    5.97        & $1.17\times10^{54}$  & CHE\\
m150vk04 & 150.0          & 0.40                     &   0.61                             &   576.3  &   13.39        & $1.51\times10^{54}$  %
         & 146.8          & 0.72                     &   1.00                             &  1559.2  &    6.00        & $1.29\times10^{54}$  & CHE\\
m150vk06 & 150.0          & 0.60                     &   0.88                             &   825.8  &   14.53        & $2.03\times10^{54}$  %
         & 139.2          & 0.72                     &   1.00                             &  1545.2  &    5.84        & $1.18\times10^{54}$  & CHE\\
\hline
m200vk02 & 200.0          & 0.20                     &   0.34                             &   310.8  &   15.40        & $1.30\times10^{54}$  %
         & 200.0          & 0.36                     &   0.57                             &   871.1  &    6.41        & $1.30\times10^{54}$  & NE \\
m200vk04 & 200.0          & 0.40                     &   0.64                             &   606.4  &   16.09        & $2.45\times10^{54}$  %
         & 194.6          & 0.65                     &   0.98                             &  1489.2  &    7.13        & $2.00\times10^{54}$  & TE\\
m200vk06 & 200.0          & 0.60                     &   0.92                             &   866.4  &   17.52        & $3.28\times10^{54}$  %
         & 185.5          & 0.66                     &   0.98                             &  1497.1  &    6.80        & $1.91\times10^{54}$  & CHE\\
\hline
m250vk02 & 250.0          & 0.20                     &   0.35                             &   332.0  &   16.85        & $1.90\times10^{54}$  %
         & 250.0          & 0.36                     &   0.60                             &   911.0  &    7.30        & $1.90\times10^{54}$  & NE \\
m250vk04 & 250.0          & 0.40                     &  0.68                              &   645.3  &   17.73        & $3.58\times10^{54}$  %
         & 242.6          & 0.65                     &  1.00                              &  1568.5  &    7.97        & $2.80\times10^{54}$  & TE\\
m250vk06 & 250.0          & 0.60                     &  0.95                              &   916.5  &   19.48        & $4.72\times10^{54}$  %
         & 232.4          & 0.66                     &  1.00                              &  1571.7  &    7.71        & $2.63\times10^{54}$  & TE\\
\hline
m300vk02 & 300.0          & 0.20                     &   0.37                             &   340.4  &   19.20        & $2.59\times10^{54}$  %
         & 300.0          & 0.36                     &   0.62                             &   948.5  &    8.07        & $2.59\times10^{54}$  & NE\\
m300vk04 & 300.0          & 0.40                     &   0.70                             &   662.0  &   20.18        & $4.85\times10^{54}$  %
         & 290.5          & 0.62                     &   1.00                             &  1570.4  &    8.75        & $3.82\times10^{54}$  & NE\\
\hline
m500vk04 & 500.0          & 0.40                     &   0.78                             &   738.3  &   26.90        & $1.13\times10^{55}$  %
         & 480.6          & 0.57                     &   1.00                             &  1600.6  &   11.44        & $8.08\times10^{54}$  & NE\\
\hline
m1000vk04& 1000.0         & 0.40                     &    0.89                            &   850.8  &   39.89        & $3.50\times10^{55}$  %
         & 956.4          & 0.48                     &    1.00                            &  1597.7  &   16.41        & $2.31\times10^{54}$  & NE\\
\hline
\end{tabularx}
\tablefoot{   
Each column has the following meaning.  $M_\mathrm{i}$: initial mass, $v_\mathrm{init}/v_\mathrm{K}$: initial rotational velocity
at the equatorial surface in units of the Keplerian value, $v_\mathrm{init}/v_\mathrm{crit}$: initial rotational velocity
at the equatorial surface in units of the critical velocity (Eq.~\ref{eq2}), 
$v_\mathrm{init}$: initial rotational velocity in units of $\mathrm{km~s^{-1}}$, 
$J_\mathrm{init}$: initial total angular momentum.  
The corresponding values on the ZAMS are indicated by the subscript ``ZAMS''. 
In the last column, the different modes of evolution are indicated by NE (normal evolution), TE (transitionary evolution) and CHE (chemically homogeneous evolution), 
as defined in Sect.~\ref{sect:evolution}. 
}
\end{minipage}
\end{sidewaystable*}

\section{Basic results}\label{sect:evolution}

\subsection{Evolution on the Hertzsprung-Russell diagram}

Figs.~\ref{fig:hr1} and~\ref{fig:hr2} show the evolutionary tracks of all of
our models in the Hertzsprung-Russell (HR) diagram.  As mentioned above, our
initially relaxed models are supported by $^{3}$He burning.  Once the evolution
starts, $^{3}$He  is very quickly destroyed and the stars undergo rapid
contraction until they reach the ZAMS.  The time scale of this initial
contraction phase ranges from $10^4$ to $3\times10^5$~yr, depending on the
initial mass.  The models with $M_\mathrm{init} =$ 10 and 15~\Msun{} are powered
by $pp$-chains for a significant fraction of the main sequence lifetime (13.3
Myr and 2 Myr, respectively, for non-rotating models).  Thereafter, the CNO
cycle becomes dominant, which is marked by the smooth turn-over toward lower
temperature in the HR diagram.  For $M_\mathrm{init} \ge $ 20~\Msun, the stars
are already supported by the CNO cycle on the ZAMS, which 
agrees well with the previous work~\citep{Marigo01, Ekstroem08}.

Compared to the models in the previous work, our non-rotating models become
redder in the HR diagram during the post-main sequence phases. As an example,
the surface temperature of the non-rotating 15 \Msun{} models becomes $\log
T_\mathrm{eff} =$ 4.3 and 4.5 in Marigo et al. and  Ekstr\"om et al.,
respectively, while it becomes as low as $\log T_\mathrm{eff} = 3.9$ in our
case.  In addition, our non-rotating models with $M \ge 150$~\Msun{} become
red-supergiant stars already during core hydrogen burning, while $M >
250$~\Msun{} was needed in Marigo et al. for this to happen.  This must result
mainly from the fact that our adopted overshooting parameter is larger: $\sim$
0.3 $H_\mathrm{P}$ and 0.2 $H_\mathrm{P}$ are used in Marigo et al.  and
Ekstr\"om et al., respectively, while 0.335 $H_\mathrm{P}$ in the present study.
The larger overshooting parameter also has an impact on the size of the stellar
core and the evolutionary time.  For example, the main sequence lifetime
($\tau_\mathrm{H}$) and the helium core size ($M_\mathrm{He-Core}$) of our
non-rotating 200~\Msun{} model are 2.40 Myr and 114~\Msun{}, respectively, (see
Table~\ref{tab:properties}), compared to 2.27 Myr and 102~\Msun{} in Ekstr\"om et al.

\begin{figure}
\begin{center}
\includegraphics[width=\columnwidth]{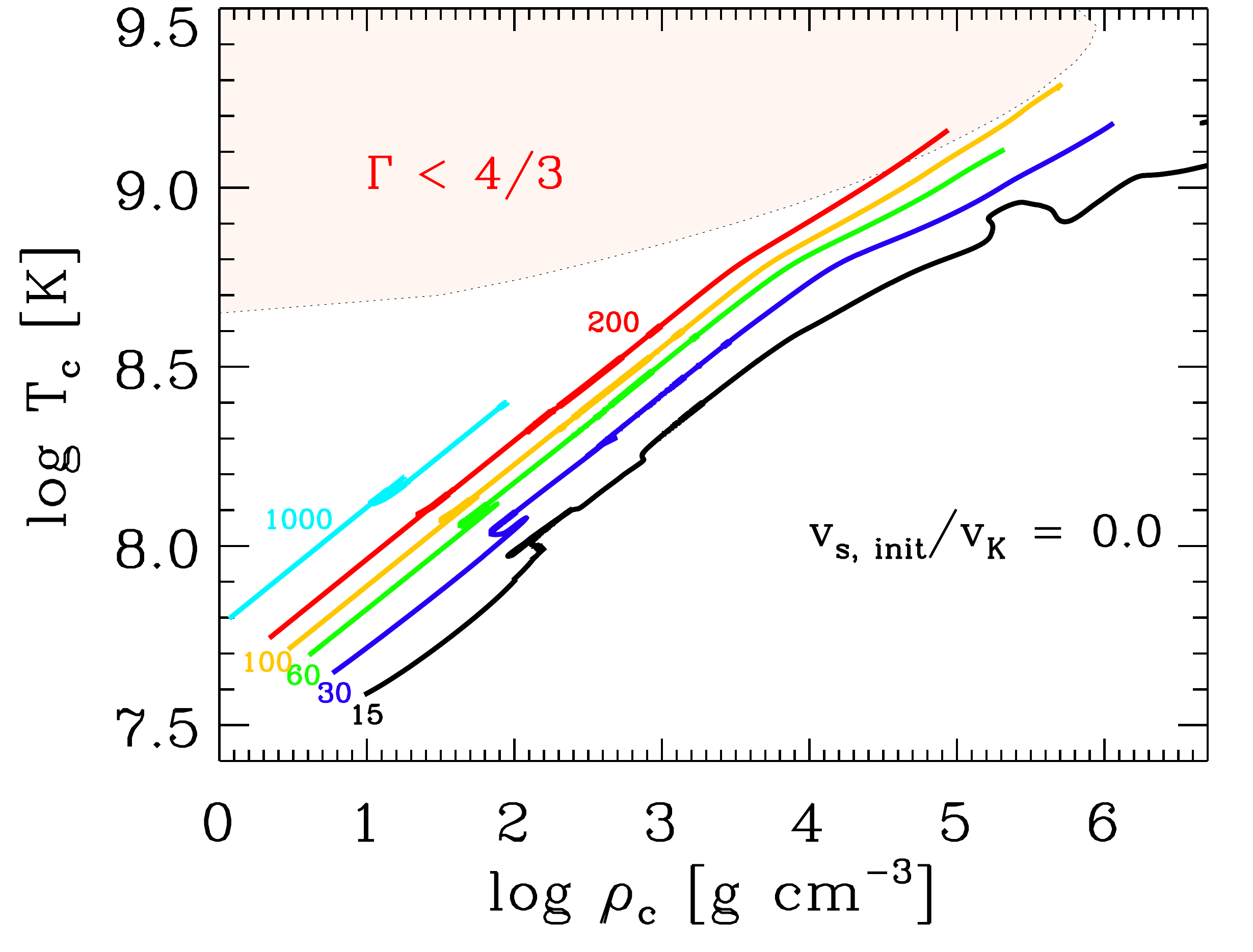}
\includegraphics[width=\columnwidth]{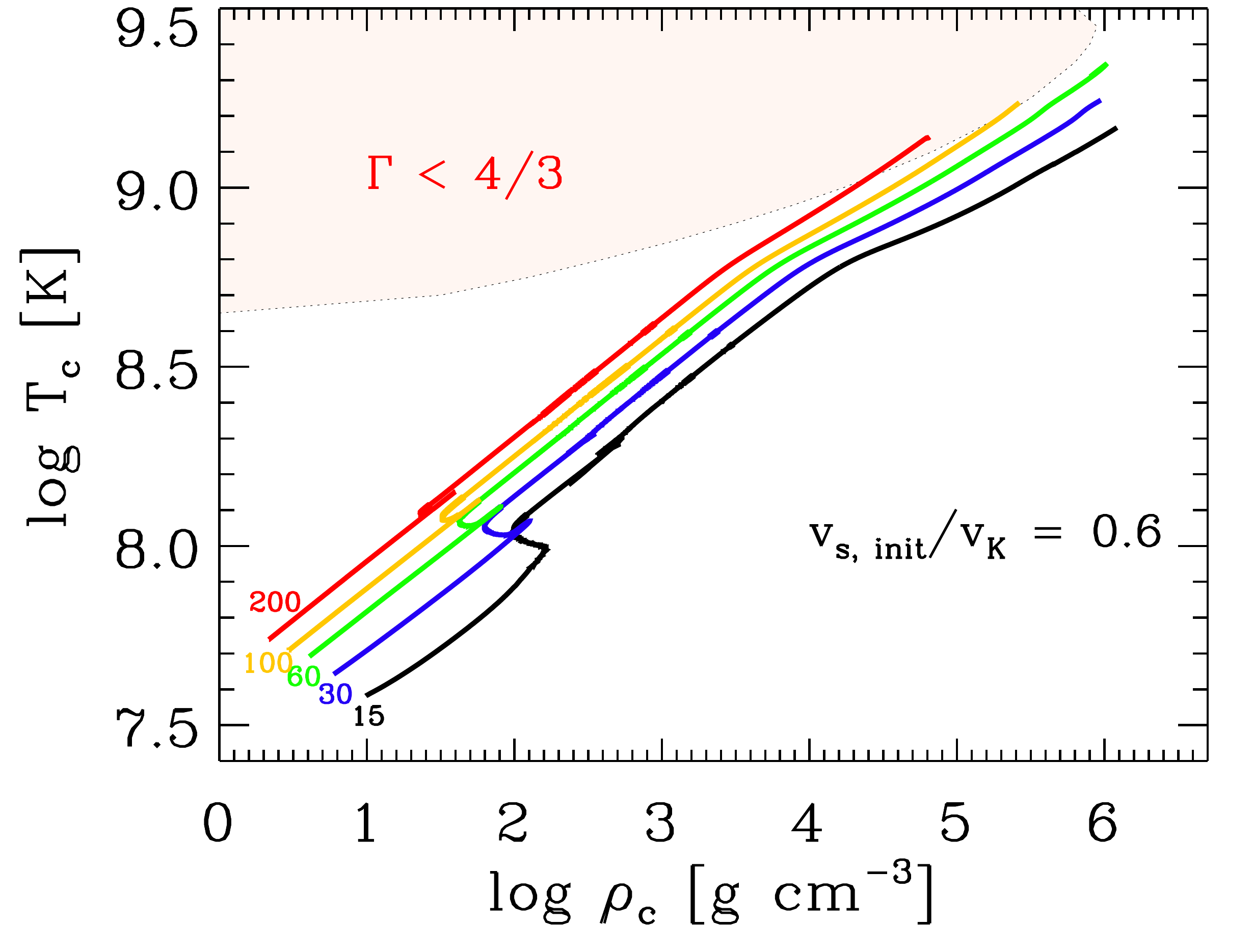}
\caption{Evolution of the central temperature and density in some model
sequences for $v_\mathrm{init}/v_\mathrm{K} = 0$ (upper panel)
and   $v_\mathrm{init}/v_\mathrm{K} = 0.6$ (lower panel). The initial
masses are marked by the labels in the figures. 
The regime for the pair-instability  is marked by pink shading}
\label{fig:rhot}
\end{center}
\end{figure}

Rotating models start at lower surface temperature and luminosity in the HR
diagram for larger $v_\mathrm{init}/v_\mathrm{K}$, reflecting the effect of
the centrifugal force on the stellar structure.  Rotationally induced mixing,
on the other hand, tends to increase the size of the convective core and the
surface helium abundance.  The latter effect becomes dominant as the star
evolves,  leading to higher luminosity and higher surface temperature than those
in the corresponding non-rotating models, for  most of the time on the main
sequence.  The evolutionary time also becomes systematically 
longer for higher initial rotational velocity because of the larger convective core. 
The central temperature 
is initially lower for  higher initial rotational velocity  because of the
effect of the centrifugal force,  but becomes higher already in the early phase
of  core hydrogen burning  than those in a corresponding lower velocity model, as the
convective core becomes larger in the rotating models  due to rotationally
induced chemical mixing.  For instance, the center of the 100~\Msun{} star
enters the pair instability regime for $v_\mathrm{init}/v_\mathrm{K} = 0.6$,
while it does not in the corresponding non-rotating sequence (see
Fig.~\ref{fig:rhot}).  The final fate of such a massive star can be
significantly influenced in this way, as discussed in Sect.~\ref{sect:fate}. 

One of the most dramatic effects of rotationally induced chemical mixing may be
the so-called chemically homogeneous evolution \citep[CHE;][]{Maeder89}, which
can be observed with the blueward evolutionary tracks in Figures~\ref{fig:hr1}
and~\ref{fig:hr2}.  We find that  CHE can occur for a certain range of the
initial mass ($15 \le M_\mathrm{init} \le 200$~\Msun) if the initial rotational
velocity is sufficiently high.  For some cases, the star follows the CHE track
for most of the main sequence phase but afterwards moves rapidly to the right
in the HR diagram, in particular during the post-main sequence phases. Such
transitionary evolution is found in Seqs.~m200vk04, m250vk04 and m250vk06, for
example.  In the present study, we define these three different evolutionary
patterns, i.e., normal evolution (NE), CHE, and transitionary evolution (TE), 
according to the following criteria.  
\begin{itemize}
\item NE: The mass fraction of helium at the surface ($Y_\mathrm{s}$) remains
smaller than 0.7 until the end of the main sequence. The star generally
moves redwards in the HR diagram throughout the evolution.  
\item CHE:
$Y_\mathrm{s}$ becomes larger than 0.8 by the end of the main sequence.  The
star generally moves bluewards throughout the evolution.  
\item TE:
$Y_\mathrm{s}$ becomes larger than 0.7 by the end of the main sequence.  The
star follows the CHE track until the end of the main sequence, but
moves redwards during the post-main sequence phases.    
\end{itemize} 
We discuss the conditions for CHE in more detail below
(Sect.~\ref{sect:che}).

\subsection{Internal structure}

\begin{figure*}
\begin{center}
\includegraphics[width=\columnwidth]{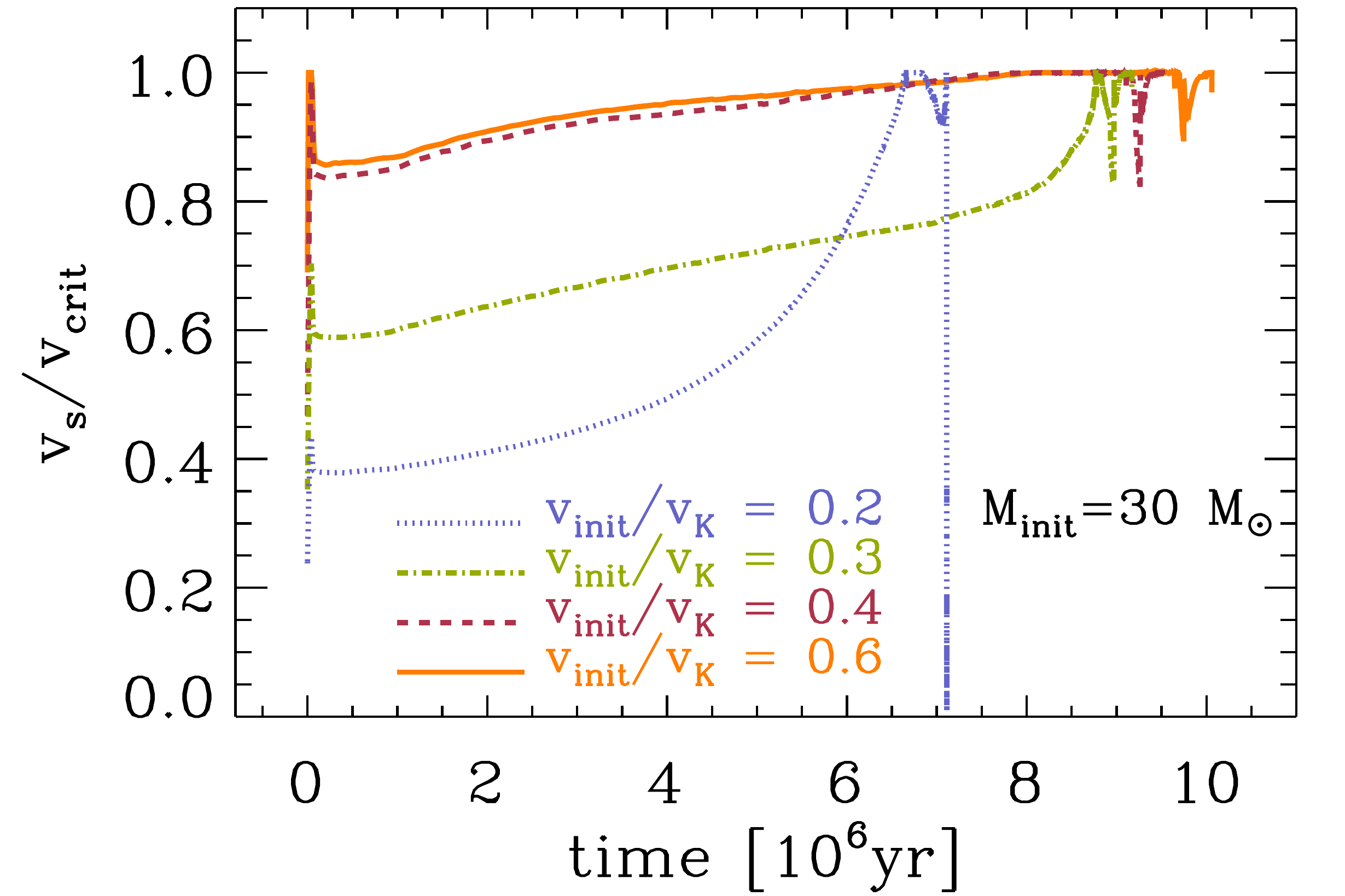}
\includegraphics[width=\columnwidth]{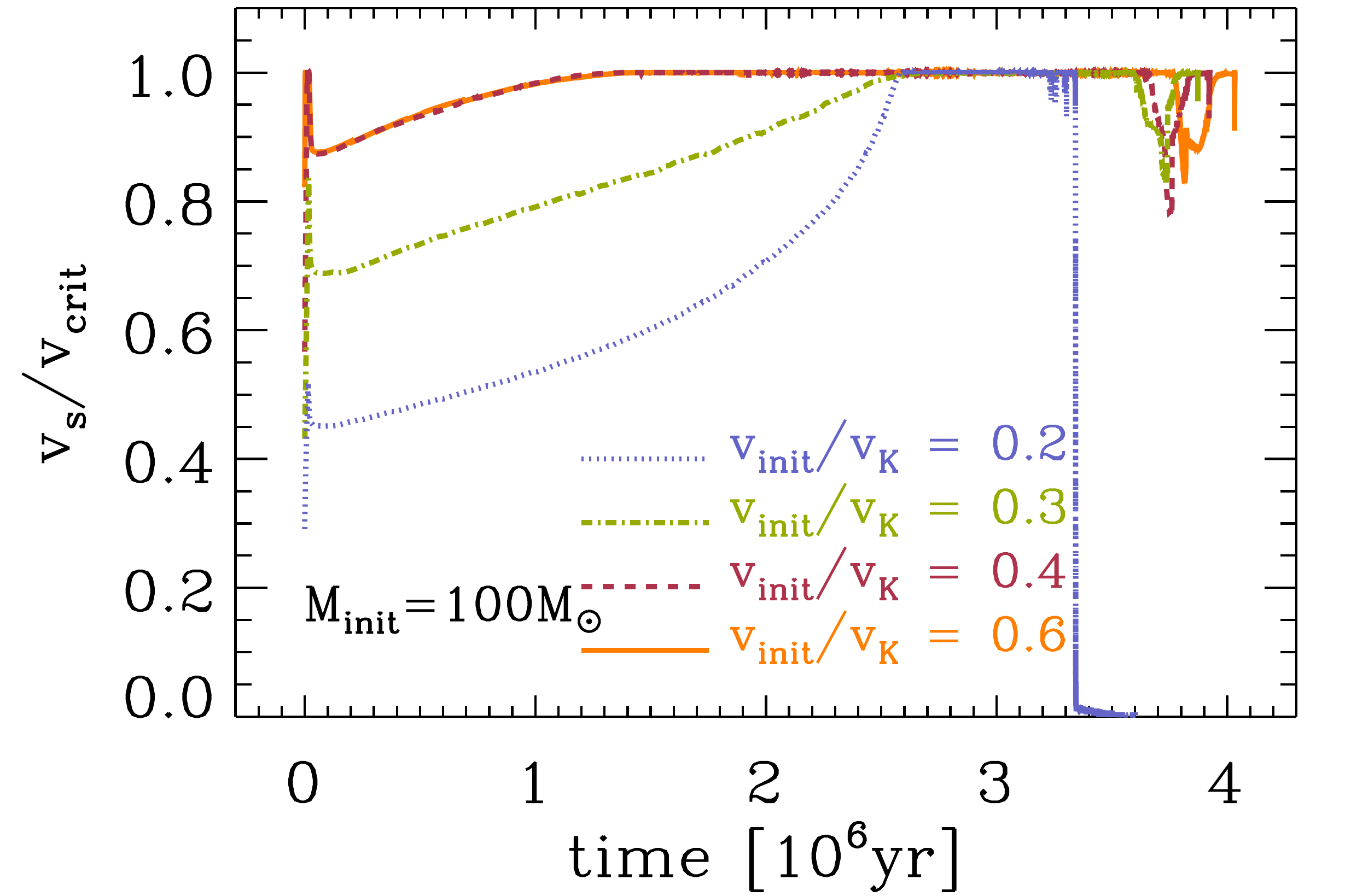}
\includegraphics[width=\columnwidth]{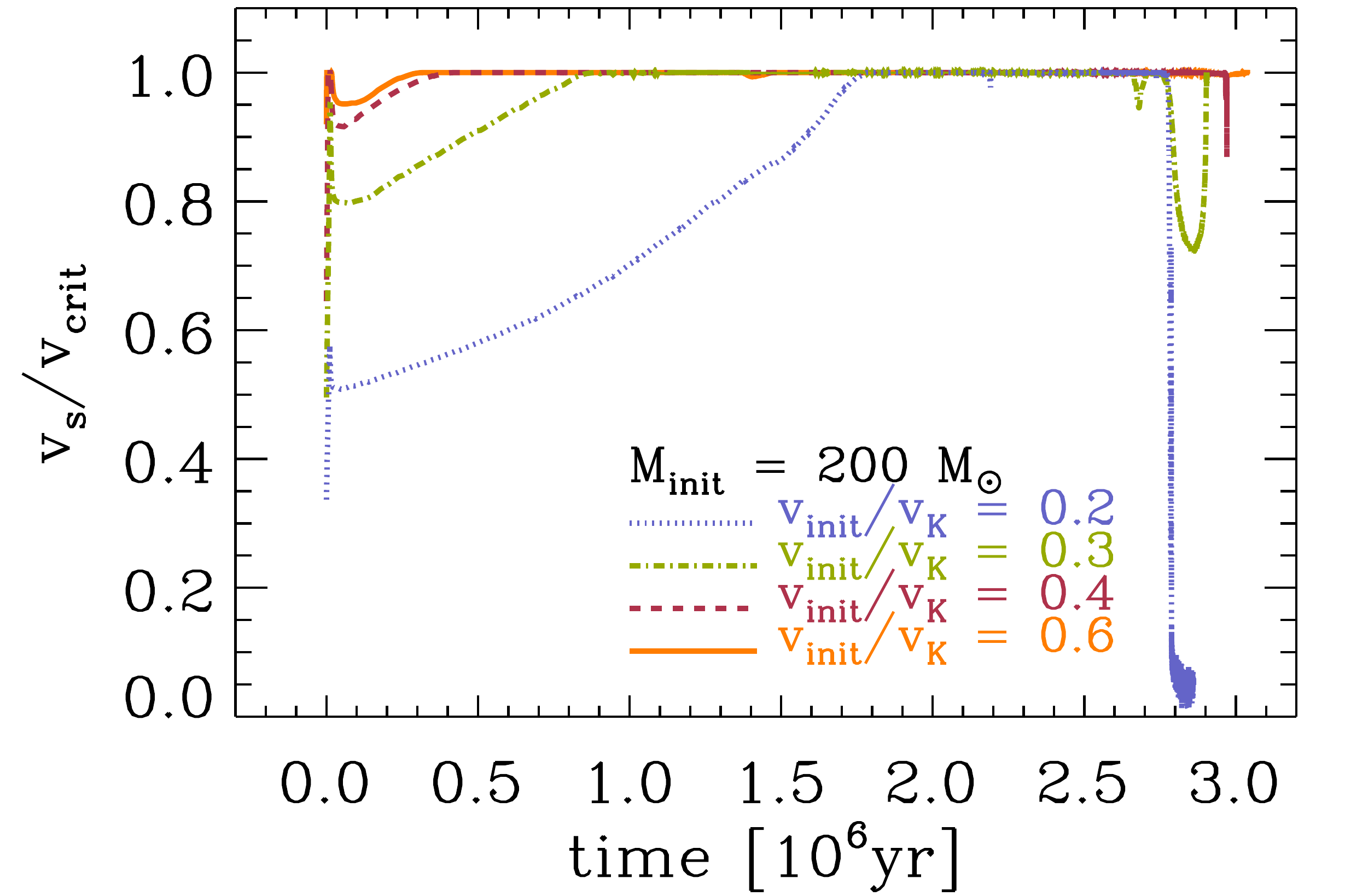}
\includegraphics[width=\columnwidth]{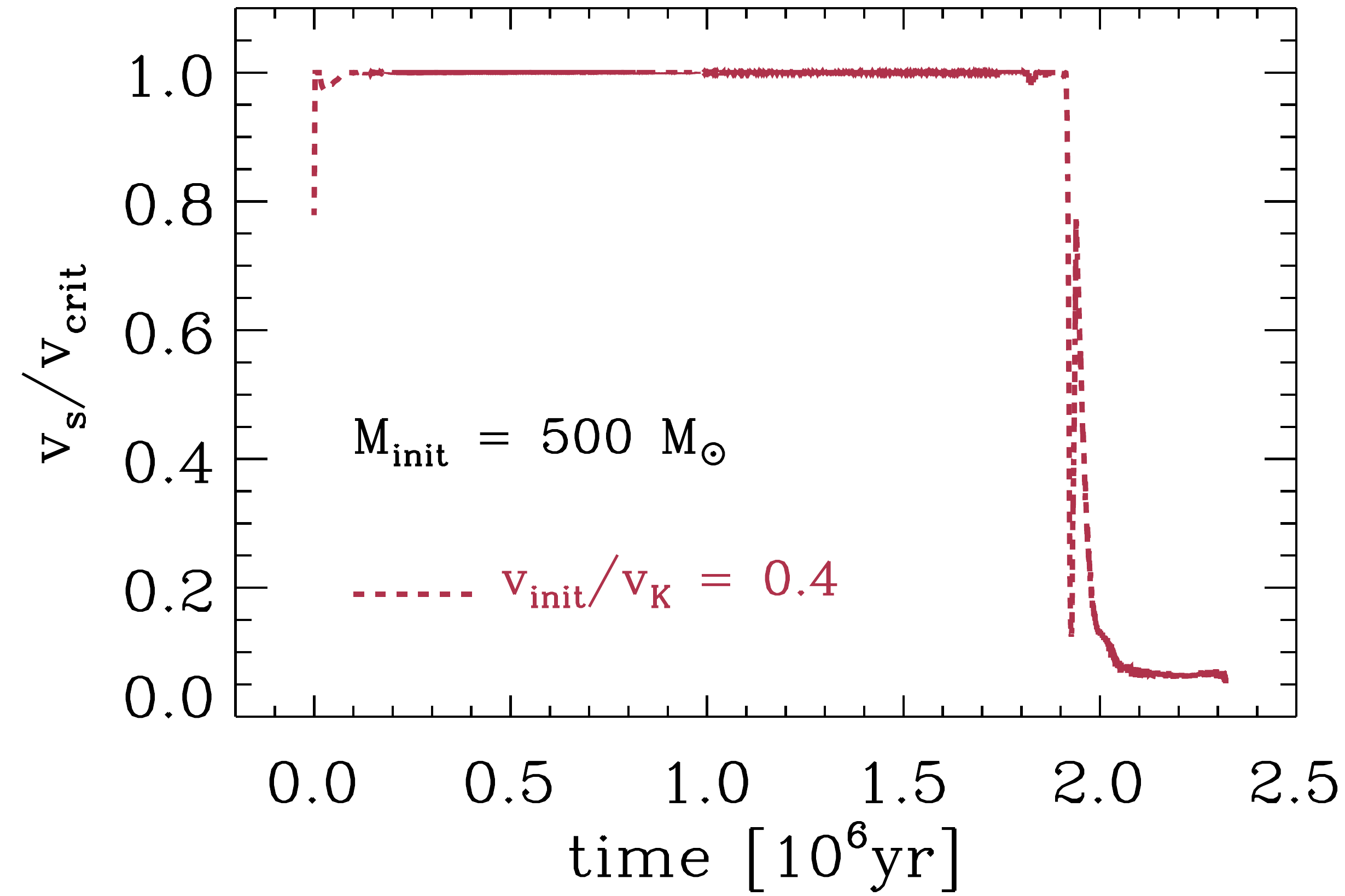}
\caption{Rotational velocity in units of the critical velocity ($v_\mathrm{crit}$) given by Eq. \ref{eq2}
as a function of time for model sequences with $M_\mathrm{init} =$ 30, 100, 200, and 500~\Msun. 
The initial  rotational velocity in units of the Keplerian value  ($v_\mathrm{K}$)
is indicated by the labels. Note that $v_\mathrm{crit} = v_\mathrm{K} \sqrt{1-\Gamma}$, where $\Gamma$
is the Eddington factor (see Eq.~\ref{eq2})}  
\label{fig:vrot}
\end{center}
\end{figure*}

Fig.~\ref{fig:kipp1} show the internal structure for model
sequences with $v_\mathrm{init}/v_\mathrm{K} =$ 0.0, 0.2, 0.4 and 0.6
\footnote{In some model sequences, the convective layers in the hydrogen envelope
during the late evolutionary stages look very noisy (m20vk02, m60vk02, and
m15vk04).  This is because the onset of convection depends sensitively  on a
minute change of the chemical gradient that is present in these layers. This
phenomenon is frequently observed in massive star models.}.  Note that the
relative size of the convective core compared to the total mass becomes larger
for higher initial mass, which is the well-known effect of higher radiation
pressure in more massive stars.  We find that the convection zone in the
envelope extends up to the near-surface layers during the post-main sequence
phases in all of our non-rotating models except for $M_\mathrm{init} =$ 15 and
20~\Msun.  This outer convective zone is not developed  for $M \le
60$~\Msun{} in \citet{Ekstroem08} for their non-rotating case. This difference
must originate from the larger overshooting parameter adopted in the present
study than in Ekstr\"om et al., as explained above.  The convective 
carbon-burning core is not developed for $M \ge 20$~\Msun{} in our models, which 
agrees well with the previous studies \citep{Chieffi04, Ekstroem08,
Heger00}. 

Interestingly, in non-rotating models of $M \ge 200$~\Msun, the inner boundary
of the convective envelope begins to  penetrate into the layers that have
previously undergone hydrogen shell burning during core helium-burning phase.
This eventually leads to a dredge-up of helium core material, and hot bottom
burning occurs afterwards where primary nitrogen is abundantly produced.  We
discuss this phenomenon below in greater detail (Sect.~\ref{sect:mixing}). 

Rotating models have a larger helium core and a smaller convective envelope
than those in the corresponding non-rotating models, which is an effect of
rotationally induced chemical mixing.   Dredge-up of helium core material by
the downward penetration of the convective hydrogen envelope does not occur any
more with rotation. The envelope remains radiative in the CHE models.

\subsection{Rotation and angular momentum}\label{sect:rotation}

Figure~\ref{fig:vrot} shows the evolution of the surface rotational velocity in
the rotating model sequences for 30, 100, 250 and 500~\Msun.  It rapidly
increases during the initial contraction phase, and decreases for a short
period when the star begins to expand on the ZAMS.  Note that in model
sequences with $v_\mathrm{init}/v_\mathrm{K} \ge 0.4$,  the critical rotation
is achieved during this initial contraction phase, leading to mass shedding
even before the star arrives on the ZAMS.  This is the reason why, in
Table~\ref{tab1}, the masses on the ZAMS for these rapidly rotating models are
somewhat smaller than the initial masses.

Soon after the ZAMS, the surface rotational velocity gradually increases again,
which is an effect of the internal transport of angular momentum.  On the main
sequence, the density of the hydrogen-burning core increases as the mean molecular
weight increases as a result of nuclear burning. Although this
tends to create differential rotation between the core and the envelope,
magnetic torques due to the Spruit-Tayler dynamo keep the star rotating
almost rigidly by transporting angular momentum from the core to the envelope.
The outer layers are spun up accordingly, and
most of our rotating models eventually reach the critical rotation on the main sequence.
The critical rotation at the surface is achieved earlier and  maintained for a
larger fraction of the evolutionary time, for a higher initial rotational
velocity.  If the star becomes a red supergiant afterwards, the expansion of
the envelope leads to a rapid decrease of the surface rotational velocity
($v/v_\mathrm{crit} < 0.1$) .  With CHE, on the other hand, it
still remains close to the critical limit ($v/v_\mathrm{crit} >
0.8$) during the post-main sequence phases.

As prescribed in Eqs.~\ref{eq1}, \ref{eq2} and \ref{eq3}, our models
lose mass when the surface velocity approaches the critical limit.  As shown
in Tables~\ref{tab1} and~\ref{tab:properties}, more mass is lost for higher initial
rotational velocity. In the case of $M_\mathrm{init} =$ 250~\Msun, for example,
the total mass ejected ($\Delta M_\mathrm{ej}$) is  only about 1.0~\Msun{} for
$v_\mathrm{init}/v_\mathrm{K} = 0.2$ (Seq. m250vk02), while it increases to about
60~\Msun{} for $v_\mathrm{init}/v_\mathrm{K} = 0.6$ (Seq. m250vk06).  Our rotating
models generally lose more mass than those of \citet{Ekstroem08} for a given
initial rotational velocity.  For instance, our 60~\Msun{} star in
Seq.~m60vk02 has a lower surface rotational velocity of
631~$\mathrm{km~s^{-1}}$ on the ZAMS (see Table~\ref{tab1}) than  that of in
Ekstr\"om et al. (800~$\mathrm{km~s^{-1}}$), but the total amount of the
ejected mass  is significantly larger ($\Delta M_\mathrm{ej} = 3.58$~\Msun)
than that of Ekstr\"om et al. ($\Delta M_\mathrm{ej} =$ 2.32~\Msun). This is
because our models reach the critical rotation earlier due to the Spruit-Tayler
dynamo than those of Ekstr\"om et al., who did not include the effect of
magnetic fields.  \citet{Marigo03}, who assumed rigid body rotation in their
models, adopted a low initial rotational velocity: $500~\mathrm{km~s^{-1}}$,
which roughly corresponds to $v_\mathrm{ZAMS}/v_\mathrm{K} \le \sim 0.1$
according to the figures ~6 and 7 in their paper. Their models therefore reach
the critical rotation only briefly near the end of the main sequence, losing
much less mass than our rotating models for a given initial mass. For example,
the 250~\Msun{} model in Marigo et al.  gives $\Delta M_\mathrm{ej} =$
3.35~\Msun{}, while we have  $\Delta M_\mathrm{ej} =$ 16.4~\Msun{} in Seq.
m250vk02, for which $v_\mathrm{ZAMS}/v_\mathrm{K} = 0.36$
(Table~\ref{tab1}).

\begin{figure*}
\begin{center}
\includegraphics[width=\columnwidth]{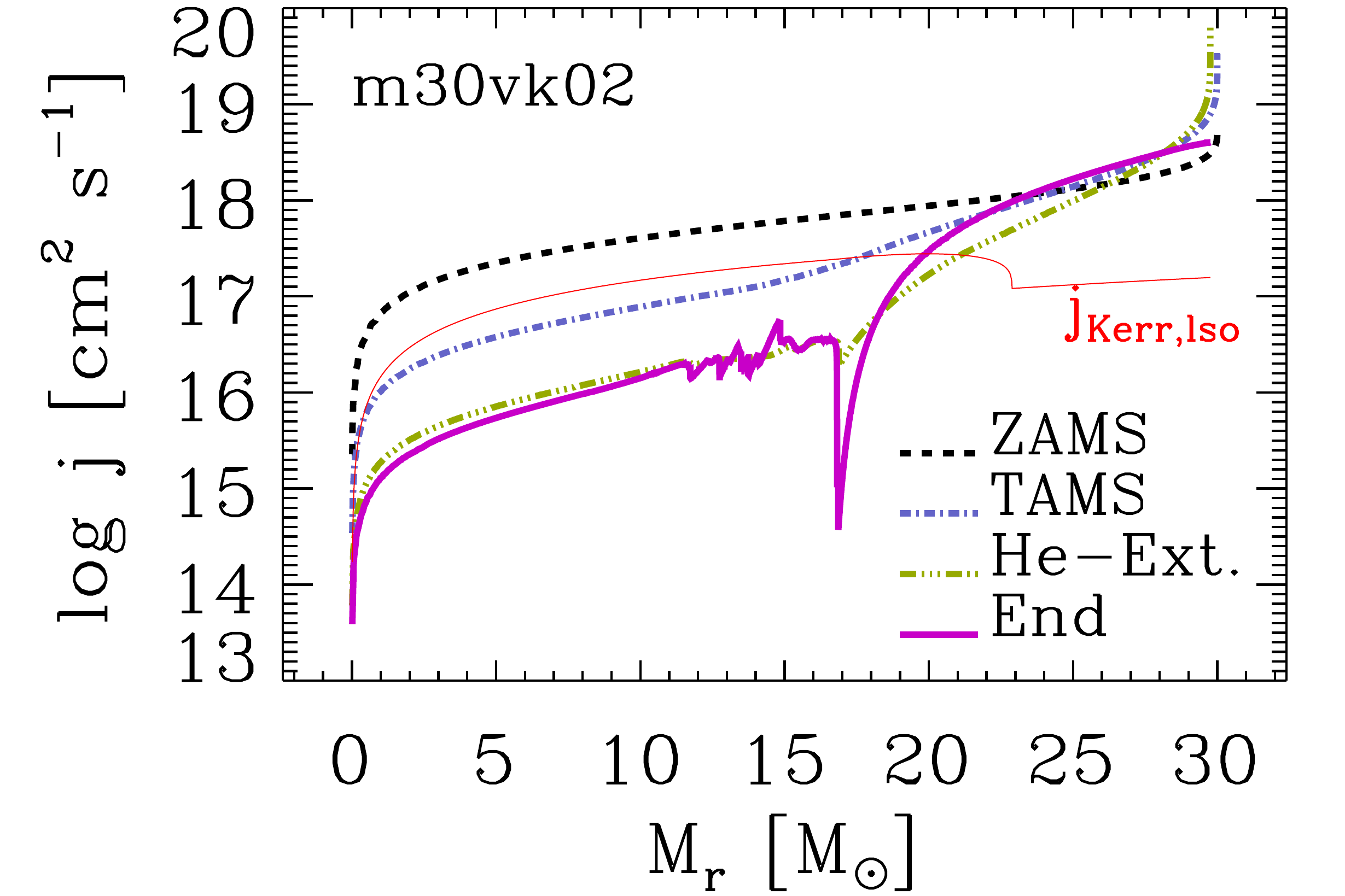}
\includegraphics[width=\columnwidth]{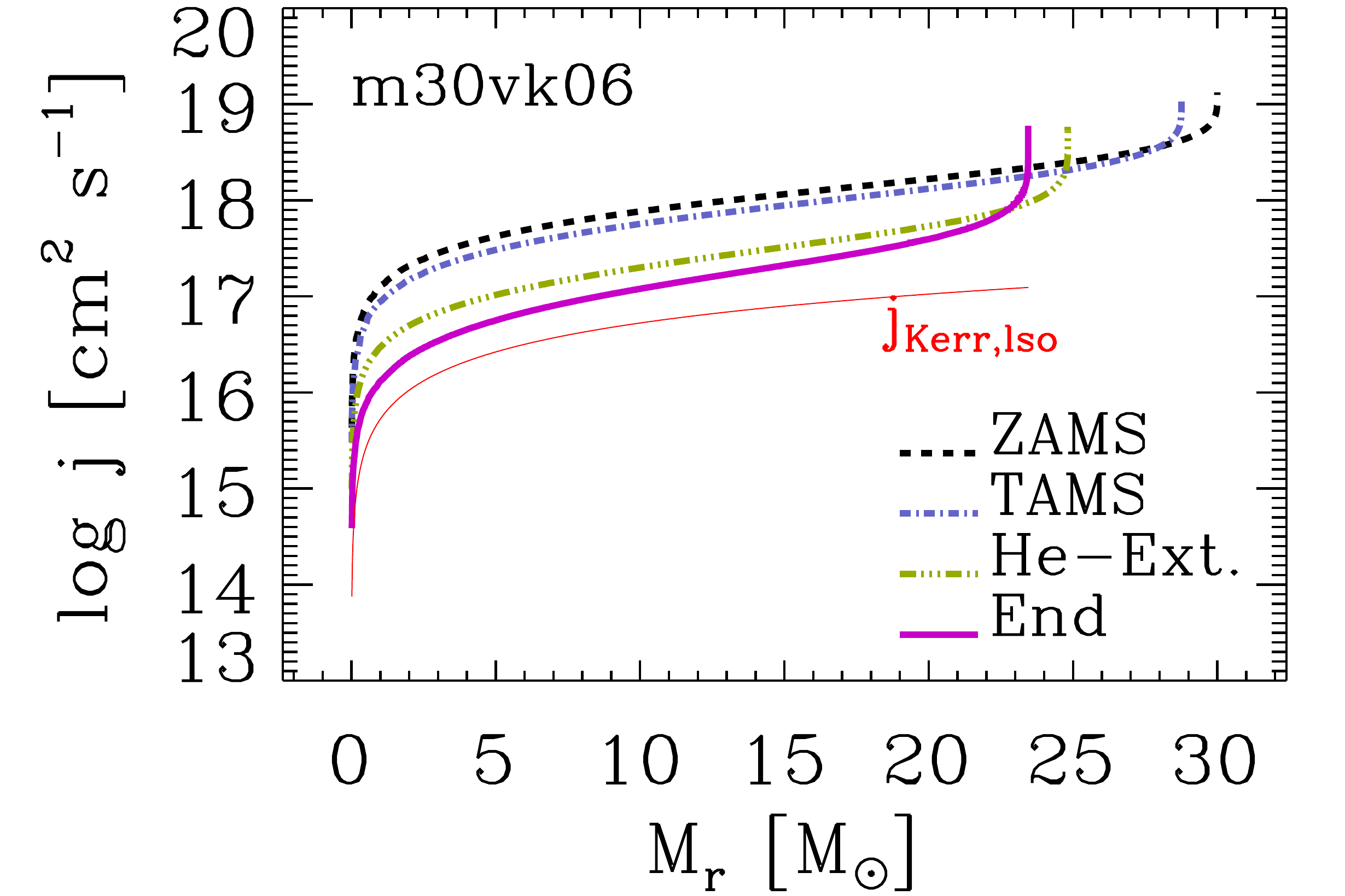}
\includegraphics[width=\columnwidth]{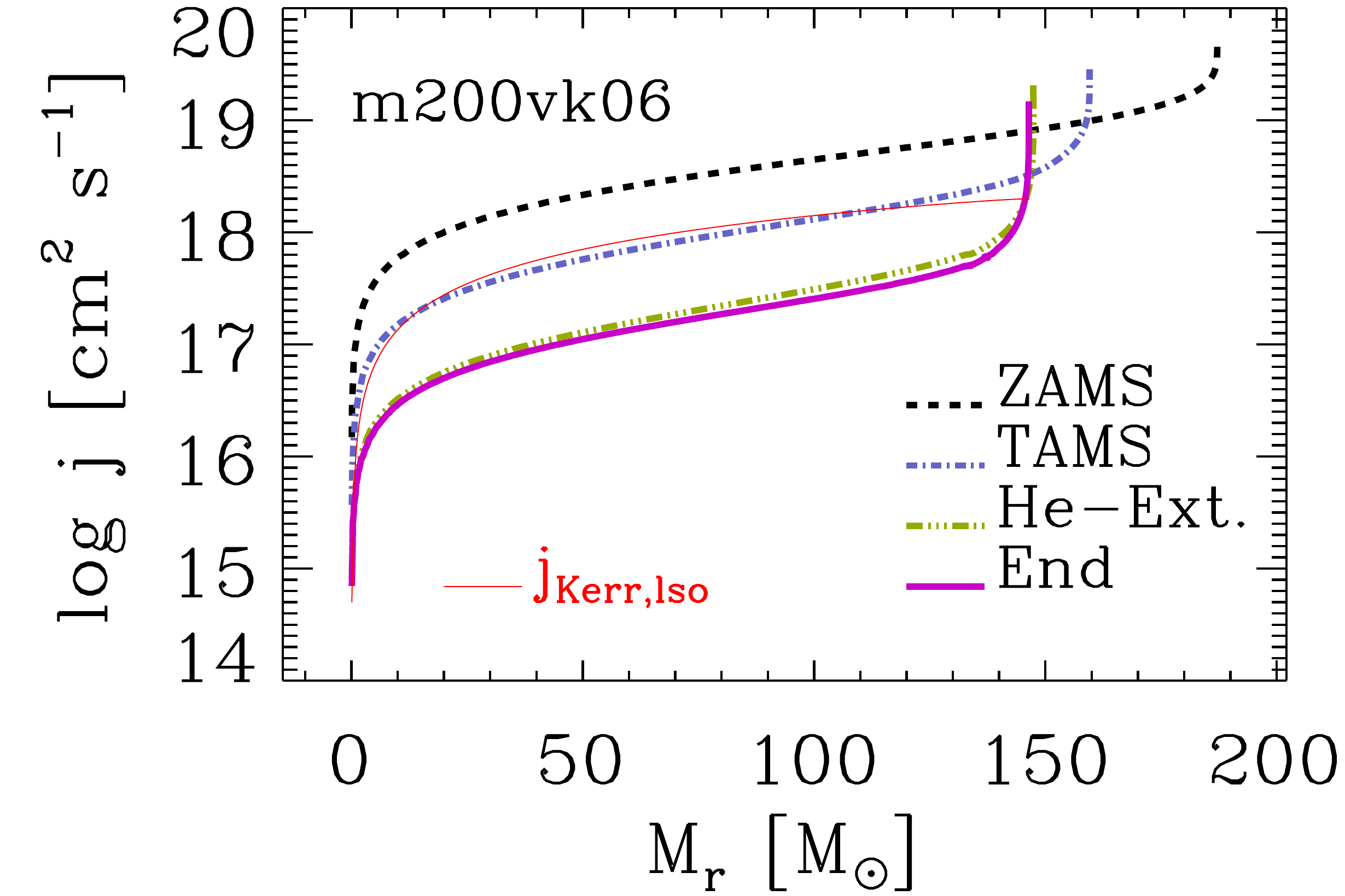}
\includegraphics[width=\columnwidth]{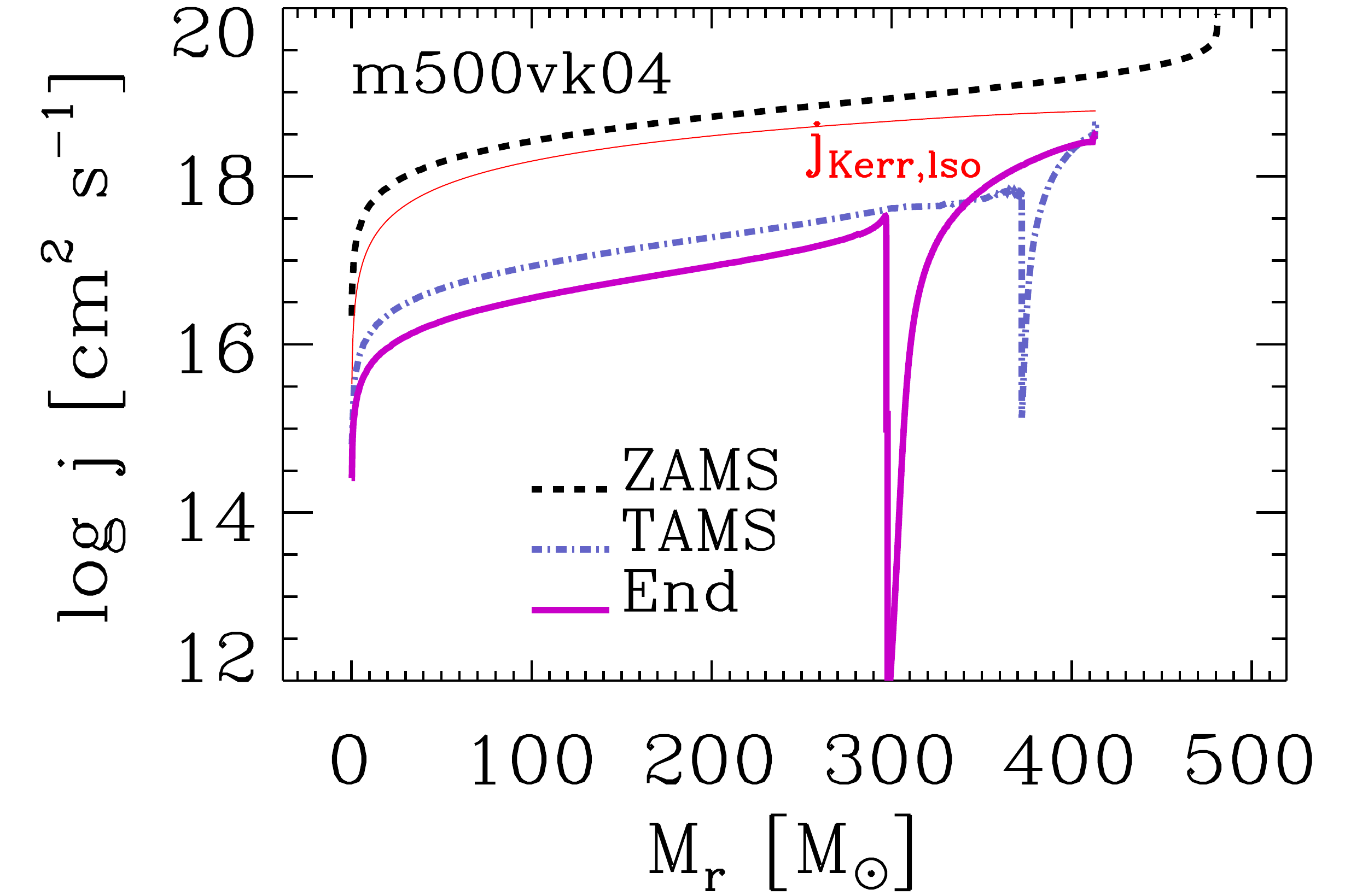}
\caption{Mean specific angular momentum over the shells as a function
of the mass coordinate at different evolutionary epochs (ZAMS, TAMS, core helium exhaustion, 
and the last calculated model) in Seqs. m30vk02, m30vk06, m200vk06 and m500vk04, 
as indicated by the labels. The thin sold line labeled $j_\mathrm{Kerr, lso}$ denotes 
the specific angular momentum
for the last stable orbit at the given mass of a black hole, assuming that all mass below
forms a rotating black hole. Here, if the contained angular momentum is higher than that of
a maximally rotating black hole, the black hole is assumed to rotate maximally.   
}  
\label{fig:jspec}
\end{center}
\end{figure*}

It is believed that the final fate of a massive star can be critically
influenced by rotation of the core~\citep[e.g.][]{Heger00}.  For example, the
so-called collapsar~\citep{Woosley93} is predicted to occur if the final
angular momentum in the core is higher than the critical limit for the
formation of an accretion disk around a rotating black hole at a given mass
($j_\mathrm{Kerr,lso}$; see also the figure caption of Fig.~\ref{fig:jspec};
\citealt{Bardeen72}). Figure~\ref{fig:jspec} shows the evolution of specific
angular momentum profile ($j_\mathrm{r}$) in four different model sequences
(m30vk02, m30vk05, m200vk06, m500vk04).  In Seq.~m30vk02, where the star
generally expands throughout the evolution, the core experiences strong braking
due to the core-envelope coupling via the magnetic torque. The final angular
momentum in the core becomes well below $j_\mathrm{Kerr,lso}$.  In
Seq.~m30vk06, on the other hands, the star of the same initial mass undergoes
CHE, to become a massive helium star by the end of the main sequence and to
avoid the core-envelope couping.  The specific angular momentum in the core
remains higher than $j_\mathrm{Kerr, lso}$ until the end of calculation (core
neon exhaustion), meaning that this model meets all the requirements for long
GRB progenitors within the collapsar scenario~\citep{Woosley93}: high angular
momentum in the core, large core mass to form a black hole, and absence of an
extended hydrogen envelope.  This implies that the CHE channel for the
formation of long GRB progenitors  can still be important for the first stars
\citep[see][for a more detailed discussion on GRB progenitors via 
CHE]{Yoon05, Yoon06, Woosley06}.  Note that the condition for magnetar
formation ($j \gsim 4\times10^{15}~\mathrm{cm^2~s^{-1}}$; e.g.,\citealt{Wheeler00}) is also satisfied in
this model.

\begin{figure}
\begin{center}
\includegraphics[width=\columnwidth]{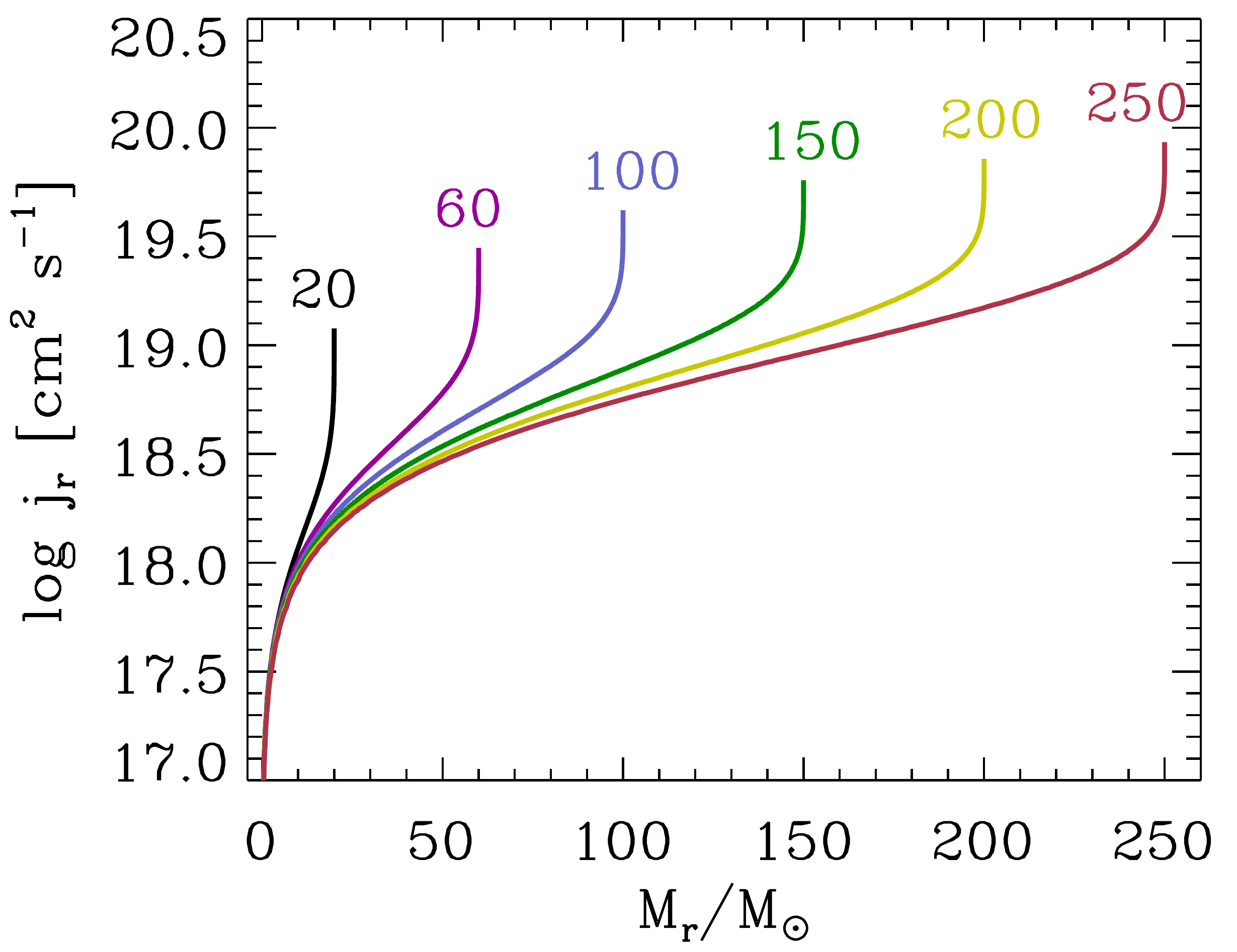}
\caption{
Specific angular momentum as a function of the mass coordinate in the initial 
models with $v_\mathrm{init}/v_\mathrm{K} = 0.6$ 
for different initial masses (20, 60, 100, 150, 200 and 250~\Msun) as indicated
by the labels.  
}  
\label{fig:jcompare}
\end{center}
\end{figure}

However, CHE does not always lead to retention of a large amount of angular
momentum in the core such that $j > j_\mathrm{Kerr, lso}$. In
Seq.~m200vk06, the specific angular momentum in the core decreases
significantly below  $j_\mathrm{Kerr, lso}$ before core helium exhaustion.  Given
the final mass of 146.43~\Msun{}, the star of this model sequence is supposed
to explode as a pair instability supernova, but even if collapse into a black
hole occurred, a GRB would not be produced in this case.  We find that the CHE
models have lower specific angular momentum in the core for larger initial
mass, in general.  This can be understood by the fact that the stellar structure of a rigidly
rotating chemically homogeneous star is such that $j_\mathrm{r}$ is smaller at
a given mass coordinate, for a given $v/v_\mathrm{K}$ as shown in
Fig.~\ref{fig:jcompare}.  In the CHE models, the surface rotational velocity
remains close to the critical rotation for most of the
lifetimes (Fig.~\ref{fig:vrot}), while near-rigid rotation is maintained until
core helium exhaustion.  Significant differential rotation
is developed only thereafter.  In addition, higher mass stars
have lower $v/v_\mathrm{K}$ at the critical rotation because of the increased
role of radiation pressure.  The general tendency for smaller $j_\mathrm{r}$
with larger mass in the CHE models results from these factors.

For $M_\mathrm{init} \ge 250$~\Msun,  CHE is not realized anymore, and
expansion of the envelope to the red supergiant phase leads to strong 
core-braking as shown Fig.~\ref{fig:jspec}. As a result, the so-called
super-collapsar (collapsar in stars of $M > 250$~\Msun{}) is not likely to
occur. We discuss below the implications of our results for the final fate of
the first stars in greater detail (Sect.~\ref{sect:fate}). 

\subsection{Ionizing flux}\label{sect:ionization}

\begin{figure}
\begin{center} 
\includegraphics[width=\columnwidth]{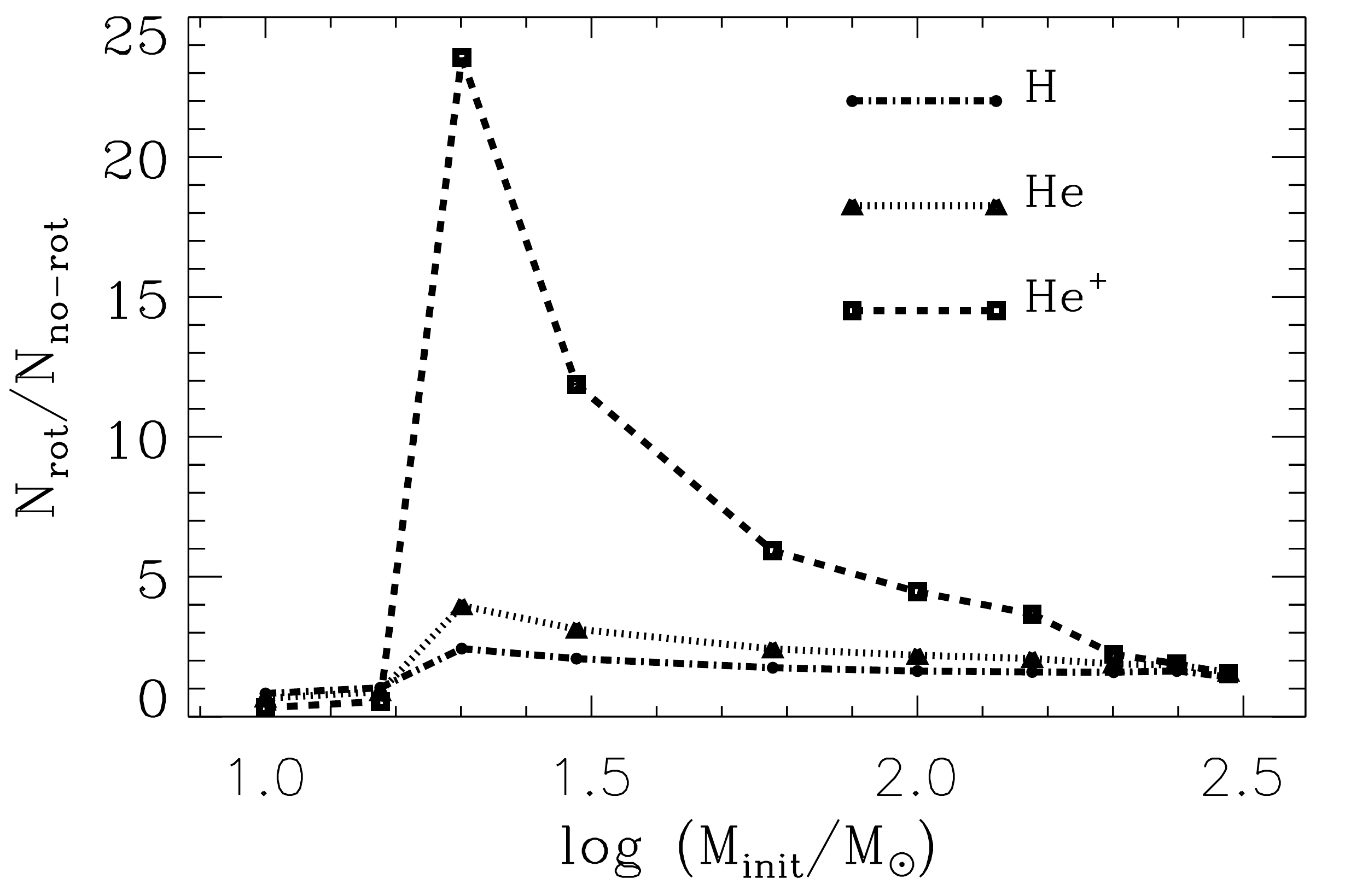}
\caption{
Ratio of the total number of ionizing photons
from non-rotating models to that from rotating models with $v_\mathrm{init}/v_\mathrm{K} = 0.4$. 
The connecting lines of filled circles, triangles and squares 
give the values for hydrogen and first and second helium  ionization, respectively. 
}
\label{fig:iflux} 
\end{center} 
\end{figure}

Detailed information on the hydrogen and helium-ionizing flux emitted from our
models is provided in Table~\ref{tab:iflux}.  Black-body radiation is assumed for
the calculation of energy and number of ionizing photons.  Compared to
\citet{Heger10}, our non-rotating models at a given mass produce  more
ionizing photons, which results from the larger overshooting parameter
adopted in our study. Rotation tends to make stars brighter, bluer and live
longer, and thus more ionizing photons are emitted than in the corresponding
non-rotating case, in general. In Fig.~\ref{fig:iflux},  the total number of
ionizing photons from the non-rotating models is compared to that from the
rotating models with $v_\mathrm{init}/v_\mathrm{K} = 0.4$, for 
hydrogen (H), neutral helium (He) and singly ionized helium (He$^+$).
The most notable change with rotation is found for ionization of He$^+$. Here, the
models with 20 - 150~\Msun{} follow CHE, and the number of  He$^+$ ionizing
photons increases by factors of 4 - 20.  For H and He, enhancement of the
number of ionizing photons is less prominent, but it is still significant for the CHE
models, showing factor of 2 - 4 increase.

\section{Effects of chemical mixing on the nucleosynthesis}\label{sect:mixing}

The heavy elements synthesised in Pop III stars can be ejected into the
surrounding medium by stellar winds and/or supernova explosions, which is
supposed to critically influence the chemical evolution of the early Universe.
Tables~\ref{tab:yield1} and~\ref{tab:yield2} present the yields for eight different isotopes
in the wind material  that has been ejected from the star
throughout the evolution and in the last computed model, respectively, in each
model sequence.  
Our non-rotating models do not experience  mass loss at all,
and therefore the wind yields of non-rotating models are not given in
Table~\ref{tab:yield1}. 
The model sequences with $M_\mathrm{init} = 1000$~\Msun{} are
not included in the tables either, because the calculations were terminated
before the end of main sequence.  
The end points of the other models are mostly
carbon burning or carbon exhaustion (see Table~\ref{tab:properties}), and
none of our models have undergone the late evolutionary phases beyond oxygen
burning.  
Therefore, in Table~\ref{tab:yield2},  neither  compact remnant mass
nor explosive nucleosynthesis is considered.  
This means that 
the amounts of $\mathrm{^{12}C}$, $\mathrm{^{16}O}$  and
$\mathrm{^{22}Ne}$  in Table~\ref{tab:yield2} should be taken as an upper limit 
for the supernova yields
since they would decrease during the latest evolutionary stages and
the supernova explosion that are not followed in this study, while the yields for
the other isotopes that are mostly located in the outer layers of the star
would not change much from the end points of our calculation.

In Table~\ref{tab:yield2}, we also give the neutron excess at the center
of the star ($\eta_\mathrm{c}$) in the last model. Here, the neutron excess is defined as 
\begin{equation}\label{eq4}
\eta = \sum (N_i - Z_i) Y_i~, 
\end{equation}
where $N_i$, $Z_i$ and $Y_i$ denote neutron number, 
proton number, and abundance of isotope $i$, respectively. 
The last column of the table gives the maximum mass
fraction of $^{14}$N (\Xnmax)  that has been achieved during the evolution. 

In the following subsections, we discuss how convection and
rotationally induced chemical mixing influence the nucleosynthesis in our
models. 

\subsection{Convective mixing}\label{sect:convmixing}

During the post-main sequence phases, convection can induce mixing of 
helium-burning products (notably $^{12}$C and $^{16}$O) into hydrogen-shell-burning
layers, boosting the CNO cycle. We find that there exist two different ways for
this mixing.  The first one is penetration of the helium convection zone
(either convective helium core or convective helium-burning shell) into the
hydrogen-burning shell, which has also been found by \citet{Ekstroem08} and
\citet{Heger10}.  The second way is penetration of the convective hydrogen
envelope into the helium core, which occurs for $M_\mathrm{init} \ge 200$~\Msun{}
in our models.  To our knowledge, the latter has not been reported in the
previous work.

Penetration of the convective helium burning shell into the
hydrogen-burning shell occurs for Seqs. m20vk02, m30vk00, m100vk00 and
m150vk02.  In the Kippenhahn diagram of Seq. m20vk02 (Fig.~\ref{fig:kipp1}), for
instance, the convection zone of the helium-burning shell continuously extends
upwards  and reaches the bottom of the hydrogen-burning shell when
$t_\mathrm{f} - t \simeq 10$~yr.  The consequent mixing of carbon into 
the hydrogen-burning shell induces boosting of the CNO cycle, and the hydrogen shell becomes
convective.  This produces a large amount of primary nitrogen. \Xnmax{} in Seq.
m20vk02 becomes as large as $10^{-3}$ (see Table~\ref{tab:yield2}), which is 1000
times larger than in the case without this mixing (i.e., \Xnmax$\simeq10^{-6}$
in Seqs. m20vk00 and m20vk03). In terms of the total yield of nitrogen
~($M_\mathrm{^{14}N}$) it is about 10000 times higher than in Seqs.~m20vk00 and
m20vk03.  However, $M_\mathrm{^{14}N}$ remains smaller than those in the
corresponding CHE models that experience strong rotationally induced chemical
mixing: e.g., $M_\mathrm{^{14}N} \simeq 10^{-3}$~\Msun{} in Seq. m20vk02,  while
$M_\mathrm{^{14}N} \simeq 10^{-2}$~\Msun{} including the wind yield
(Tables~\ref{tab:yield1} and~\ref{tab:yield2}). 

 In these sequences, a small
amount of hydrogen is also mixed into the convective region of the
helium-burning shell.  Given the large mass fraction of carbon and the fairly high
temperature ($\sim 2\times10^{8}$~K) in this region, the nuclear time scale
for hydrogen burning  becomes shorter than the convective turn-over time scale.
This rapid nuclear burning in convective layers cannot be properly described
with our code because the nuclear network and the transport equation are
calculated in succession. 
This leads to an overestimate of the energy generation rate here since the hydrogen mass
fraction at the bottom of the convection zone above the helium burning shell
can become artificially higher than it should be.  However, we find that the
overall amount of nitrogen and energy produced in this region is small compared to that of
the hydrogen-burning shell. 
In addition, our calculations are terminated soon after this mixing occurs, and
this uncertainty should not affect the main conclusions of the present study.

Penetration of the convective helium-core into the hydrogen-burning shell can
be most prominently observed in Seqs. m100vk00 and  m150vk02.  Here, the
convective helium core grows and reaches the bottom of the hydrogen-burning
shell, as shown in Fig.~\ref{fig:kipp1}.  In this case, a significant
fraction of the newly produced primary nitrogen in the hydrogen shell source is
mixed into the convective helium core, enhancing the neutron excess in the core, 
which is discussed below in Sect.~\ref{sect:neutron}.  

\begin{figure}
\begin{center}
\includegraphics[width=\columnwidth]{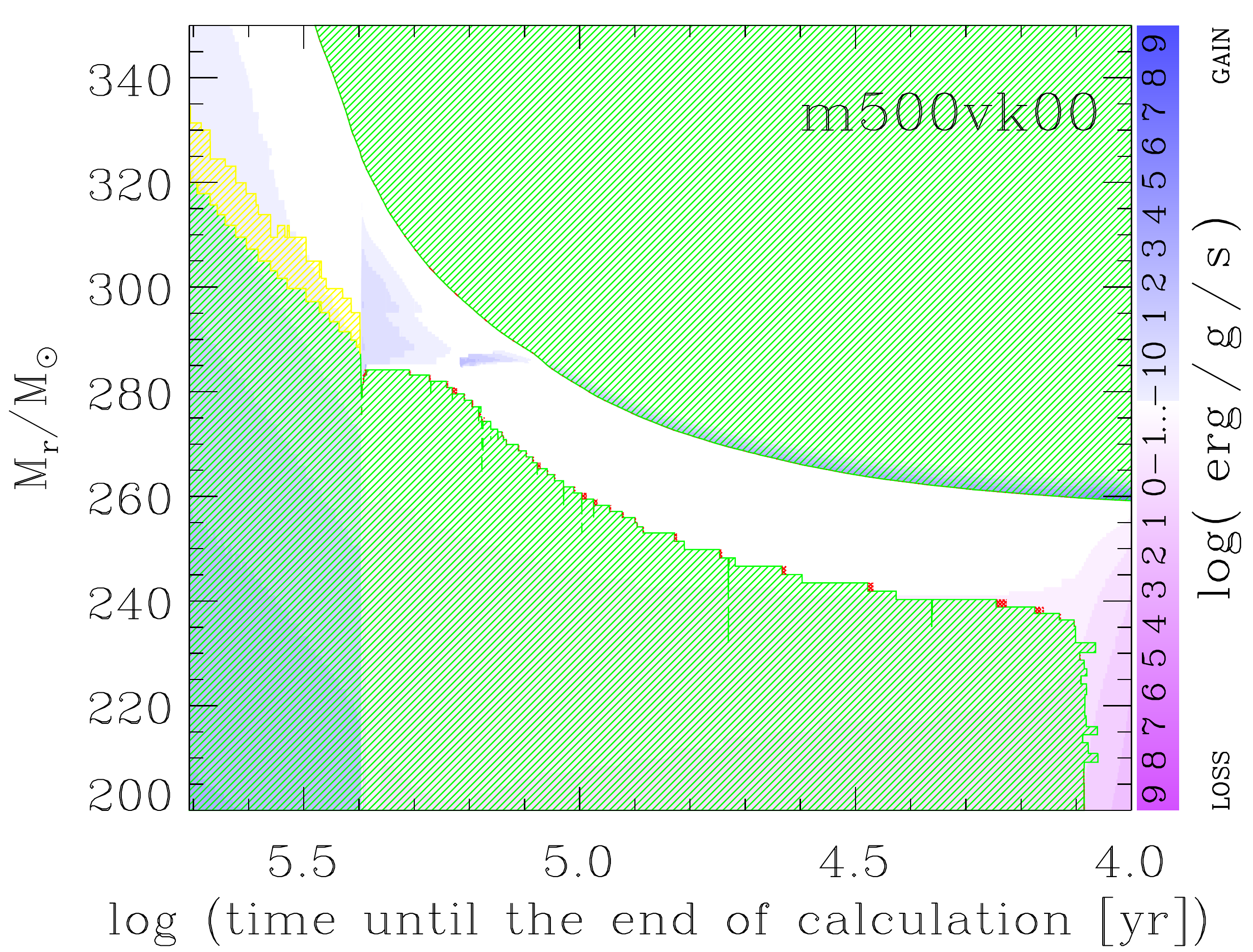}
\caption{
Kippenhahn diagram for Seq.~m500vk00, which is same as
in Fig.~\ref{fig:kipp1} but zoomed up for $200 \le M_\mathrm{r}/M_\odot \le 350$, 
and $ 4.0 \le \log(t_\mathrm{f} - t) \le 5.7$. Here $t_\mathrm{f}$ 
is the evolutionary time at the end of calculation. 
}  
\label{fig:kipp500-zoom}
\end{center}
\end{figure}

\begin{figure}
\begin{center}
\includegraphics[width=\columnwidth]{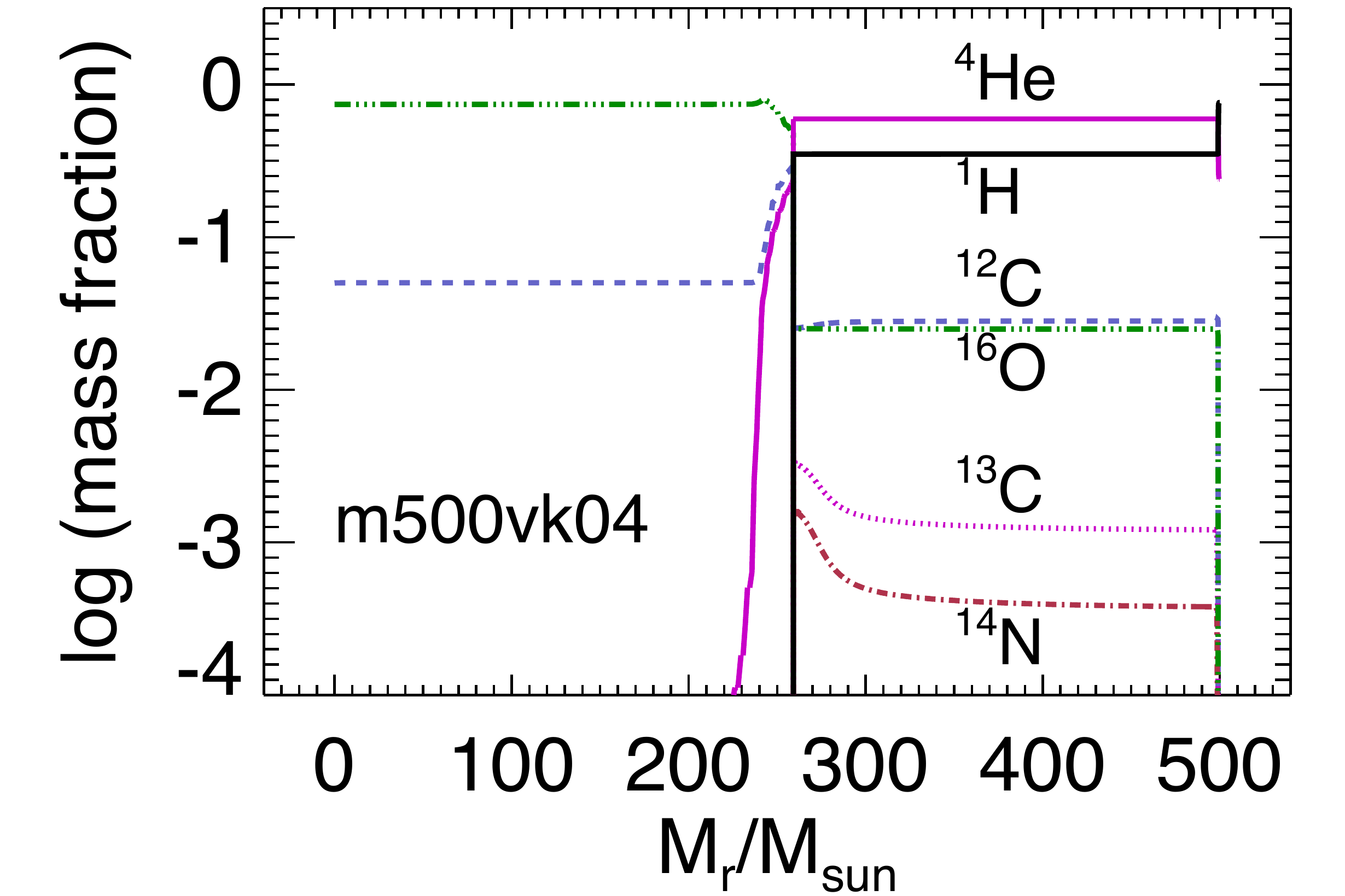}
\caption{Mass fraction of different isotopes in the last model of Seq.~m500vk00. 
}  
\label{fig:chem500}
\end{center}
\end{figure}

On the other hand, our non-rotating models with $M_\mathrm{init} \ge 200$~\Msun{}
experience dredge-up of the helium core material into the hydrogen-burning
shell by convection in the hydrogen envelope (see Fig.~\ref{fig:kipp1}).  As an
example, the detailed structure around the boundary between the helium core and
the hydrogen envelope is shown in Fig.~\ref{fig:kipp500-zoom}, for Seq.
m500vk00.  The hydrogen envelope becomes convective already on the main
sequence and the convection zone progressively extends downwards.  During core
helium burning, it eventually penetrates the layers that previously belonged to
the convective helium core.  A large amount of carbon and oxygen, which has
been produced by core helium burning, is thus mixed into the hydrogen envelope,
boosting the CNO cycle at the bottom of the convection zone.
Fig.~\ref{fig:chem500} shows that the mass fraction of the CNO elements in the
500 star model becomes as large as 0.03, after this dredge-up history.

We find that this convective dredge-up of the helium core material becomes more
efficient for more massive stars, in general.  The total yields of $^{13}$C and
$^{14}$N increase from $1.5\times10^{-3}$~\Msun{}  and
$3.71~\times10^{-4}$~\Msun{} in 200~\Msun{} model to 0. 0.34~\Msun{} and
0.12~\Msun{} in 500~\Msun{} model, respectively.  Note that in our non-rotating
models, the outermost layers of about 0.1~\Msun{} remain radiative even when
the most part of the hydrogen envelope becomes convective.  Since chemical
mixing cannot occur beyond the convection zone without rotation, the initial
values are maintained for the chemical composition at the surface.  This is the
reason for the sudden change in the mass fractions of the isotopes near the
surface in Fig.~\ref{fig:chem500}.   No mass loss occurs in our non-rotating
model sequences, given our wind prescription that allows stellar winds only
with surface-enrichment of heavy elements (Sect.~\ref{sect:wind}).  In reality,
massive stars  may become unstable to the pulsational instability, which may
cause mass loss \citep{Baraffe05}.  Pulsation caused by partial ionization of
hydrogen may be particularly important during the red supergiant phase, when the
above-discussed convective dredge-up can occur efficiently.  Therefore, we
cannot exclude the possibility of significant mass loss from non-rotating
massive Pop III stars, and the consequent enrichment of the CNO elements into
the circumstellar material.  Because pulsation-driven winds from RSG stars must be
slow, it should have interesting consequences in the chemical evolution history
as in the case for massive AGB stars \citep[e.g.,][]{Ventura01}.  Even formation
of dusts around such Pop III RSG stars might be possible, given the high
abundance of CNO elements in the envelope (Fig.~\ref{fig:chem500}). This question
deserves more studies.

\subsection{Rotationally induced mixing}\label{sect:rotmixing}

\begin{figure}
\begin{center}
\includegraphics[width=\columnwidth]{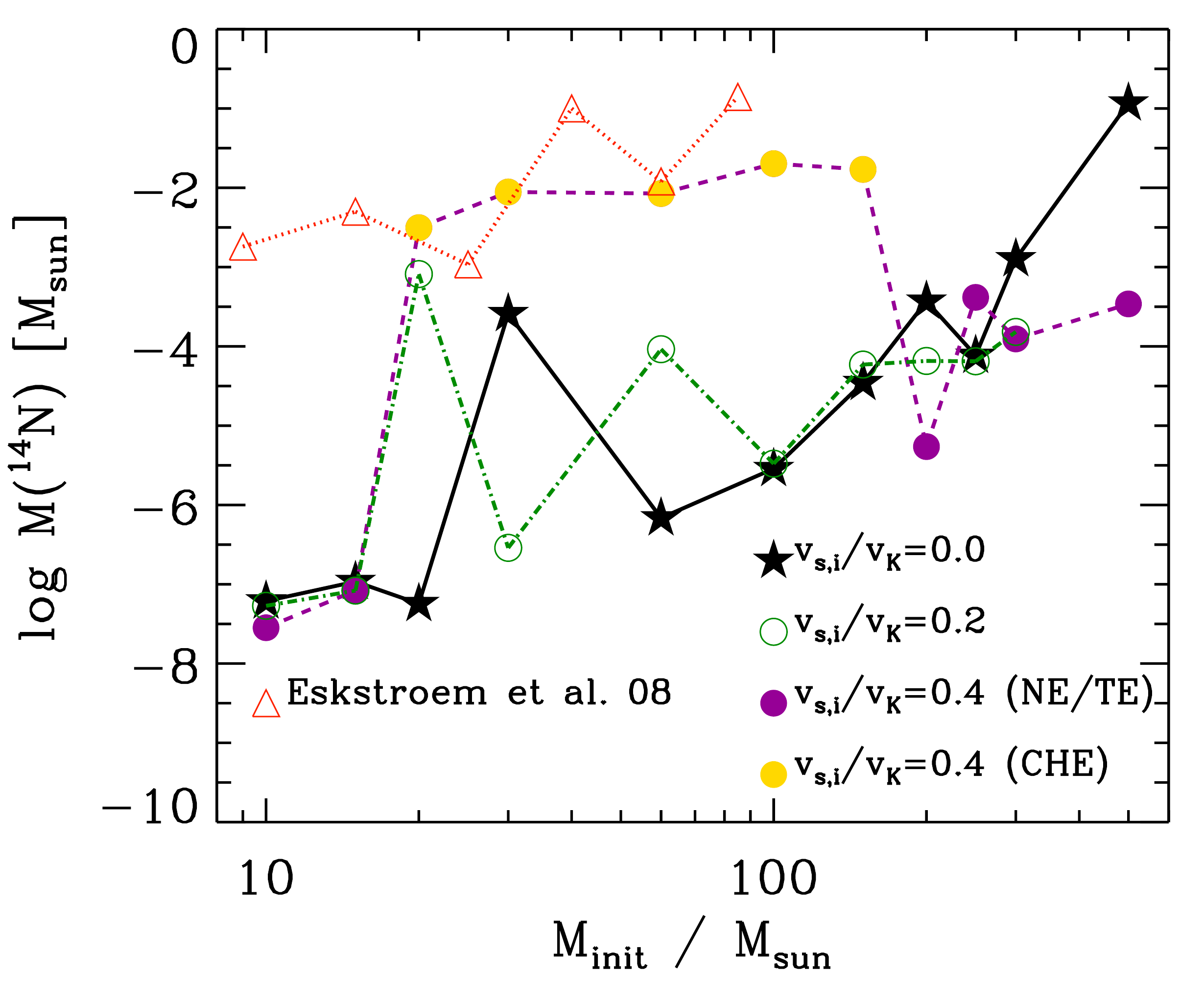}
\caption{ Total yields of nitrogen as a function of the initial mass
for $v_\mathrm{init}/v_\mathrm{K} =$ 0.0 (star symbol), 0.2 (open circle) and
0.4 (filled circle). For $v_\mathrm{init}/v_\mathrm{K} = 0.4$,  
the CHE models are marked with a yellow color, while the NE/TE models with a purple color.  
The connecting line with open triangles gives the nitrogen yields of 
the rotating models by \citet{Ekstroem08}. 
}  
\label{fig:nitrogen}
\end{center}
\end{figure}

The significance of rotationally induced chemical mixing on nucleosynthesis
varies according to the initial rotational velocity for a given initial mass.
As shown in Fig.~\ref{fig:nitrogen} (see also Tables~\ref{tab:yield1}
and~\ref{tab:yield2}), in most of the slowly rotating models that follow NE, the
total yield of primary nitrogen is not higher by more than a factor of 10 than
in the corresponding non-rotating case, if the star does not experience the
encroachment of the helium convection zone into the hydrogen shell source.  For
instance,   Seq.~m100vk00 gives $M_\mathrm{^{14}N} = 2.87\times10^{-6}$~\Msun,
while Seq.~m100vk02 gives $M_\mathrm{^{14}N} = 3.29\times10^{-6}$~\Msun{}.  For
models with $M_\mathrm{init} =$ 30, 200, 300 and 500~\Msun{},
$M_\mathrm{^{14}N}$ is even larger in non-rotating sequences than in the
corresponding sequences with $v_\mathrm{init}/v_\mathrm{K} = 0.2$.   This is
because these non-rotating massive stars undergo the convective mixing discussed
above: penetration of the convection zone of the helium-burning shell into the
hydrogen-burning shell for $M_\mathrm{init} =$  30~\Msun{}, and  the convective
dredge-up of the helium core material for $M_\mathrm{init} =$  200, 300 and
500~\Msun.  On the other hand, Seqs. m20vk02 and m60vk02 give exceptionally
large  $M_\mathrm{^{14}N}$ ($8.14\times10^{-4}$ and $9.21\times10^{-5}$~\Msun,
respectively), compared to the other cases with the same initial velocity.
This is not because of the rotational mixing, but because of the convective
mixing between the convective helium shell and the hydrogen shell source. 

\begin{figure}
\begin{center}
\includegraphics[width=\columnwidth]{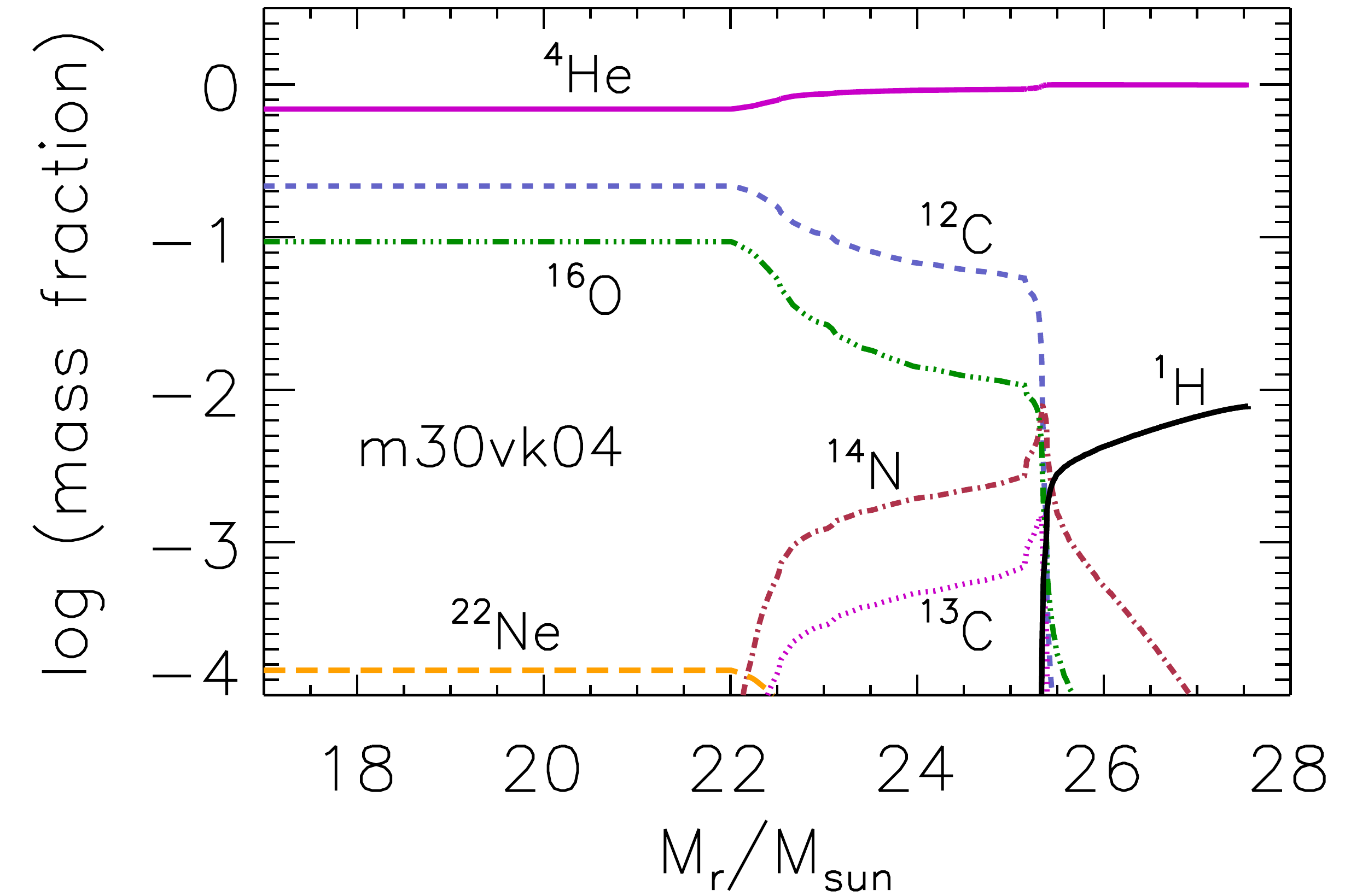}
\caption{Mass fraction of different isotopes in Seq.~m30vk04, 
when primary nitrogen is abundantly produced during core helium burning. 
}  
\label{fig:chem30}
\end{center}
\end{figure}

The CHE models produce more primary nitrogen by several orders of magnitude
than in the corresponding non-rotating or slowly rotating case, in most cases.
For instance, Seq.~m30vk04 gives $M_\mathrm{^{14}N}$ = 0.024~\Msun, compared to
$1.19\times10^{-7}$~\Msun{} in Seq.~m30vk00.  Although they evolve almost
chemically homogeneously throughout the main sequence, some amount of hydrogen
is still left in the outermost layers, which undergoes hydrogen shell burning
during the core helium-burning phase (See
Fig.~\ref{fig:kipp1} and \ref{fig:chem30}).  Chemical mixing
induced by rotation between the helium core and the hydrogen shell then leads
to abundant production of primary $^{14}$N and $^{13}$C, as well as $^{22}$Ne
(see Fig.~\ref{fig:chem30} for such an example).

The total yields of primary nitrogen in the rotating models by
\citet{Ekstroem08} are compared with ours in Fig.~\ref{fig:nitrogen}. 
In their
study, where magnetic torques are not considered for the transport of
angular momentum, none of the models follows CHE, and
the production of primary nitrogen is enhanced mainly by chemical mixing
due to the shear instability during the core helium-burning phase, where a
strong degree of differential rotation is developed across the boundary between
the helium core and the hydrogen envelope.  In our models with the
Spruit-Tayler dynamo, the role of the shear instability is negligible because 
near rigid-body rotation is maintained until the end of core helium burning. In our
NE models, chemical mixing by the Eddington-Sweet (ES) circulations becomes
rather inefficient during the post-main sequence phase, because the strong
chemical stratification between the helium-burning core and the hydrogen
envelope significantly slows down the ES circulations.  This explains the weak
production of primary nitrogen in our rotating NE models compared with the rotating
models by Ekstr\"om et al.  In the CHE models, however, the inhibition of the ES
circulations by the chemical stratification is much less significant and the
total yields of primary nitrogen are comparable to those of Ekstr\"om et al. 
The yields of other isotopes like $^{13}$C, $^{18}$O and $^{22}$Ne also
increase by several orders of magnitude with CHE. 

As shown in the Kippenhahn diagrams in Fig.~\ref{fig:kipp1}, 
the CHE models with $M_\mathrm{init} \le 150$~\Msun{} lose the hydrogen-burning
shell layers before the end of the core helium-burning phase. Note that the
mass loss here is dominated by the mechanical winds, which must be slow. Slow
ejecta of hydrogen-burning products  via either winds from massive AGB stars or
mechanical winds from massive single or binary  stars  are often invoked to
explain some chemical anomalies observed in globular clusters, characterized by
enrichment of He, N, Na and Al, and depletion of C and O \citep{Ventura01,
Denissenkov03, Prantzos06, Decressin07, demink09}.  As shown in
Table~\ref{tab:yield1}, the wind materials of our CHE models are indeed marked by
enrichment of He and N, but the amounts of C and O may be too large to explain
the anomalies. 
We will address the
implications of these models for the chemical evolution in the early Universe
in a future paper.

\subsection{Neutron excess}\label{sect:neutron}

The neutron excess (Eq.~\ref{eq4}) in the stellar core is one of the most
important factors that determine the nucleosynthesis of a supernova explosion.
This is particularly the case for the pair-instability supernova (PISN).
Detailed calculations by \citet{Heger02} showed remarkable deficiency of
odd-charge elements in the yields of PISNe from Pop III stars, which results
from very low neutron excess. Despite the theoretical expectation of frequent
PISN explosions in the early Universe,  such a peculiar abundance pattern has not
been found in the observed very metal-poor stars,  which are believed to
conserve the signatures of the nucleosynthesis in the first generations of
stars.  It should be noted, however, that \citet{Heger02}  only calculated pure
helium star models, ignoring the complicated history of chemical mixing that is
discussed above. It is therefore important to investigate whether or not the
convective and/or rotational mixing can significantly increase the neutron
excess in Pop III PISN progenitors.

In Table~\ref{tab:yield2}, we show the value of the central neutron excess 
($\eta_\mathrm{c}$) at the end of calculation for each evolutionary sequence.
In our calculations, the major source for the neutron excess is the
$^{14}\mathrm{N}(\alpha, \gamma)^{18}\mathrm{F}(e^+, \nu)^{18}\mathrm{O}$
reaction, which can occur in helium-burning layers.   This means that $\eta_c$
in our models is determined by the amount of primary nitrogen mixed into the
helium-burning core, and  does not include any excess beyond helium
burning.  On the other hand, \citet{Heger02} include many more reactions that
can further increase the neutron excess during the advanced stages beyond helium
burning (e.g., the
$\mathrm{^{20}Ne}(p,\gamma)\mathrm{^{21}Na}(e^+,\nu)\mathrm{^{21}Ne}$ reaction
during  carbon burning and many other weak processes). The implication is that
the problem with the deficiency of odd-charged elements would not be solved by
chemical mixing of primary nitrogen, unless the values of $\eta_c$ in our
models were significantly  higher than those of Heger \& Woosley. 

As discussed below (Sect.~\ref{sect:fate}),  PISNe are expected for  $120 \lsim
M_\mathrm{init}/M_\odot \lsim 240$ with  NE, and  $84 \lsim
M_\mathrm{init}/M_\odot \lsim 200$ with CHE. For this mass range, the neutron
excess in our models is limited to $\sim 10^{-5}$,  which is much lower than
those of the Pop III star models by \citet{Heger02} ($\eta_\mathrm{c} \sim
10^{-4}$).  Therefore, we conclude that PISNe from our models would not result
in a significantly different nucleosynthesis compared to that of Woosley \&
Heger's models.  It should be investigated in future if this
conclusion  would still hold with non-magnetic models, where strong mixing due
to the shear instability would occur during the post-main sequence
phases~\citep{Ekstroem08}.

\section{Conditions for chemically homogeneous evolution}\label{sect:che}

\begin{figure*}
\begin{center}
\includegraphics[width=0.9\textwidth]{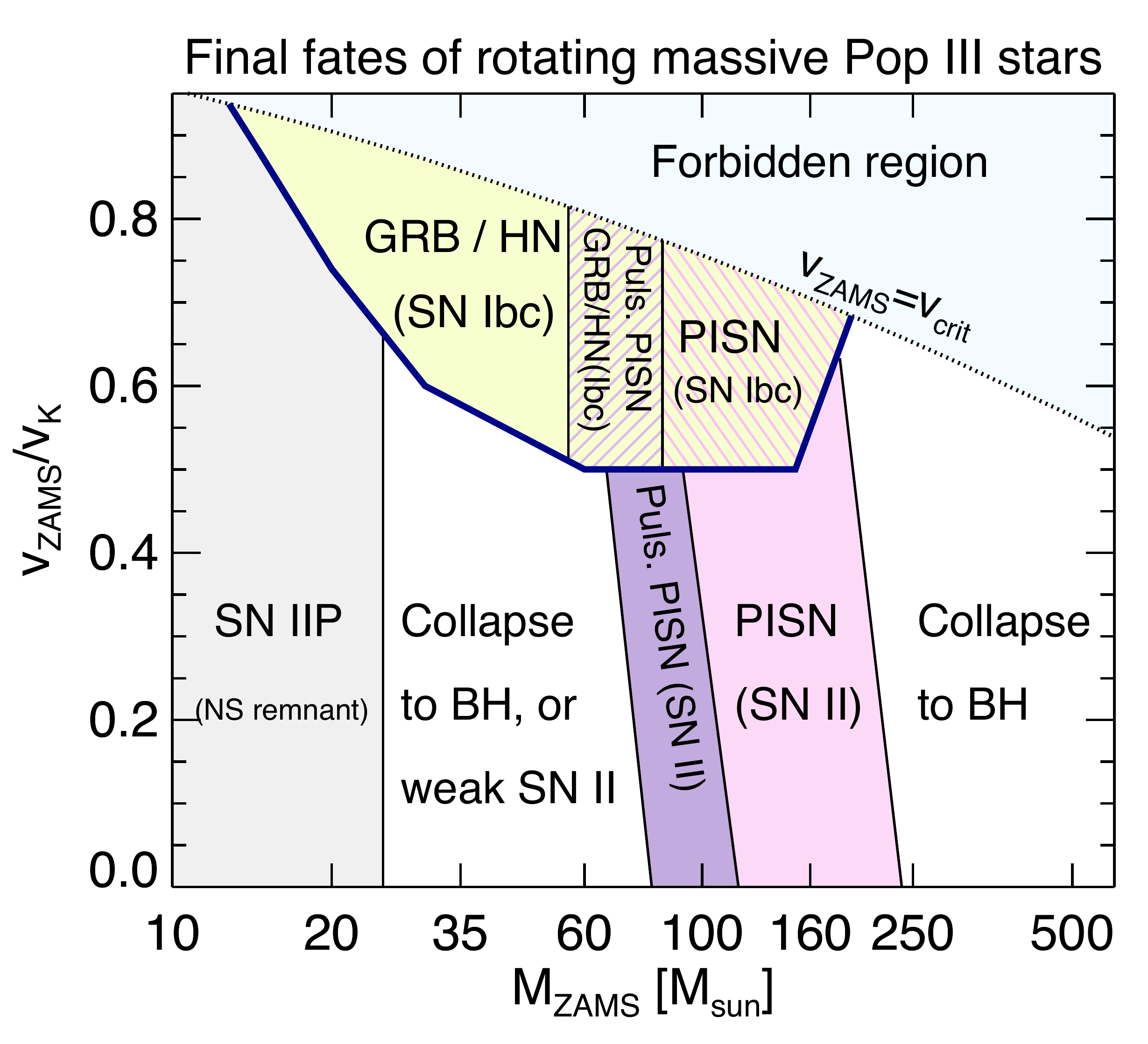}
\caption{
Phase diagram for the final fates of  massive Pop III stars, in the plane of
the mass and the fraction of the Keplerian value of the equatorial rotational
velocity on the zero-age main sequence.  The dotted line denotes the 
limit above which the rotational velocity at the surface
would exceed the critical rotation velocity ($v_\mathrm{crit}$ in
Eq.~\ref{eq2}).  The thick solid line marks the boundary between the regimes
for the chemically homogeneous evolution (CHE), and the non-CHE evolution (normal evolution or transitionary
evolution; see Sect.~\ref{sect:evolution}). The regions for different final fates 
are divided by thin solid lines. Each acronym has the following meaning.
SN IIP: Type IIP supernova, NS: neutron star, BH: black hole, SN II: type II supernova, 
PISN: pair-instability supernova, Puls. PISN: pulsational pair-instability supernova, 
GRB: gamma-ray bursts, HN: hypernova, SNIbc: type Ib or Ic supernova. 
}
\label{fig:final} 
\end{center} 
\end{figure*}

The CHE has important consequences not only for the nucleosynsthesis, but also
for the final fate of massive Pop III stars (Sect.~\ref{sect:fate}).  We give
the boundary for CHE in Fig.~\ref{fig:final} in the plane spanned by the
initial mass and the surface rotational velocity \footnote{Note
that in this figure, we use the values on the ZAMS for rotational velocity and
mass (i.e., $v_\mathrm{ZAMS}$ and $M_\mathrm{ZAMS}$) given in
Table~\ref{tab1}, instead of the initial values.} (see Sect.~\ref{sect:fate}
for more detailed explanation on the figure).  Our results indicate that 
CHE would not occur for $M_\mathrm{ZAMS} \lsim 13$~\Msun{} and $M_\mathrm{ZAMS}
\gsim 190$~\Msun{}. 
The limiting rotational velocity for 
CHE continuously decreases from  $M_\mathrm{ZAMS} = 13$~\Msun{} to about
60~\Msun{}, and rapidly increases from about 150~\Msun{} until it reaches the
forbidden region at $M_\mathrm{ZAMS} \simeq 190$~\Msun.  The mass dependency of
the limiting $v_\mathrm{ZAMS}$ for CHE can be understood as
follows.

In our models, chemical mixing is dominated by Eddington-Sweet (ES) circulations. 
According to \citet{Kippenhahn74}, who we follow for the prescription
of ES circulations in our models, 
the circulation velocity in a radiative region is given as 
\begin{align}\label{eq5}
v_\mathrm{e} &  :=  \left(\frac{\omega_r}{\omega_{\mathrm{K},r}}\right)^2 v_\mathrm{e,K}, \nonumber \\ \mathrm{where} &~~ \nonumber\\
v_\mathrm{e,K} &:=  \frac{\nabla_\mathrm{ad}}{\delta (\nabla_\mathrm{ad} - \nabla)}%
          \frac{L_r}{GM_r} %
         \left(\frac{2(\epsilon_\mathrm{n}+\epsilon_{\nu})r^2}{L_r} - \frac{2r^2}{M_r} %
               - \frac{3}{4\pi\rho r}\right)~. 
\end{align}
Here $\omega_r$ is the angular velocity at a radius $r$, $\omega_{\mathrm{K},r}$
the local Keplerian value, $\epsilon_\mathrm{n}$ the nuclear
energy generation rate,  $\epsilon_\nu$ the energy loss rate due to neutrino emission, 
and $L_r$ the local luminosity. The other symbols have their usual meaning. 
Note that $v_\mathrm{e}$ becomes $v_\mathrm{e,K}$ when $\omega_r = \omega_{\mathrm{K},r}$.

The chemical stratification tends to slow down the circulation velocity, 
and a correction is made for the actual circulation velocity in our models \citep{Heger00}:
\begin{equation}\label{eq6}
v_\mathrm{ES}  = \max{(|v_\mathrm{e}| - |v_\mu|, 0)}, 
\end{equation}
where 
\begin{equation}\label{eq7}
v_\mu : = \frac{H_P}{\tau_{\mathrm{KH},r}}\frac{\phi \nabla_\mu}{\delta(\nabla - \nabla_\mathrm{ad})}~. 
\end{equation}
Here $\tau_{\mathrm{KH},r}$ is the local Kelvin-Helmoltz time-scale at a radius $r$.

For the discussion below, we define three different time scales:
\begin{equation}\label{eq8}
\tau_\mathrm{ES,K} :=\int_{M_\mathrm{rad}} \frac{r}{v_\mathrm{e,K}} dM_r / M_\mathrm{rad}~, 
\end{equation}
\begin{equation}\label{eq9}
\tau_\mathrm{ES,0} :=\int_{M_\mathrm{rad}} \frac{r}{v_\mathrm{e}} dM_r / M_\mathrm{rad}~, 
\end{equation}
and
\begin{equation}\label{eq10}
\tau_\mathrm{ES} :=\int_{M_\mathrm{rad}} \frac{r}{v_\mathrm{ES}} dM_r / M_\mathrm{rad}~. 
\end{equation}
Here $M_\mathrm{rad}$ denotes the total mass of the radiative layers in the envelope. 
Note that $\tau_\mathrm{ES,K}$ does not depend on the rotational velocity, and thus
can be used as a ``susceptibility" of a fluid to ES circulations, for a given
thermodynamic condition.  On the other hand, $\tau_\mathrm{ES, 0}$ represents
the ES circulation time scale for chemically homogeneous layers  since the
effect of a chemical stratification is not included in $v_\mathrm{e}$.
Given that $\tau_\mathrm{ES,0} \approx
\overline{(\omega_{\mathrm{K,r}}/\omega_r)^2} \tau_\mathrm{ES,K}$ (see
Eq.~\ref{eq5}), the ratio $\tau_\mathrm{ES,0}/\tau_\mathrm{ES,K}$  gives
information on how the ES circulation is influenced by rotation in stars. For
example, loss of angular momentum and/or expansion of a star would result in
 an increase of $\tau_\mathrm{ES,0}/\tau_\mathrm{ES,K}$.  Finally,
$\tau_\mathrm{ES}$  is the real ES circulation time that considers  the
stabilizing effect of a chemical stratification.  The ratio
$\tau_\mathrm{ES}/\tau_\mathrm{ES,0}$ can be used as a measure of chemical
inhomogeneity inside a star: $\tau_\mathrm{ES}/\tau_\mathrm{ES,0} = 1$ for
$\nabla_\mu = 0$ and $\tau_\mathrm{ES}/\tau_\mathrm{ES,0} > 1$ otherwise. 

It should be noted that the actual time scale of ES circulations in stellar
evolution models is much larger than the above defined $\tau_\mathrm{ES}$, as a
result of the calibration of the mixing efficiency with observational data
(see Sect.~\ref{sect:method}).  Therefore, our discussion below should be
considered as only indicative. 
 
\begin{figure}
\begin{center}
\includegraphics[width=\columnwidth]{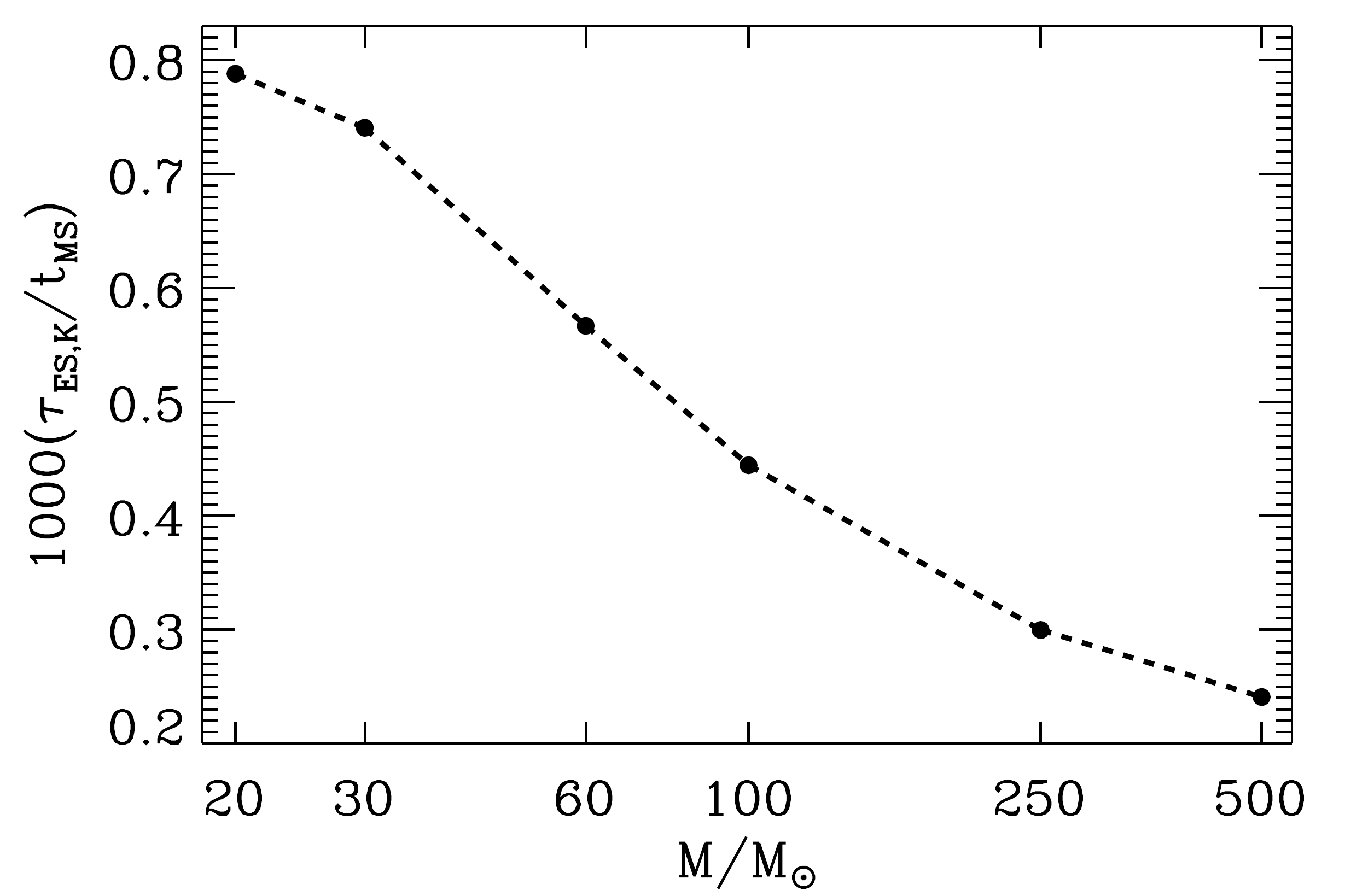}
\caption{$\tau_\mathrm{ES, K}$ in Eq.~\ref{eq8}, divided
by the main sequence lifetime ($t_\mathrm{MS}$)
and multiplied by 1000, as
a function of the initial mass. 
The ZAMS models of 20, 30, 60, 100, 250 and 500~\Msun{}
were used for the evaluation of $\tau_\mathrm{ES, K}$. 
}
\label{fig:estimea}
\end{center}
\end{figure}

We present the ratio of $\tau_\mathrm{ES,K}$ to the main sequence lifetime
($t_\mathrm{MS}$) as a function of initial mass in Fig.~\ref{fig:estimea}, to
investigate how thermodynamic conditions vary according to the initial mass in
favor/disfavor of ES circulations. Here, $\tau_\mathrm{ES, K}$ is estimated on
the ZAMS for each mass. The figure shows that
$\tau_\mathrm{ES,K}/t_\mathrm{MS}$  is systematically smaller for a higher mass
star. This is related to  higher radiation pressure and lower density in a higher mass star.
This means that, for a given initial rotation velocity,  CHE should be favored
in a higher mass star.  This may explain why the limiting rotation velocity for
CHE continuously decreases with increasing mass from 13~\Msun{} to about 60~\Msun{} in Fig.~\ref{fig:final}. 

\begin{figure}
\begin{center}
\includegraphics[width=\columnwidth]{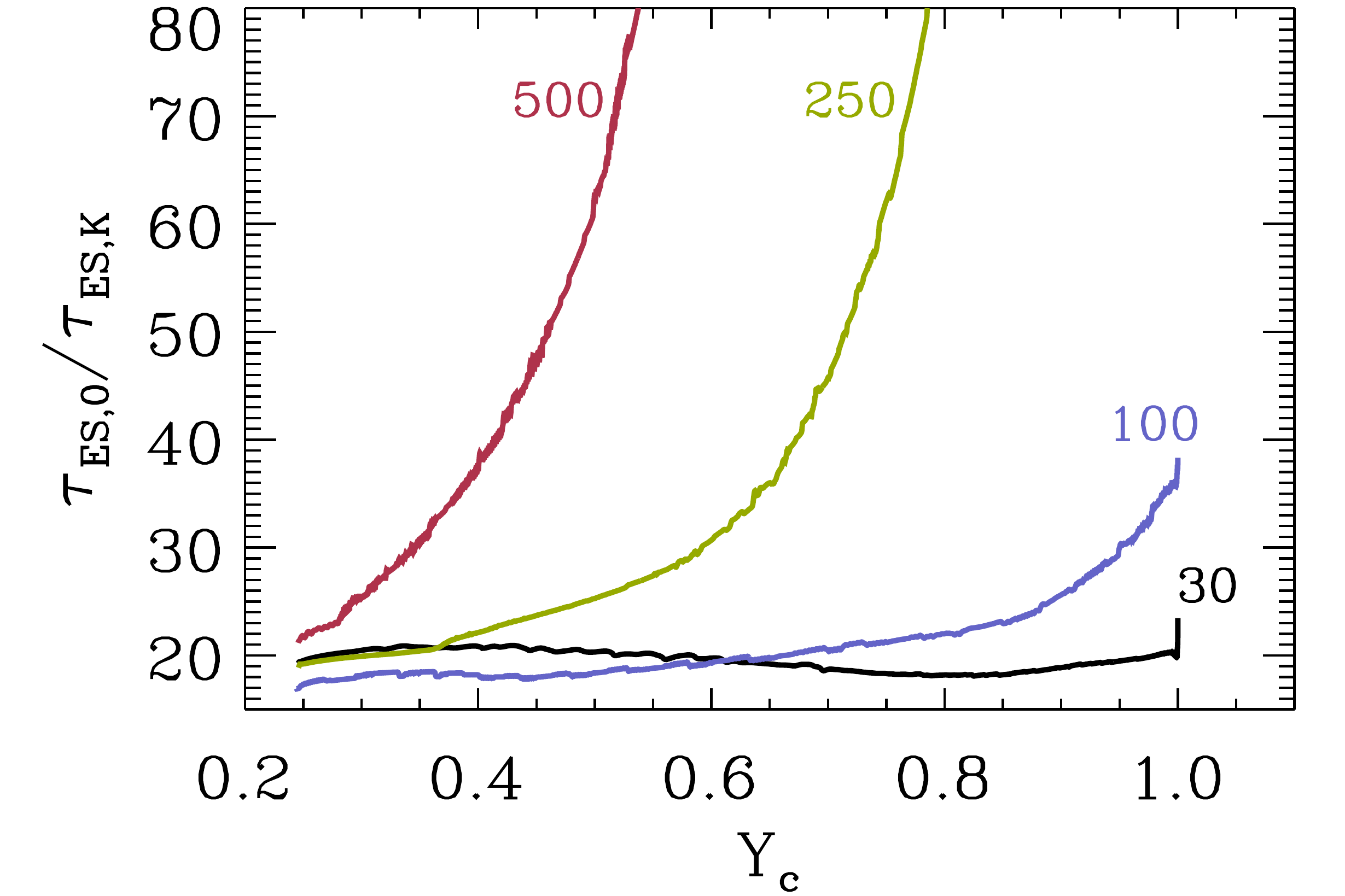}
\includegraphics[width=\columnwidth]{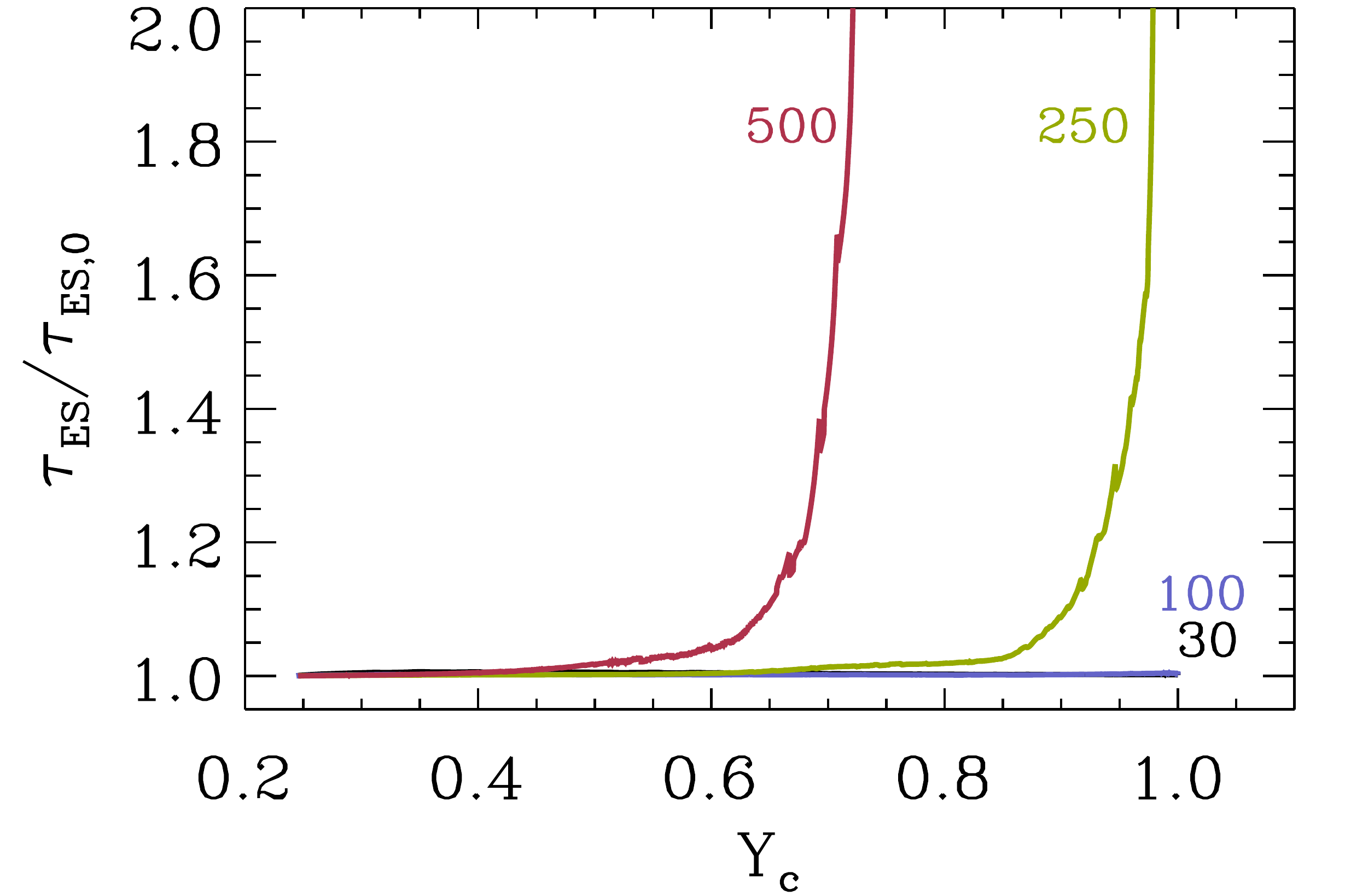}
\caption{
\emph{Upper panel:}
The evolution of $\tau_\mathrm{ES,0}/\tau_\mathrm{ES,K}$ (see Eqs.~\ref{eq8}
and \ref{eq9})
in the model sequences  m30vk03, m100vk04, m250vk04 and m500vk04. 
Only main sequence models are considered here. 
\emph{Bottom panel:}
The evolution of $\tau_\mathrm{ES}/\tau_\mathrm{ES,0}$ (see Eqs.~\ref{eq9} and \ref{eq10})
in the corresponding model sequences. 
}
\label{fig:estime}
\end{center}
\end{figure}

However, the influence of mass loss becomes important for very massive stars:
As discussed in Sect.~\ref{sect:rotation}, more massive stars can reach the
critical rotation in an earlier evolutionary phase, losing more mass and
angular momentum. This effect is reflected in the evolution of
$\tau_\mathrm{ES,0}/\tau_\mathrm{ES,K}$ that is shown in the upper panel of
Fig.~\ref{fig:estime} for some model sequences with
$v_\mathrm{init}/v_\mathrm{K} = 0.4$.  
$\tau_\mathrm{ES,0}/\tau_\mathrm{ES,K}$ increases very rapidly in 250~\Msun{}
and 500~\Msun{} stars, compared to those in lower mass stars.  For example, in
500~\Msun{} star, $\tau_\mathrm{ES,0}/\tau_\mathrm{ES,K}$ increases by a factor
of two from its initial value, when the central helium mass fraction ($Y_c$)
becomes about 0.4.  Since the change of the radius during this period is only
moderate (from 12~\Rsun{} to 15~\Rsun{}), the rapid increase in
$\tau_\mathrm{ES,0}/\tau_\mathrm{ES,K}$ results mainly from the loss of angular
momentum due to centrifugally-driven mass loss.

\begin{figure}
\begin{center}
\includegraphics[width=\columnwidth]{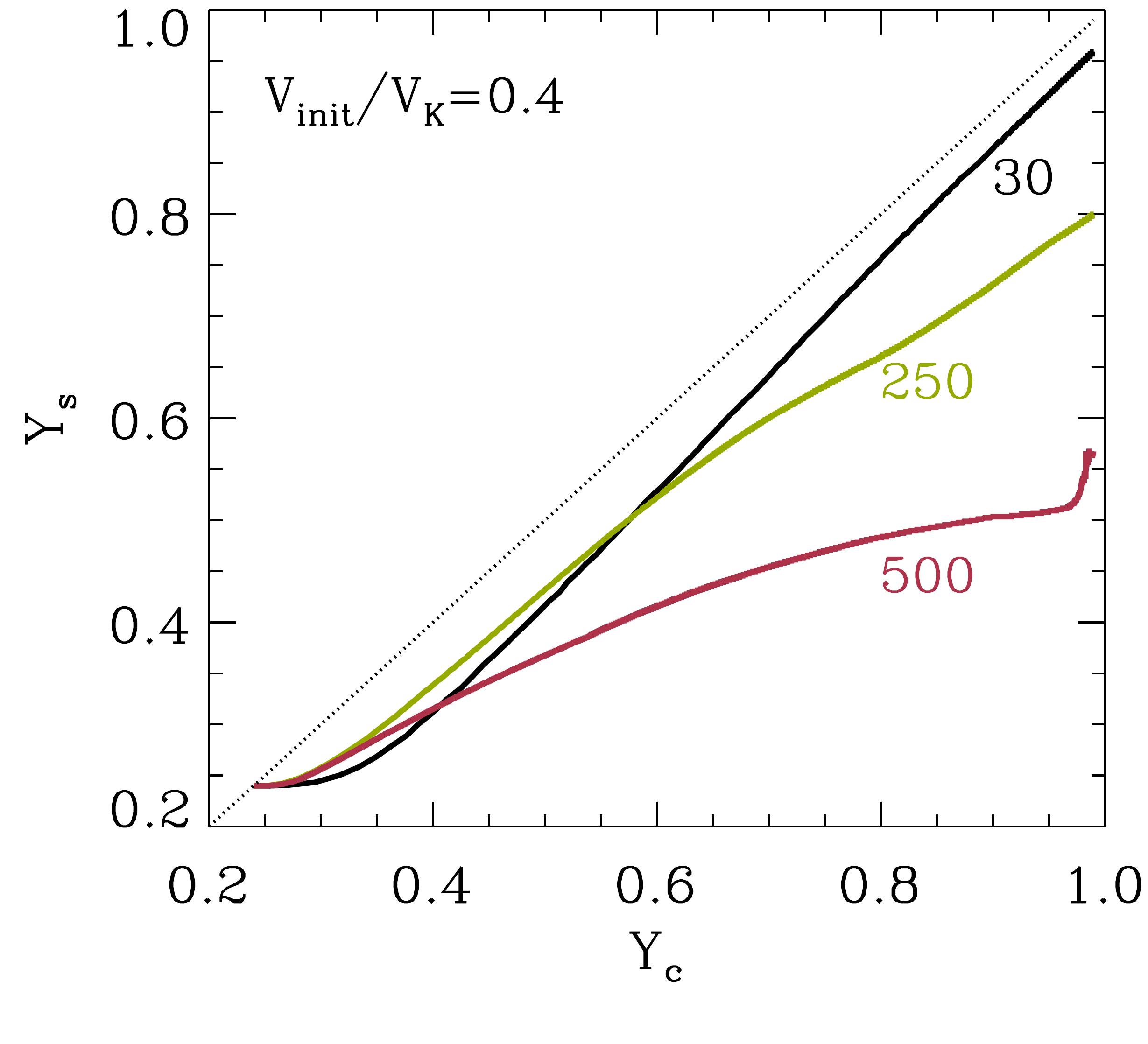}
\caption{
Mass fraction of helium at the surface ($Y_\mathrm{s}$) 
as a function of the mass fraction of helium at the center ($Y_\mathrm{c}$)
during the main sequence for model sequences m30vk04, m250vk04 and m500vk04. 
The thin dotted line corresponds to the case of perfect chemical homogeneity.  
} 
\label{fig:ysyc} 
\end{center} 
\end{figure}

The evolution of $\tau_\mathrm{ES}/\tau_\mathrm{ES,0}$ is shown in the bottom
panel of Fig.~\ref{fig:estime}. For 30~\Msun{} and 100~\Msun{} models,
$\tau_\mathrm{ES}/\tau_\mathrm{ES,0} \simeq 1.0$ is maintained until the end
of main sequence, as a result of CHE. For 250~\Msun{} and  500~\Msun{},
$\tau_\mathrm{ES}/\tau_\mathrm{ES,0}$ starts deviating from 1.0 at
$Y_\mathrm{c} \simeq 0.6$ and $Y_\mathrm{c} \simeq 0.4$, respectively.  This
indicates that although these stars initially follow CHE,   chemical mixing
due to the ES circulation becomes inefficient later as they progressively lose
large amounts of angular momentum. This interpretation is confirmed by
Fig.~\ref{fig:ysyc}, which shows the evolution of surface helium mass fraction
($Y_\mathrm{s}$) as a function of central helium mass fraction
($Y_\mathrm{c}$), for model sequences of m30vk04, m250vk04 and m500vk04.  Note
that $Y_\mathrm{s}$ is larger in 250~\Msun{} and 500~\Msun{} stars until
$Y_\mathrm{c}\simeq 0.6$ and 0.4, respectively, than in 30~\Msun{}.  This
implies that chemical mixing is initially more efficient in those very massive
stars than in 30~\Msun{} star, although they eventually deviate from CHE because of
angular momentum loss. 

This means that without angular momentum loss,  the CHE regime in
Fig.~\ref{fig:final} would extend to larger masses. Specifically,  given
that $\tau_\mathrm{ES} \propto (v_\mathrm{K}/v_\mathrm{ZAMS})^2
\tau_\mathrm{ES, K}$, the result shown in Fig.~\ref{fig:estimea} implies the
limiting $v_\mathrm{ZAMS}/v_\mathrm{K}$ for CHE would decrease from 0.74 for
20~\Msun{} to about 0.4 for 500~\Msun{} in Fig.~\ref{fig:final}, if there were
no loss of angular momentum. 

From the above analysis, we draw the following conclusions. 

\begin{itemize}
\item Thermodynamic conditions are generally more favorable for chemical mixing by the ES circulation in more massive stars. 
Therefore, the limiting rotational velocity for CHE can decrease with increasing mass, as long as the ES circulation
is not significantly slowed down through angular momentum loss. 
\item  Critical rotation can be more easily  achieved  in more massive stars, since their surface luminosity
is close to the Eddington limit. The consequent rapid loss of mass and angular
momentum eventually tends to inhibit CHE in very massive stars ($ > \sim 200~M_\odot$),
even if rotationally induced chemical mixing is initially very efficient.
\end{itemize}

\section{Implications for the final fate}\label{sect:fate}

In Fig.~\ref{fig:final}, we present a phase diagram of the expected final fates
of rotating massive Pop III stars in the plane of the mass and the rotational
velocity in units of the Keplerian value.  Note that here we used the ZAMS
values instead of the initial values for mass and rotational velocity (see
Table~\ref{tab1} and discussion in Sect.~\ref{sect:grid}).  To determine the 
boundaries for different regimes, we use interpolated values 
between the grid points of our calculated models, when it was needed. 

The forbidden region in the figure is the regime where the surface rotational
velocity exceeds the critical limit $v_\mathrm{crit}$ (Eq.~\ref{eq2}). Any star
in this region would be unstable, experiencing mass eruption, and forced to
come back to the regime of sub-critical rotation.  The CHE regime
($v_\mathrm{CHE} \le v_\mathrm{ZAMS} \le v_\mathrm{crit}$) is marked by a
thick solid line. Both NE and TE regimes are below this limit.  Because our model
grid is not fine enough to clearly identify the TE regime, we do not distinguish
the two in the non-CHE regime in the figure. 
In any case, the parameter space for  TE in this plane should be very
small, located in the vicinity of the border of the CHE regime.

\subsection{Core collapse supernovae}

In the NE regime, the lower mass limit for black hole (BH) formation is simply
assumed to be 25~\Msun.  We did not follow the evolution up to iron core
formation, and  we cannot discuss how rotation would change the
pre-supernova structure, which may affect this limit.  However, in all 
models in the NE regime, the amounts of angular momentum in the core fall below
the limit for magnetar  (i.e., $j \gsim 4\times10^{15}~\mathrm{cm^2~s^{-1}}$ in
the innermost 1.4~\Msun{}; e.g., \citealt{Wheeler00}) or collapsar formation
(i.e., $j \gsim 4\times10^{16}~\mathrm{cm^2~s^{-1}}$ in the layers that would
be accreted to the black hole; \citealt{Woosley93}), by the end of core helium
burning.  This means that a jet-driven supernova or hypernova-like event
powered by rotation may not occur in stars having hydrogen-rich envelopes,
which have been discussed in ~\citet{Heger03} in detail. Type IIP supernovae
via usual core-collapse to a neutron star are expected for 8~\Msun{} $\lsim
M_\mathrm{ZAMS} \lsim$ 25~\Msun{}, and weak fall-back SNe or direct collapse to
BH would occur for the other mass range outside the pair-instability regimes. 

The light curve,  luminosity,  and chemical yields of a SN IIP are critically
affected by the radius of the progenitor star at the pre-supernova stage. Our
10~\Msun{} models expand to the red-supergiant phase, having the final radii of
$R_\mathrm{f} \simeq 450 R_\odot$.  15 and 20 \Msun{} models remain relatively blue
up to core carbon exhaustion, having $R_\mathrm{f} \simeq 130...150~R_\odot$
and $R_\mathrm{f} \simeq 100~R_\odot$, respectively.  We find these values do
not change much with rotation in the NE regime, but they may still be affected
by uncertain factors such as core overshooting and production of primary nitrogen,
as discussed in Sects.~\ref{sect:evolution} and ~\ref{sect:mixing}.

\subsection{Pair-instability supernovae}\label{sect:pisn}

In the present study, we did not follow the explosive nuclear burning caused by
pair instability and thus our models do not provide an exact boundary for the
PISN regime.  For the phase diagram of Fig.~\ref{fig:final}, we simply adopted
the mass limits given by \citet{Heger02}: pulsational pair-instability SNe
(puls-PISN) for helium core masses of $40 \le M_\mathrm{He,core}/M_\odot \le
64$, and pair-instability SNe (PISN) for $64 \le M_\mathrm{He,core}/M_\odot \le
133$. Note that these limits are based on non-rotating stellar models, and that
the effect of the centrifugal force
can lead to an upward-shift of each boundary with rotation \citep{Glatzel85}.
Because in our potential PISN progenitor models the rotational velocity in the
innermost layers does not exceed a few percent of the local Keplerian values
even in the case of CHE, the role of rotation may be minor. But more
detailed studies should be carried out for clarifying the uncertainty
\footnote{Note added in proof: 
a paper by \citet{Chatzopoulos12} has recently been posted on astro-ph. 
Their result on the mass range for pair instability supernovae from rotating Pop III stars
agrees well with ours. 
}. 

In the NE regime, the mass limits for puls-PISN and PISN decrease with
rotational velocity, because the helium core mass increases with rotation.  In
the CHE regime, they do not change much with rotation, since in this case, the
whole star is transformed to a helium star by complete mixing by the end of the
main sequence.  The progenitors of PISNe in the NE regime have a very extended
hydrogen envelope with $R_\mathrm{f} \simeq 3000...5000~R_\odot$, depending on
the mass.  As shown in the recent calculations by \citet{Kasen11}, these PSNE
would be marked by a long plateau phase of the light curve (100 - 400 days), and
strong hydrogen lines (absorption and/or emission) in the spectra.  In the CHE
regime,  the progenitors of puls-PISNe and PISNe do not have any
hydrogen. The total amount helium at the pre-SN stage  is  small, ranging from
0.3 to 0.7~\Msun{} (Table~\ref{tab:yield2}).  A PISN from a CHE model is thus likely
to appear as a SN 2007bi-like event, which is a very bright SN Ic
\citep{Gal-Yam09}, if the final mass of the progenitor is high enough ($\gsim
90$~\Msun) to produce a large amount of nickel \citep{Heger02, Kasen11}.

\subsection{Gamma-ray bursts}\label{sect:grb}

If specific angular momentum in the CO core is higher than $j_\mathrm{K, lso}$,
and if the helium core mass is not within the range of PISNe, a collapsar may
occur.  In the NE regime, no collapsar formation is expected because 
angular momentum is too low in the core, while the conditions for a collapsar are
fulfilled for stars of $13\lsim M_\mathrm{ZAMS}/M_\odot \lsim 84$, in the CHE
regime.  The innermost cores ($\sim 1.4$~\Msun) in the stars in the same regime
also have high angular momentum to produce a magnetar ($j > \sim 5.4
\times10^{15}~\mathrm{cm~s^{-1}}$).  Therefore, stars in this regime may be
considered as GRB progenitors within both the collapsar scenario
\citep{Woosley93, Macfadyen99} and the magnetar scenario
\citep[e.g.,][]{Duncan92, Wheeler00, Bucciantini09} \footnote{Explosions as an
energetic supernova without making a GRB might be another possibility for a
star in this regime, as implied by the numerical studies of \citet{Burrows07}
and \citet{Dessart08} on magnetically driven explosions caused by the
magneto-rotational instability during core-collapse}. 

Interestingly, both puls-PISN and GRB are expected for $56 \lsim
M_\mathrm{ZAMS}/M_\odot \lsim 84$ in the CHE regime.  In this case, pulsational
pair-instability that occurs after carbon burning in the core may eject several
solar masses, probably episodically \citep{Woosley07}.  The chemical
composition of the puls-PISN ejecta would be characterized by large amounts of
carbon and oxygen, and some helium ($\sim 0.3$~\Msun).  Only about several
years thereafter, a GRB would be produced when the core collapses.  
Therefore,
the GRB jet should interact with  the very dense ejecta, which might result in
an extremely luminous afterglow if the GRB jet were energetic enough.  
\citet{Heger03} also suggested the possibility of occurrence of both
puls-PISN and GRB from the same progenitor for very low metallicity. 
However,  in their scenario,  the puls-PISN is supposed to eject the
hydrogen-rich envelope from the progenitor, while in our scenario based on
the CHE models, no hydrogen would be contained in the ejecta of the puls-PISN.  The
accompanied supernova (if it were not out-shined by the GRB afterglow)  would not
appear as a SN 2006gy-like event \citep[luminous type IIn][]{Smith07} but
rather look like a Quimby-type supernova \citep{Quimby11} or a SN Ibn 
if helium were excited as a result of the interaction \citep{Foley07, Pastorello07}. 

The GRB progenitors with $12 \lsim M_\mathrm{ZAMS}/M_\odot \lsim 56$ may not
suffer the pulsational pair-instability. In these models, however, the critical
rotation is maintained at the surface for a significant fraction of the
evolutionary time, which leads to significant mass loss (Fig.~\ref{fig:vrot}).
In particular, the mass loss rate becomes very high during the last $10^4
-10^5$ yrs, because of the rapid decrease of the radius of the star
(Fig.~\ref{fig:hr1}).  The mass loss would predominantly occur via a decretion
disk in this case, and radiation driven winds along the poles must be very weak
or negligible (see Sect.~\ref{sect:method}).  This may lead to a very complex,
highly aspherical structure of the circumstellar medium, which might be
imprinted on the afterglow \citep[see][for a related study]{Marle08}.  

Recently several authors investigated possible observational signatures from
collapsars in very massive stars with $M > 250$~\Msun{} \citep{Komissarov10,
Meszaros10, Suwa11, Toma11}.  Our result implies that a GRB-like event from
such a super-collapsar may not occur. First of all,  CHE is inhibited for very
massive stars (Sect.~\ref{sect:che}), and consequently  they are expected to have a very
extended hydrogen envelope, from which a jet from the central engine may not break
out easily.  Even if  CHE were realized, the final angular momentum in such a
massive star would not be high enough to produce a GRB (see discussion in
Sect.~\ref{sect:rotation}).  \citet{Suwa11} argued that for very massive stars,
accretion of the long-lived hydrogen envelope would result in jet-breakout even
if they had a supergiant envelope.  However, our 250, 300 and 500~\Msun{}
models show that the specific angular momentum would become well below
$j_\mathrm{K,lso}$ for the entire mass range including the hydrogen envelope,
at the pre-collapse stage (Fig.~\ref{fig:jspec}). Direct collapse of the whole
star into a BH without any remarkable visible event is the most likely outcome
in this case.  Although this conclusion is based on our assumption of
the presence of weak-seed magnetic fields in Pop III stars, our preliminary results
show that necessary conditions for a super-collapsar still cannot be easily fulfilled
even without the assumption of magnetic fields (Yoon et al., in preparation).

According to theoretical studies on the formation of the first stars in the
early Universe,  Pop III stars may have two different classes in terms of
initial mass function \citep{Johnson06, Yoshida07, Bromm09}: the 
first-generation stars formed via collapse of neutral primordial gas in isolated
dark-matter minihalos (Pop III.1),  and the stars formed in an environment
affected by photo-ionization from PopIII.1 stars (Pop III.2).  Formation of Pop
III.1 and Pop III.2 stars is governed by different cooling agents (H$_2$ and HD
molecules, respectively) and Pop III.2 stars are supposed to have much lower
initial masses ($\sim 40$~ \Msun) than Pop III.1 stars ($\sim 500$~\Msun). In
this framework,  our result implies that the initial mass function  of Pop
III.2 stars would be more favorable for GRB production than that of Pop III.1
stars.

\section{Conclusions}\label{sect:conclusion}

We presented a new grid of massive Pop III star models,  
which include the effects of rotation on the stellar structure and mass loss,
rotationally induced chemical mixing, and the transport of angular momentum.
The most up-to-date calibration for the mixing efficiency parameters by
\citet{Brott11} was adopted, and the presence of initial magnetic seed fields in
Pop III stars was assumed such that the transport of angular momentum is
dominated by magnetic torques due to the Spruit-Tayler dynamo \citep{Spruit02}.
We summarize the main conclusions from the analyses of the model grid as
follows.

\begin{enumerate}

\item Because of the larger overshooting parameter adopted here than in the previous work, our
non-rotating models become redder on the HR diagram, and stars with
$M_\mathrm{init} = 10$~\Msun{} and  $M_\mathrm{init} \ge 30$~\Msun{} become
red-supergiants, even when they do not experience a boosting of the CNO cycle
caused by mixing in the hydrogen shell source
(Sect.~\ref{sect:evolution};Figs.~\ref{fig:hr1} and \ref{fig:hr2}). \\ 

\item We find, for the first time, that dredge-up of helium core materials into
the convective hydrogen envelope occurs as the bottom of the envelope moves down
during core helium burning in non-rotating stars with $M_\mathrm{init} \ge
200$~\Msun{} (Sect.~\ref{sect:mixing}; Fig.~\ref{fig:kipp500-zoom}).  This
leads to mixing of large amounts of carbon and oxygen into the  convective
hydrogen envelope, and abundant production of primary nitrogen
(Fig.~\ref{fig:chem500}) .  The stars are already in the red-supergiant phase
when the dredge-up occurs. The hydrogen envelope at this stage must be unstable to
pulsational instability driven by partial ionization of hydrogen.  If
pulsations caused mass loss at this stage,  it would have interesting
consequences such as enrichment of CNO elements and other 
hydrogen-burning products into the circumstellar region, and
formation of dusts. \\

\item  The so-called chemically homogeneous evolution (CHE) is expected
for a certain range of masses ($13 \lsim M_\mathrm{ZAMS}/M_\odot \lsim 190$)
and rotational velocities (Sects.~\ref{sect:evolution} and \ref{sect:che};
Fig.~\ref{fig:final}). 
On one hand, thermodynamic conditions in more massive stars are more favorable
for chemical mixing due to Eddington-Sweet circulations.  On the other hand,
more massive stars can reach the critical rotation at an earlier stage of the
evolution to induce rapid loss of angular momentum,  and   Eddington Sweet
circulations can be significantly slowed down.  As a result, the limiting
rotational velocity for CHE decreases from $M_\mathrm{ZAMS} = 13$~\Msun{} to
about 60~\Msun, but increases from $\sim$ 150~\Msun{} to $\sim$190~\Msun{}. The
CHE does not occur any more for more massive stars ($M_\mathrm{ZAMS} >
190$~\Msun). \\

\item  The rotating models are kept rotating almost rigidly until the end of
core helium burning, which results from magnetic torques due to the
Spruit-Tayler dynamo (Sect.~\ref{sect:rotation}).  The
surface can reach the critical rotation during the evolution, and undergo mass
loss via a centrifugally driven wind,  which may form a decretion disk, if the
initial rotational velocity is sufficiently high (Fig.~\ref{fig:vrot}). In this
case, mass loss occurs mostly on the main sequence, in which the wind material
is mainly composed of hydrogen and helium, if the star does not undergo the
CHE.  With CHE, however, significant mass loss can occur even during the
post-main sequence phases, which may lead to significant enrichment of CNO
elements in the surrounding medium.  \\

\item  Chemical mixing by convection and/or rotation may lead to an increase of
the neutron excess in the stellar core during the helium-burning phase
(Sect.~\ref{sect:neutron}).   The most notable enhancement is observed in the
CHE models.  However, this effect is still weak 
compared to the expected increase of the neutron excess during the more
advanced nuclear burning stages, and it would have little
impact in the nucleosynthesis of PISNe. 
\\

\item The mass limits for PISN may decrease with rotation, as the core size
increases with rotationally induced mixing (Sect.~\ref{sect:pisn};
Fig.~\ref{fig:final}). In particular, stars of $84 \lsim
M_\mathrm{ZAMS}/M_\odot \lsim 190$ may produce hydrogen-free PISNe (type Ibc),
via CHE. Such PISNe may look like the type Ic SN 2007bi \citep{Gal-Yam09},
which is believed to originate from the pair instability. \\

\item If a Pop III star does not undergo CHE, the core cannot retain enough
angular momentum to produce an explosion powered by rotation like a jet-driven
supernova or gamma-ray burst (Sects.~\ref{sect:rotation} and \ref{sect:grb}).
With CHE,  conditions for gamma-ray burst (GRB) progenitors may be
fulfilled for $13\lsim M_\mathrm{ZAMS}/M_\odot \lsim 84$.  All of our GRB
progenitor models suffer mass loss caused by rotation until the end. This may lead to formation
of a complicated aspherical circumstellar structure, which should significantly
deviate from the typical density profile around a hot star ($\rho \propto
r^{-2}$).  On the other hand, super-collapsars from $M > 250$~\Msun{} are not
likely to occur. \\ 

\item For the mass range of $56\lsim M_\mathrm{ZAMS}/M_\odot \lsim 84$ in the
CHE regime, both pulsational PISN and GRB may occur one after another from the
same progenitor (Fig.~\ref{fig:final}).  In this case, the GRB jet would
penetrate a very  dense and massive medium  that has been formed shortly before
core-collapse by the ejecta of the pulsational PISN, and an extremely bright
afterglow may be produced  as a consequence.

\end{enumerate}

\begin{sidewaystable*}[h!]
\begin{minipage}[t][0mm]{\textwidth}
\caption{Some model properties. }\label{tab:properties}
\centering
\begin{tabularx}{\linewidth}{c | >{\centering\arraybackslash}X >{\centering\arraybackslash}X >{\centering\arraybackslash}X >{\centering\arraybackslash}X | %
>{\centering\arraybackslash}X >{\centering\arraybackslash}X >{\centering\arraybackslash}X >{\centering\arraybackslash}X >{\centering\arraybackslash}X %
>{\centering\arraybackslash}X >{\centering\arraybackslash}X >{\centering\arraybackslash}X >{\centering\arraybackslash}X >{\centering\arraybackslash}X c }\hline
                    & $T_\mathrm{c,f}$      & $\tau_\mathrm{H}$     & $\tau_\mathrm{f}$      & $M_\mathrm{f}$      & $M_\mathrm{He-Core}$     & $M_\mathrm{CO-Core}$ &%
$Y_\mathrm{s, TAMS}$& $Y_\mathrm{s, f}$  & $\log X_\mathrm{s,C}$ & $\log X_\mathrm{s, N}$& $\log X_\mathrm{s, O}$ & $\log J_\mathrm{f}$ &  $j_\mathrm{CO,f}$ & Mode \\
                    & [$10^8$~K]            & [$10^6$~yr]           & [$10^6$~yr]            & [$M_\odot$]         & [$M_\odot$]               & [$M_\odot$]          &%
                    &                    &                       &                       &                        &  [erg~s]             & [$\mathrm{cm^2~s^{-1}}$] &  \\
\hline
 m10vk00 &   16.53 &   24.76 &   26.11 &   10.00 &    2.76 &    2.08 &    0.24 &    0.24 &  $-\infty$   &  $-\infty$   &  $-\infty$   &    $-\infty$ &      0.00 & NE\\
 m10vk02 &   12.76 &   25.14 &   26.58 &   10.00 &    2.64 &    2.00 &    0.25 &    0.30 &  $-13.57$ &  $-11.67$ &  $-13.43$ &   51.78 &      0.63 & NE\\
 m10vk04 &   12.37 &   25.84 &   27.41 &    9.84 &    2.35 &    1.78 &    0.29 &    0.32 &  $-12.71$ &  $-10.85$ &  $-12.64$ &   51.95 &      0.51 & NE\\
 m10vk06 &   12.38 &   32.79 &   34.27 &    9.41 &    2.60 &    1.95 &    0.36 &    0.38 &  $-12.78$ &  $-10.95$ &  $-12.75$ &   51.89 &      0.60 & NE\\
 m10vk08 &   11.15 &   34.20 &   35.76 &    9.11 &    2.40 &    1.80 &    0.35 &    0.38 &  $-12.98$ &  $-11.15$ &  $-12.95$ &   51.88 &      0.53 & NE \\
\hline
 m15vk00 &   13.28 &   13.64 &   14.35 &   15.00 &    5.26 &    3.65 &    0.24 &    0.24 &    $-\infty$ &    $-\infty$ &    $-\infty$ &    $-\infty$ &      0.00 & NE \\
 m15vk02 &   22.93 &   14.58 &   15.28 &   15.00 &    5.31 &    3.72 &    0.27 &    0.27 &  $-13.20$ &  $-11.07$ &  $-12.64$ &   52.11 &      1.70 & NE\\
 m15vk04 &   18.89 &   15.81 &   16.52 &   14.45 &    4.76 &    3.44 &    0.31 &    0.31 &  $-12.86$ &  $-10.76$ &  $-12.36$ &   52.16 &      1.52 & NE \\
 m15vk06 &   14.71 &   24.08 &   24.68 &   11.51 &   11.51 &    9.46 &    0.96 &    0.42 &  $ -1.69$ &  $ -2.64$ &  $ -0.26$ &   51.43 &     61.05 & CHE\\
\hline
 m20vk00 &   10.85 &    9.65 &   10.20 &   20.00 &    7.73 &    5.89 &    0.24 &    0.24 &    $-\infty$ &    $-\infty$ &    $-\infty$ &    $-\infty$ &      0.00 & NE \\
 m20vk02 &    9.59 &   10.02 &   10.56 &   20.00 &    7.14 &    5.55 &    0.27 &    0.27 &  $-12.47$ &  $-10.38$ &  $-11.98$ &   52.34 &      3.11 & NE \\
 m20vk03 &    8.82 &   11.84 &   12.33 &   19.41 &    8.46 &    6.61 &    0.39 &    0.39 &  $-11.45$ &  $ -9.41$ &  $-11.05$ &   52.23 &      4.02 & NE \\
 m20vk04 &   16.46 &   14.72 &   15.20 &   15.93 &   15.93 &   13.52 &    0.97 &    0.31 &  $ -0.78$ &  $ -2.71$ &  $ -0.29$ &   51.71 &    101.61 & CHE \\
 m20vk06 &   13.35 &   15.79 &   16.30 &   15.19 &   15.19 &   12.40 &    0.97 &    0.36 &  $ -1.74$ &  $ -2.76$ &  $ -0.21$ &   51.71 &     98.81 & CHE \\
\hline
 m30vk00 &   23.05 &    6.37 &    6.82 &   30.00 &   12.42 &   11.18 &    0.24 &    0.24 &  $-\infty$   &  $-\infty$   &  $-\infty$   &    $-\infty$ &      0.00 & NE \\
 m30vk02 &   14.75 &    6.66 &    7.11 &   29.76 &   12.74 &   11.17 &    0.27 &    0.28 &  $-11.64$ &   $-9.66$ &  $-11.31$ &   52.56 &      7.07 & NE\\
 m30vk03 &   20.78 &    8.77 &    9.17 &   24.45 &   24.45 &   21.47 &    0.96 &    0.25 &  $ -1.62$ &   $-2.86$ &  $ -0.14$ &   52.03 &    153.40 & CHE \\
 m30vk04 &   19.13 &    9.10 &    9.51 &   23.45 &   23.45 &   20.31 &    0.97 &    0.28 &  $ -1.83$ &   $-2.89$ &  $ -0.16$ &   52.01 &    146.09 & CHE\\
 m30vk06 &   17.55 &    9.64 &   10.06 &   22.31 &   22.31 &   19.32 &    0.97 &    0.23 &  $ -1.61$ &   $-2.94$ &  $ -0.13$ &   51.97 &    142.41 & CHE \\
\hline
 m60vk00 &   12.79 &    3.88 &    4.22 &   60.00 &   30.14 &   28.49 &    0.24 &    0.24 &    $-\infty$ &    $-\infty$ &    $-\infty$ &    $-\infty$ &      0.00 & NE \\
 m60vk02 &   11.74 &    4.45 &    4.76 &   56.42 &   37.78 &   35.32 &    0.59 &    0.62 &  $-10.26$ &   $-8.32$ &  $-10.03$ &   52.43 &     21.80 & NE\\
 m60vk03 &   19.79 &    4.82 &    5.14 &   46.79 &   46.79 &   42.85 &    0.97 &    0.15 &  $ -1.35$ &   $-3.70$ &  $ -0.10$ &   52.41 &    211.63 & CHE \\
 m60vk04 &   20.00 &    4.95 &    5.28 &   44.79 &   44.79 &   40.87 &    0.97 &    0.16 &  $ -1.34$ &   $-3.90$ &  $ -0.11$ &   52.39 &    209.16 & CHE\\
 m60vk06 &   20.40 &    5.19 &    5.52 &   42.11 &   42.11 &   38.49 &    0.97 &    0.17 &  $ -1.39$ &   $-3.86$ &  $ -0.11$ &   52.35 &    205.64 & CHE\\
\hline
\end{tabularx}
\tablefoot{
Each column has the following meaning. 
$T_\mathrm{c}$: central temperature at the end of the calculation, 
$\tau_\mathrm{H}$ : duration for core hydrogen burning, 
$\tau_\mathrm{f}$ : evolutionary time from the beginning to the end of the calculation, 
$M_\mathrm{f}$ : final mass,  
$M_\mathrm{He-Core}$ : size of the helium core at the end of the calculation 
 (the upper boundary is defined as the location where the hydrogen mass fraction becomes less than 0.01), 
$M_\mathrm{CO-Core}$ : size of the carbon-oxygen core at the end of the calculation
 (the upper boundary is defined as the location where the helium mass fraction becomes less than 0.01) ,
$Y_\mathrm{s,TAMS}$ : surface helium mass fraction at the terminal age of the main sequence, 
$Y_\mathrm{s,f}$ : surface helium mass fraction at the end of the calculation,
$X_\mathrm{s,C}$ : surface carbon mass fraction at the end of the calculation,
$X_\mathrm{s,N}$ : surface nitrogen mass fraction at the end of the calculation,
$X_\mathrm{s,O}$ : surface oxygen mass fraction at the end of the calculation,
$J_\mathrm{f}$ : total angular momentum at the end of the calculation,
$j_\mathrm{3M,f}$ : mean specific angular momentum of the innermost $3~M_\odot$ at the end of the calculation,
$j_\mathrm{CO,f}$ : mean specific angular momentum of the carbon-oxygen core at the end of the calculation.
The different modes of evolution are indicated by NE (normal evolution), TE (transitionary evolution) and CHE (chemically homogeneous evolution), 
as defined in Sect.~\ref{sect:evolution}. 
}
\end{minipage}
\end{sidewaystable*}

\begin{sidewaystable*}[h!]
\begin{minipage}[t][ 0mm]{\textwidth}
{\bf Table~\ref{tab:properties}.} Continued
\vskip 1cm
\centering
\begin{tabularx}{\linewidth}{c | >{\centering\arraybackslash}X >{\centering\arraybackslash}X >{\centering\arraybackslash}X >{\centering\arraybackslash}X | %
>{\centering\arraybackslash}X >{\centering\arraybackslash}X >{\centering\arraybackslash}X >{\centering\arraybackslash}X >{\centering\arraybackslash}X %
>{\centering\arraybackslash}X >{\centering\arraybackslash}X >{\centering\arraybackslash}X >{\centering\arraybackslash}X >{\centering\arraybackslash}X c }\hline
                    & $T_\mathrm{c,f}$      & $\tau_\mathrm{H}$     & $\tau_\mathrm{f}$      & $M_\mathrm{f}$      & $M_\mathrm{He CORE}$     & $M_\mathrm{CO CORE}$ &%
 $Y_\mathrm{s, TAMS}$  & $Y_\mathrm{s, f}$  & $\log X_\mathrm{s,C}$ & $\log X_\mathrm{s, N}$& $\log X_\mathrm{s, O}$ & $\log J_\mathrm{f}$ &  $j_\mathrm{CO,f}$ & Mode \\
                    & [$10^8$~K]            & [$10^6$~yr]           & [$10^6$~yr]            & [$M_\odot$]         & [$M_\odot$]               & [$M_\odot$]          &%
                    &                       &                       &                       &                        &  [erg~s]            & [$\mathrm{cm^2~s^{-1}}$] &  \\
\hline
 m100vk00 &   18.94 &    3.02 &    3.30 &  100.00 &   53.44 &   51.93 &    0.24 &    0.24 &   $-\infty$  &   $-\infty$  &   $-\infty$  &    $-\infty$ &       0.00 & NE\\
 m100wk02 &   17.33 &    3.34 &    3.60 &   92.52 &   70.65 &   65.81 &    0.51 &    0.53 &  $-10.33$ &   $-8.38$ &  $-10.15$ &   52.36 &      28.23 & NE \\
 m100vk03 &   16.51 &    3.59 &    3.88 &   76.51 &   76.51 &   70.96 &    0.95 &    0.18 &   $-0.96$ &   $-3.96$ &  $-0.15$ &   52.68 &     251.28 & CHE\\
 m100vk04 &   16.75 &    3.64 &    3.93 &   74.40 &   74.40 &   68.78 &    0.96 &    0.23 &   $-1.10$ &   $-3.51$ &  $-0.16$ &   52.68 &     255.10 & CHE\\
 m100vk06 &   17.01 &    3.74 &    4.04 &   70.43 &   70.43 &   64.79 &    0.96 &    0.19 &   $-0.98$ &   $-3.46$ &  $-0.16$ &   52.64 &     243.29 & CHE\\
\hline 
 m150vk00 &    3.96 &    2.61 &    2.88 &  150.00 &   84.25 &   82.22 &    0.24 &    0.24 &  $-\infty$   &   $-\infty$  &  $-\infty$   &    $-\infty$ &       0.00 & NE \\
 m150vk02 &   14.92 &    2.82 &    3.10 &  137.50 &  101.73 &   99.10 &    0.47 &    0.48 &  $-10.34$ &   $-8.39$ & $ -10.17$ &   52.63 &      35.32 & NE\\
 m150vk03 &   14.43 &    2.96 &    3.21 &  121.95 &  118.93 &  109.95 &    0.87 &    0.96 &  $ -8.15$ &   $-6.33$ & $  -8.17$ &   52.73 &     156.77 & CHE \\
 m150vk04 &   14.06 &    3.02 &    3.28 &  112.54 &  112.54 &  105.71 &    0.90 &    0.22 &  $ -0.64$ &   $-3.20$ & $  -0.26$ &   52.90 &     290.06 & CHE \\
 m150vk06 &   14.03 &    3.09 &    3.35 &  105.06 &  105.06 &   97.93 &    0.92 &    0.13 &  $ -0.74$ &   $-3.69$ & $  -0.17$ &   52.83 &     261.79 & CHE \\
\hline 
 m200vk00 &    9.55 &    2.40 &    2.65 &  200.00 &  107.32 &   97.71 &    0.24 &    0.24 &  $-\infty$   &  $-\infty$   &  $-\infty$   &    $-\infty$ &       0.00 & NE \\
 m200vk02 &   13.97 &    2.61 &    2.86 &  183.96 &  129.37 &  112.26 &    0.59 &    0.69 &  $ -9.73$ &  $ -7.91$ &  $ -9.69$ &   52.91 &      27.01 & NE\\
 m200vk04 &   24.96 &    2.73 &    2.97 &  159.99 &  140.33 &  130.42 &    0.81 &    0.86 &  $ -9.52$ &  $ -7.65$ &  $ -9.51$ &   52.55 &      60.52 & TE\\
 m200vk06 &   13.72 &    2.79 &    3.04 &  146.43 &  141.07 &  131.57 &    0.86 &    0.96 &  $ -9.08$ &  $ -7.24$ &  $ -9.13$ &   52.86 &     173.13 & CHE \\
\hline 
 m250vk00 &    9.02 &    2.28 &    2.53 &  250.00 &  140.87 &  128.85 &    0.24 &    0.24 &  $-\infty$   &  $-\infty$   &  $-\infty$   &    $-\infty$ &       0.00 & NE \\
 m250vk02 &   13.29 &    2.36 &    2.61 &  229.23 &  157.87 &  153.98 &    0.49 &    0.49 &  $-10.23$ &  $ -8.30$ & $ -10.07$ &   53.10 &      43.19 & NE\\
 m250vk04 &    4.33 &    2.50 &    2.73 &  199.22 &  179.31 &  169.59 &    0.81 &    0.88 &  $ -9.45$ &  $ -7.59$ & $  -9.65$ &   52.64 &      68.58 & TE \\
 m250vk06 &    9.70 &    2.53 &    2.77 &  190.07 &  166.72 &  157.28 &    0.78 &    0.85 &  $ -9.55$ &  $ -7.69$ & $  -9.69$ &   52.66 &      58.72 & TE \\

\hline 
 m300vk00 &    7.85 &    2.18 &    2.43 &  300.00 &  161.16 &  148.47 &    0.24 &    0.24 &  $-\infty$   &  $-\infty$   &  $-\infty$   &    $-\infty$ &       0.00 & NE\\
 m300vk02 &   12.89 &    2.27 &    2.51 &  276.18 &  174.88 &  162.23 &    0.52 &    0.54 &  $-10.10$ &   $-8.17$ &   $-9.92$ &   53.25 &      43.13 & NE\\
 m300vk04 &   10.96 &    2.39 &    2.63 &  245.69 &  189.10 &  173.31 &    0.68 &    0.80 &  $ -9.42$ &   $-7.63$ &   $-9.45$ &   53.00 &      48.70 & NE\\
\hline 
 m500vk00 &    9.63 &    2.01 &    2.25 &  500.00 &  259.16 &  239.49 &    0.24 &    0.24 &  $-\infty$   &  $-\infty$   &  $-\infty$   &    $-\infty$ &       0.00 &  NE\\
 m500vk04 &    7.99 &    2.07 &    2.32 &  413.26 &  297.04 &  281.31 &    0.55 &    0.58 &  $-10.05$&  $-8.09$ &  $-10.14$ &   53.45 &      64.82 & NE \\
\hline 
 m100wk00\footnotemark &  2.51 &   1.84 &    1.88 & 1000.00 &  622.23 &    n/a  &    0.24 &    0.24 &  $-\infty$   &  $-\infty$   &  $-\infty$   &    $-\infty$ &      0.00 & NE\\
m1000vk04\footnotemark  & 1.32 &   1.54 &    1.54 &  808.29 &  n/a    &    n/a  &    n/a  &    0.52 &  $-10.09$ &   $-8.14$ &  $-10.26$ &   53.95 &     n/a  & NE \\
\hline 
\end{tabularx}
\addtocounter{footnote}{-2}
\stepcounter{footnote}\footnotetext{1. The calculation was terminated before the end of core helium exhaustion}
\stepcounter{footnote}\footnotetext{2. The calculation was terminated before the end of the main sequence}
\end{minipage}
\end{sidewaystable*}

\begin{table*}[h!]
\caption{Total energy (E), total number (N), and average wavelength ($\bar{\lambda}$) of the ionizing photons
for hydrogen  (H), neutral helium (He) and singly ionized helium (He$^+$) during the evolution
for each model sequence. 
}\label{tab:iflux}
\begin{tabularx}{\linewidth}
{c | >{\centering\arraybackslash}X >{\centering\arraybackslash}X >{\centering\arraybackslash}X
     >{\centering\arraybackslash}X >{\centering\arraybackslash}X >{\centering\arraybackslash}X
     >{\centering\arraybackslash}X >{\centering\arraybackslash}X >{\centering\arraybackslash}X} 
\hline  
        & E($\gamma_\mathrm{H}$)& N($\gamma_\mathrm{H}$) & $\bar{\lambda} (\gamma_\mathrm{H}$) & %
          E($\gamma_\mathrm{He}$)& N($\gamma_\mathrm{He}$) & $\bar{\lambda} (\gamma_\mathrm{He}$) & %
          E($\gamma_\mathrm{He^+}$)& N($\gamma_\mathrm{He^+}$) & $\bar{\lambda}(\gamma_\mathrm{He^+}$) \\
        & [erg]     &           & [\AA]     &    [erg] &            & [\AA]     &     [erg] &           & [\AA]    \\
\hline
m10vk00 &  1.69E+52 &  5.51E+62 &  648.51 &  3.59E+51 &  7.62E+61 &  421.82 &  1.22E+49 &  1.30E+59 &  211.04 \\
m10vk02 &  1.61E+52 &  5.28E+62 &  653.63 &  3.23E+51 &  6.89E+61 &  423.67 &  9.31E+48 &  9.91E+58 &  211.43 \\
m10vk04 &  1.36E+52 &  4.55E+62 &  667.55 &  2.32E+51 &  5.01E+61 &  428.98 &  3.87E+48 &  4.14E+58 &  212.65 \\
m10vk06 &  1.72E+52 &  5.73E+62 &  663.56 &  3.08E+51 &  6.64E+61 &  427.52 &  6.06E+48 &  6.47E+58 &  212.26 \\
m10vk08 &  1.56E+52 &  5.24E+62 &  667.37 &  2.68E+51 &  5.78E+61 &  428.85 &  4.59E+48 &  4.91E+58 &  212.54 \\
\hline
m15vk00 &  3.46E+52 &  1.06E+63 &  608.94 &  1.05E+52 &  2.15E+62 &  405.57 &  1.24E+50 &  1.30E+60 &  207.25 \\
m15vk02 &  3.61E+52 &  1.11E+63 &  612.28 &  1.07E+52 &  2.19E+62 &  406.87 &  1.16E+50 &  1.21E+60 &  207.59 \\
m15vk04 &  3.46E+52 &  1.09E+63 &  625.25 &  9.20E+51 &  1.91E+62 &  412.56 &  6.67E+49 &  7.01E+59 &  208.95 \\
m15vk06 &  1.10E+53 &  2.88E+63 &  518.98 &  5.56E+52 &  1.00E+63 &  358.72 &  5.81E+51 &  5.49E+61 &  187.98 \\
\hline
m20vk00 &  5.66E+52 &  1.67E+63 &  586.19 &  2.03E+52 &  4.04E+62 &  395.80 &  4.20E+50 &  4.33E+60 &  204.93 \\
m20vk02 &  5.70E+52 &  1.69E+63 &  588.77 &  2.01E+52 &  4.02E+62 &  396.95 &  3.91E+50 &  4.03E+60 &  205.25 \\
m20vk03 &  7.69E+52 &  2.25E+63 &  580.70 &  2.86E+52 &  5.68E+62 &  395.21 &  5.86E+50 &  6.05E+60 &  205.22 \\
m20vk04 &  1.63E+53 &  4.06E+63 &  495.57 &  9.05E+52 &  1.60E+63 &  351.31 &  1.09E+52 &  1.02E+62 &  186.83 \\
m20vk06 &  1.55E+53 &  3.89E+63 &  499.29 &  8.48E+52 &  1.50E+63 &  352.48 &  9.95E+51 &  9.37E+61 &  187.08 \\
\hline
m30vk00 &  1.05E+53 &  2.95E+63 &  558.54 &  4.45E+52 &  8.58E+62 &  383.51 &  1.63E+51 &  1.66E+61 &  202.00 \\
m30vk02 &  1.06E+53 &  3.02E+63 &  563.56 &  4.39E+52 &  8.49E+62 &  384.82 &  1.53E+51 &  1.56E+61 &  202.31 \\
m30vk03 &  2.61E+53 &  6.15E+63 &  468.83 &  1.60E+53 &  2.75E+63 &  342.02 &  2.25E+52 &  2.11E+62 &  186.35 \\
m30vk04 &  2.56E+53 &  6.11E+63 &  473.67 &  1.55E+53 &  2.68E+63 &  343.86 &  2.10E+52 &  1.97E+62 &  186.46 \\
m30vk06 &  2.42E+53 &  5.81E+63 &  476.28 &  1.45E+53 &  2.51E+63 &  344.72 &  1.95E+52 &  1.83E+62 &  186.42 \\
\hline
m60vk00 &  2.58E+53 &  6.76E+63 &  521.37 &  1.31E+53 &  2.41E+63 &  366.30 &  9.02E+51 &  8.98E+61 &  197.72 \\
m60vk02 &  4.09E+53 &  1.04E+64 &  503.97 &  2.24E+53 &  4.09E+63 &  363.14 &  1.63E+52 &  1.63E+62 &  197.76 \\
m60vk03 &  5.41E+53 &  1.20E+64 &  442.55 &  3.62E+53 &  6.04E+63 &  331.81 &  6.03E+52 &  5.63E+62 &  185.67 \\
m60vk04 &  5.27E+53 &  1.18E+64 &  446.09 &  3.48E+53 &  5.83E+63 &  333.05 &  5.70E+52 &  5.32E+62 &  185.62 \\
m60vk06 &  4.93E+53 &  1.11E+64 &  448.25 &  3.24E+53 &  5.44E+63 &  333.95 &  5.22E+52 &  4.88E+62 &  185.70 \\
\hline
m100vk00 &  4.63E+53 &  1.18E+64 &  504.36 &  2.52E+53 &  4.54E+63 &  357.71 &  2.26E+52 &  2.23E+62 &  195.39 \\
m100vk02 &  5.20E+53 &  1.23E+64 &  469.73 &  3.23E+53 &  5.67E+63 &  348.89 &  3.44E+52 &  3.37E+62 &  194.88 \\
m100vk03 &  9.12E+53 &  1.98E+64 &  432.21 &  6.29E+53 &  1.04E+64 &  327.59 &  1.12E+53 &  1.05E+63 &  185.58 \\
m100vk04 &  8.79E+53 &  1.92E+64 &  434.24 &  6.02E+53 &  9.95E+63 &  328.32 &  1.06E+53 &  9.94E+62 &  185.56 \\
m100vk06 &  8.31E+53 &  1.82E+64 &  435.35 &  5.67E+53 &  9.38E+63 &  328.70 &  9.98E+52 &  9.31E+62 &  185.45 \\
\hline
m150vk00 &  7.03E+53 &  1.75E+64 &  494.17 &  3.97E+53 &  7.04E+63 &  352.50 &  4.11E+52 &  4.01E+62 &  193.98 \\
m150vk02 &  8.88E+53 &  2.19E+64 &  489.55 &  5.11E+53 &  9.08E+63 &  353.13 &  5.17E+52 &  5.06E+62 &  194.31 \\
m150vk03 &  1.30E+54 &  2.87E+64 &  440.71 &  8.78E+53 &  1.48E+64 &  334.53 &  1.32E+53 &  1.26E+63 &  190.80 \\
m150vk04 &  1.28E+54 &  2.79E+64 &  431.75 &  8.87E+53 &  1.46E+64 &  327.87 &  1.57E+53 &  1.47E+63 &  186.17 \\
m150vk06 &  1.23E+54 &  2.66E+64 &  430.85 &  8.49E+53 &  1.40E+64 &  327.17 &  1.52E+53 &  1.42E+63 &  185.76 \\
\hline
m200vk00 &  9.35E+53 &  2.30E+64 &  489.40 &  5.37E+53 &  9.46E+63 &  349.92 &  5.96E+52 &  5.80E+62 &  193.22 \\
m200vk02 &  1.41E+54 &  3.45E+64 &  485.24 &  8.29E+53 &  1.47E+64 &  351.78 &  8.54E+52 &  8.35E+62 &  194.33 \\
m200vk04 &  1.59E+54 &  3.65E+64 &  454.98 &  1.04E+54 &  1.78E+64 &  341.18 &  1.33E+53 &  1.29E+63 &  193.04 \\
m200vk06 &  1.59E+54 &  3.53E+64 &  442.45 &  1.07E+54 &  1.81E+64 &  335.60 &  1.56E+53 &  1.50E+63 &  191.39 \\
\hline
m250vk00 &  1.11E+54 &  2.70E+64 &  484.51 &  6.48E+53 &  1.13E+64 &  347.65 &  7.59E+52 &  7.36E+62 &  192.68 \\
m250vk02 &  1.36E+54 &  3.31E+64 &  483.27 &  8.03E+53 &  1.41E+64 &  349.36 &  8.87E+52 &  8.64E+62 &  193.59 \\
m250vk04 &  1.89E+54 &  4.40E+64 &  462.38 &  1.20E+54 &  2.08E+64 &  344.63 &  1.43E+53 &  1.39E+63 &  193.83 \\
m250vk06 &  1.76E+54 &  4.10E+64 &  462.31 &  1.12E+54 &  1.94E+64 &  344.48 &  1.34E+53 &  1.30E+63 &  193.73 \\
\hline
m300vk00 &  1.35E+54 &  3.30E+64 &  484.31 &  7.91E+53 &  1.38E+64 &  347.22 &  9.41E+52 &  9.11E+62 &  192.43 \\
m300vk02 &  1.89E+54 &  4.61E+64 &  485.29 &  1.11E+54 &  1.96E+64 &  351.77 &  1.15E+53 &  1.12E+63 &  194.04 \\
m300vk04 &  2.00E+54 &  4.69E+64 &  465.91 &  1.26E+54 &  2.19E+64 &  346.47 &  1.43E+53 &  1.40E+63 &  194.14 \\
\hline
m500vk00 &  2.15E+54 &  5.20E+64 &  480.96 &  1.27E+54 &  2.21E+64 &  345.26 &  1.59E+53 &  1.53E+63 &  191.81 \\
m500vk04 &  2.45E+54 &  6.00E+64 &  487.16 &  1.42E+54 &  2.52E+64 &  351.39 &  1.49E+53 &  1.45E+63 &  194.33 \\
\hline
\end{tabularx}
\end{table*}

\begin{table*}[h!]
\caption{Yields of stellar winds from rotating models, in units of solar mass}\label{tab:yield1}
\begin{tabularx}{\linewidth}
{c | >{\centering\arraybackslash}X >{\centering\arraybackslash}X >{\centering\arraybackslash}X %
     >{\centering\arraybackslash}X >{\centering\arraybackslash}X >{\centering\arraybackslash}X %
     >{\centering\arraybackslash}X >{\centering\arraybackslash}X }
\hline
         &   $^1$H    &   $^4$He   &  $^{12}$C  &  $^{13}$C  & $^{14}$N  &  $^{16}$O  &   $^{18}$O  & $^{22}$Ne  \\
\hline
m10vk02  &  0.00E+00  &  0.00E+00  &  0.00E+00  &  0.00E+00  &  0.00E+00  &  0.00E+00  &  0.00E+00  &  0.00E+00 \\
m10vk04  &  1.11E-01  &  4.47E-02  &  2.56E-22  &  8.29E-23  &  3.79E-20  &  1.11E-21  &  6.85E-28  &  8.26E-58 \\
m10vk04  &  4.05E-01  &  1.83E-01  &  9.18E-17  &  2.97E-17  &  1.29E-14  &  3.50E-16  &  1.88E-22  &  1.02E-22 \\
m10vk06  &  6.37E-01  &  2.55E-01  &  1.24E-20  &  4.02E-21  &  1.82E-18  &  5.26E-20  &  3.20E-26  &  6.37E-36 \\
\hline
m15vk02 &  0.00E+00  &  0.00E+00  &  0.00E+00  &  0.00E+00  &  0.00E+00  &  0.00E+00  &  0.00E+00  &  0.00E+00 \\
m15vk04 &  3.79E-01  &  1.67E-01  &  7.36E-14  &  2.37E-14  &  9.36E-12  &  2.39E-13  &  1.14E-19  &  2.53E-20 \\
m15vk06 &  5.32E-01  &  2.45E+00  &  2.46E-02  &  2.08E-03  &  9.79E-03  &  4.76E-01  &  3.59E-06  &  2.47E-04 \\
\hline
m20vk02 &  0.00E+00  &  0.00E+00  &  0.00E+00  &  0.00E+00  &  0.00E+00  &  0.00E+00  &  0.00E+00  &  0.00E+00 \\
m20vk03 &  3.58E-01  &  2.31E-01  &  2.06E-12  &  6.64E-13  &  2.29E-10  &  5.21E-12  &  2.00E-18  &  6.26E-18 \\
m20vk04 &  3.80E-02  &  2.03E+00  &  6.08E-02  &  1.84E-03  &  7.22E-03  &  1.36E-01  &  1.98E-06  &  2.35E-04 \\
m20vk06 &  6.90E-01  &  3.23E+00  &  6.28E-02  &  5.29E-03  &  1.19E-02  &  8.07E-01  &  3.80E-05  &  8.70E-04 \\
\hline
m30vk02 &  1.77E-01  &  6.43E-02  &  3.34E-13  &  1.08E-13  &  3.63E-11  &  8.40E-13  &  3.24E-19  &  6.72E-19 \\
m30vk03 &  5.61E-02  &  3.92E+00  &  6.22E-02  &  7.92E-03  &  2.52E-02  &  1.47E+00  &  2.30E-05  &  2.28E-03 \\
m30vk04 &  1.41E-01  &  4.90E+00  &  4.94E-02  &  7.68E-03  &  2.24E-02  &  1.43E+00  &  1.30E-05  &  1.88E-03 \\
m30vk06 &  1.26E+00  &  4.98E+00  &  1.92E-02  &  6.81E-04  &  2.06E-02  &  1.40E+00  &  1.53E-06  &  2.23E-03 \\
\hline
m60vk02 &  1.52E+00  &  2.06E+00  &  1.61E-10  &  5.17E-11  &  1.43E-08  &  2.83E-10  &  8.85E-17  &  9.24E-16 \\
m60vk03 &  2.08E-01  &  9.51E+00  &  9.72E-02  &  1.01E-02  &  3.45E-02  &  3.34E+00  &  3.59E-06  &  4.36E-03 \\
m60vk04 &  1.02E+00  &  1.08E+01  &  9.62E-02  &  8.32E-03  &  2.21E-02  &  3.26E+00  &  3.11E-06  &  3.64E-03 \\
m60vk06 &  3.87E+00  &  1.72E+01  &  1.36E-01  &  1.80E-02  &  4.20E-02  &  4.39E+00  &  9.97E-06  &  5.64E-03 \\
\hline
m100vk02 &  3.73E+00  &  3.75E+00  &  3.12E-10  &  9.96E-11  &  2.67E-08  &  4.49E-10  &  1.49E-16  &  1.77E-15 \\
m100vk03 &  9.12E-01  &  1.74E+01  &  6.18E-01  &  1.53E-02  &  3.30E-02  &  4.53E+00  &  2.54E-05  &  2.90E-03 \\
m100vk04 &  2.97E+00  &  1.84E+01  &  3.20E-01  &  1.96E-02  &  4.04E-02  &  3.82E+00  &  1.04E-05  &  2.78E-03 \\
m100vk06 &  6.80E+00  &  1.87E+01  &  3.27E-01  &  2.72E-03  &  4.91E-02  &  3.71E+00  &  3.97E-07  &  2.49E-03 \\
\hline
m150vk02 &  6.62E+00  &  5.88E+00  &  5.57E-10  &  1.77E-10  &  4.73E-08  &  7.85E-10  &  2.60E-16  &  3.64E-15 \\
m150vk03 &  4.65E+00  &  2.34E+01  &  1.61E-08  &  5.14E-09  &  1.11E-06  &  1.34E-08  &  4.44E-15  &  5.52E-11 \\
m150vk04 &  7.24E+00  &  2.67E+01  &  1.35E+00  &  1.56E-02  &  4.88E-02  &  2.05E+00  &  4.60E-07  &  3.25E-04 \\
m150vk06 &  1.22E+01  &  2.76E+01  &  1.81E+00  &  1.71E-02  &  4.66E-02  &  3.23E+00  &  8.78E-07  &  6.80E-04 \\
\hline
m200vk02 &  7.22E+00  &  8.82E+00  &  1.25E-09  &  4.02E-10  &  1.01E-07  &  1.72E-09  &  4.98E-16  &  9.36E-15 \\
m200vk04 &  1.38E+01  &  2.62E+01  &  5.80E-09  &  1.86E-09  &  4.54E-07  &  5.49E-09  &  1.67E-15  &  4.82E-14 \\
m200vk06 &  1.96E+01  &  3.39E+01  &  1.02E-08  &  3.27E-09  &  7.48E-07  &  8.42E-09  &  2.78E-15  &  8.95E-14 \\
\hline
m150vk02 &  1.16E+01  &  9.13E+00  &  1.06E-09  &  3.35E-10  &  8.26E-08  &  1.27E-09  &  3.98E-16  &  6.60E-15 \\
m250vk04 &  1.99E+01  &  3.08E+01  &  6.70E-09  &  2.17E-09  &  5.19E-07  &  4.11E-09  &  2.60E-15  &  5.75E-14 \\
m250vk06 &  2.75E+01  &  3.24E+01  &  5.65E-09  &  1.81E-09  &  4.59E-07  &  3.77E-09  &  2.27E-15  &  4.82E-14 \\
\hline
m300vk02 &  1.25E+01  &  1.13E+01  &  1.42E-09  &  4.56E-10  &  1.21E-07  &  2.09E-09  &  6.06E-16  &  1.16E-14 \\
m300vk04 &  2.80E+01  &  3.63E+01  &  6.13E-09  &  1.98E-09  &  4.98E-07  &  4.21E-09  &  2.49E-15  &  5.46E-14 \\
\hline
m500vk04 &  5.53E+01  &  3.14E+01  &  1.06E-08  &  1.18E-09  &  2.64E-07  &  2.58E-08  &  1.36E-15  &  3.01E-13 \\
\hline
\end{tabularx}
\end{table*}

\begin{table*}[h!]
\caption{Yields in the last calculated model for each model sequence, in units of solar mass. 
The wind yields are not included here. 
$\eta_\mathrm{c}$ denotes the neutron excess at the center of the last model. 
\Xnmax is the maximum mass fraction of $^{14}$N that has been achieved during the evolution. 
}\label{tab:yield2}
\begin{tabularx}{\linewidth}
{c | >{\centering\arraybackslash}X >{\centering\arraybackslash}X >{\centering\arraybackslash}X >{\centering\arraybackslash}X %
     >{\centering\arraybackslash}X >{\centering\arraybackslash}X >{\centering\arraybackslash}X >{\centering\arraybackslash}X %
     >{\centering\arraybackslash}X  >{\centering\arraybackslash}X}
\hline
         &   $^1$H  &  $^4$He & $^{12}$C & $^{13}$C & $^{14}$N & $^{16}$O & $^{18}$O & $^{22}$Ne& $\eta_\mathrm{c}$ & $X^\mathrm{max}(^{14}\mathrm{N})$ \\
\hline
m10vk00 & 4.90E+00 & 2.92E+00 & 1.84E-01 & 2.44E-09 & 6.13E-08 & 1.04E+00 & 8.18E-09 & 1.53E-07 & 4.69E-08 & 6.98E-07 \\
m10vk02 & 4.94E+00 & 2.95E+00 & 1.80E-01 & 2.39E-09 & 5.35E-08 & 1.05E+00 & 8.18E-09 & 1.96E-07 & 4.45E-08 & 6.47E-07  \\
m10vk04 & 5.04E+00 & 2.96E+00 & 9.21E-02 & 1.39E-09 & 2.83E-08 & 8.84E-01 & 7.27E-09 & 8.48E-08 & 3.95E-08 & 6.26E-07 \\
m10vk06 & 4.17E+00 & 3.19E+00 & 1.74E-01 & 2.10E-09 & 4.49E-08 & 1.06E+00 & 9.97E-09 & 1.95E-07 & 4.24E-08 & 6.25E-07 \\
m10vk08 & 4.12E+00 & 3.11E+00 & 1.50E-01 & 1.79E-09 & 3.59E-08 & 9.42E-01 & 7.18E-09 & 1.30E-07 & 3.16E-08 & 5.86E-07 \\
\hline
m15vk00 & 6.75E+00 & 4.35E+00 & 3.57E-01 & 2.82E-09 & 1.08E-07 & 2.12E+00 & 7.02E-08 & 3.23E-07 & 6.56E-08 & 1.10E-06 \\
m15vk02 & 6.46E+00 & 4.49E+00 & 3.21E-01 & 2.63E-09 & 8.32E-08 & 1.71E+00 & 7.69E-08 & 3.81E-07 & 8.33E-08 & 8.50E-07  \\
m15vk04 & 6.03E+00 & 4.70E+00 & 3.32E-01 & 5.16E-09 & 1.05E-07 & 2.13E+00 & 9.26E-08 & 4.62E-07 & 7.95E-08 & 8.42E-07  \\
m15vk06 & 1.99E-20 & 6.24E-01 & 1.99E-01 & 1.55E-04 & 6.13E-04 & 9.79E+00 & 6.09E-06 & 4.81E-03 & 6.90E-05 & 1.43E-02  \\
\hline
m20vk00 & 8.26E+00 & 5.55E+00 & 1.04E+00 & 4.57E-07 & 5.69E-08 & 4.69E+00 & 1.36E-07 & 2.72E-06 & 1.36E-07 & 9.60E-07   \\
m20vk02 & 8.08E+00 & 5.96E+00 & 1.37E+00 & 4.85E-05 & 8.16E-04 & 4.16E+00 & 1.47E-06 & 3.17E-06 & 1.22E-07 & 1.06E-03 \\
m20vk03 & 5.67E+00 & 6.69E+00 & 1.80E+00 & 7.00E-07 & 1.21E-07 & 5.12E+00 & 2.51E-07 & 4.30E-06 & 1.41E-07 & 1.21E-06 \\
m20vk04 & 1.97E-20 & 4.55E-01 & 9.89E-01 & 2.80E-04 & 1.30E-03 & 1.33E+01 & 5.44E-04 & 3.67E-03 & 1.58E-04 & 1.40E-02 \\
m20vk06 & 1.05E-20 & 5.61E-01 & 2.51E-01 & 3.80E-04 & 1.24E-03 & 1.29E+01 & 4.66E-04 & 2.86E-03 & 1.25E-04 & 1.16E-02  \\
\hline
m30vk00 & 1.07E+01 & 7.96E+00 & 3.37E-01 & 1.13E-05 & 2.58E-04 & 9.21E+00 & 1.44E-09 & 3.49E-06 & 1.48E-06 & 1.35E-02  \\
m30vk02 & 1.00E+01 & 8.00E+00 & 5.22E-01 & 2.90E-08 & 2.88E-07 & 1.04E+01 & 1.17E-07 & 6.13E-06 & 1.10E-06 & 1.57E-04 \\
m30vk03 & 1.85E-20 & 4.18E-01 & 3.99E-01 & 2.96E-04 & 1.27E-03 & 2.08E+01 & 4.92E-04 & 2.37E-03 & 1.69E-04 & 1.48E-02 \\
m30vk04 & 2.03E-20 & 4.98E-01 & 3.45E-01 & 3.15E-04 & 1.17E-03 & 1.98E+01 & 4.78E-04 & 2.40E-03 & 1.41E-04 & 1.61E-02  \\
m30vk06 & 1.96E-20 & 2.76E-01 & 4.29E-01 & 2.60E-05 & 7.65E-04 & 1.91E+01 & 1.99E-04 & 1.50E-03 & 1.70E-04 & 1.22E-02  \\
\hline
m60vk00 & 1.67E+01 & 1.41E+01 & 1.75E+00 & 7.61E-08 & 6.85E-07 & 2.49E+01 & 8.30E-07 & 5.29E-06 & 1.86E-06 & 2.38E-04  \\
m60vk02 & 5.40E+00 & 1.44E+01 & 2.05E+00 & 3.34E-05 & 9.21E-05 & 3.09E+01 & 1.59E-05 & 5.34E-06 & 8.26E-07 & 1.47E-04 \\
m60vk03 & 2.18E-20 & 2.81E-01 & 9.48E-01 & 6.44E-05 & 2.31E-04 & 3.84E+01 & 3.00E-05 & 6.88E-04 & 1.16E-04 & 1.11E-02  \\
m60vk04 & 2.68E-20 & 3.02E-01 & 9.25E-01 & 5.45E-05 & 1.41E-04 & 3.70E+01 & 1.94E-05 & 6.07E-04 & 9.32E-05 & 7.88E-03  \\
m60vk06 & 2.67E-20 & 2.90E-01 & 7.89E-01 & 6.66E-05 & 1.43E-04 & 3.49E+01 & 2.60E-05 & 7.60E-04 & 1.05E-04 & 7.91E-03  \\
\hline
m100vk00 & 2.52E+01 & 2.29E+01 & 2.68E+00 & 1.77E-06 & 2.87E-06 & 4.32E+01 & 3.50E-07 & 7.31E-06 & 6.14E-06 & 9.96E-03  \\
m100vk02 & 4.25E+00 & 2.17E+01 & 3.84E+00 & 6.25E-07 & 3.29E-06 & 5.42E+01 & 5.16E-06 & 9.58E-05 & 7.13E-07 & 1.75E-04  \\
m100vk03 & 2.01E-20 & 4.51E-01 & 2.71E+00 & 8.30E-05 & 1.64E-04 & 6.09E+01 & 3.37E-06 & 3.84E-04 & 4.80E-05 & 7.42E-03 \\
m100vk04 & 2.00E-18 & 5.97E-01 & 2.26E+00 & 2.54E-04 & 5.43E-04 & 5.91E+01 & 3.50E-05 & 6.49E-04 & 5.57E-05 & 9.11E-03 \\
m100vk06 & 4.52E-19 & 5.58E-01 & 2.50E+00 & 5.10E-05 & 7.45E-04 & 5.65E+01 & 4.77E-05 & 5.43E-04 & 5.17E-05 & 1.37E-02 \\
\hline
m150vk00 & 3.24E+01 & 3.54E+01 & 5.27E+00 & 1.52E-06 & 3.49E-05 & 6.67E+01 & 1.18E-08 & 1.26E-05 & 2.23E-06 & 1.10E-03 \\
m150vk02 & 9.37E+00 & 2.72E+01 & 5.12E+00 & 7.97E-07 & 5.89E-05 & 7.96E+01 & 3.13E-05 & 2.44E-04 & 1.19E-05 & 3.72E-03  \\
m150vk03 & 8.32E-02 & 9.86E+00 & 8.78E+00 & 1.56E-04 & 1.15E-03 & 8.82E+01 & 4.89E-04 & 5.23E-05 & 1.35E-06 & 1.62E-03 \\
m150vk04 & 1.56E-18 & 7.19E-01 & 7.96E+00 & 3.48E-04 & 1.45E-03 & 8.86E+01 & 2.23E-06 & 1.53E-04 & 1.20E-05 & 1.40E-02  \\
m150vk06 & 1.28E-18 & 3.99E-01 & 6.83E+00 & 9.79E-05 & 3.37E-04 & 8.36E+01 & 1.11E-06 & 9.69E-05 & 1.72E-05 & 1.07E-02 \\
\hline
m200vk00 & 3.97E+01 & 5.30E+01 & 8.49E+00 & 1.51E-03 & 3.71E-04 & 8.54E+01 & 2.58E-11 & 1.83E-05 & 6.36E-07 & 3.44E-03    \\
m200vk02 & 1.42E+01 & 4.48E+01 & 7.04E+00 & 2.89E-06 & 6.54E-05 & 9.74E+01 & 1.90E-05 & 7.52E-05 & 1.67E-06 & 4.46E-04 \\
m200vk04 & 1.96E+00 & 2.56E+01 & 7.92E+00 & 1.96E-06 & 5.44E-06 & 1.02E+02 & 6.20E-05 & 1.60E-04 & 4.76E-07 & 7.90E-05 \\
m200vk06 & 1.75E-01 & 1.29E+01 & 9.54E+00 & 3.91E-05 & 5.10E-04 & 1.03E+02 & 6.00E-04 & 5.11E-05 & 1.02E-06 & 3.96E-04  \\
\hline
m250vk00 & 4.27E+01 & 6.66E+01 & 1.06E+01 & 1.24E-03 & 7.68E-05 & 1.10E+02 & 3.26E-10 & 2.37E-05 & 7.10E-07 & 1.70E-04  \\
m250vk02 & 2.19E+01 & 5.31E+01 & 7.81E+00 & 7.93E-06 & 6.48E-05 & 1.18E+02 & 1.12E-05 & 2.94E-04 & 3.96E-06 & 1.76E-03   \\
m250vk04 & 1.28E+00 & 2.67E+01 & 1.02E+01 & 7.17E-05 & 4.15E-04 & 1.31E+02 & 3.92E-09 & 3.84E-06 & 6.69E-07 & 2.84E-04  \\
m250vk06 & 2.06E+00 & 2.90E+01 & 9.13E+00 & 9.13E-06 & 1.94E-04 & 1.22E+02 & 1.48E-04 & 1.11E-05 & 6.83E-07 & 3.52E-04   \\
\hline
m300vk00 & 5.33E+01 & 8.25E+01 & 1.27E+01 & 1.50E-02 & 1.27E-03 & 1.26E+02 & 4.67E-10 & 4.29E-05 & 6.03E-07 & 9.36E-03 \\
m300vk02 & 3.00E+01 & 8.01E+01 & 9.45E+00 & 2.62E-05 & 1.52E-04 & 1.27E+02 & 8.47E-05 & 3.19E-05 & 4.10E-07 & 1.40E-04  \\
m300vk04 & 4.19E+00 & 4.20E+01 & 9.89E+00 & 2.21E-06 & 1.24E-04 & 1.43E+02 & 1.74E-04 & 9.39E-06 & 5.29E-07 & 2.18E-04  \\
\hline
m500vk00 & 8.46E+01 & 1.45E+02 & 2.22E+01 & 3.41E-01 & 1.17E-01 & 1.97E+02 & 2.30E-09 & 3.57E-05 & 1.95E-07 & 8.59E-03 \\
m500vk04 & 2.63E+01 & 9.08E+01 & 1.47E+01 & 5.33E-04 & 3.43E-04 & 2.17E+02 & 6.56E-11 & 4.34E-06 & 3.92E-06 & 1.78E-03 \\
\hline
\end{tabularx}
\end{table*}

\vskip 10cm

\begin{acknowledgements}
We are grateful to the anonymous referee for many invaluable criticisms, which have led to a significant improvement of the paper. 
\end{acknowledgements}

\pagebreak

\begin{appendix}

\section{Kippenhan diagram}

\begin{figure*}
\begin{center}
\includegraphics[width=\columnwidth]{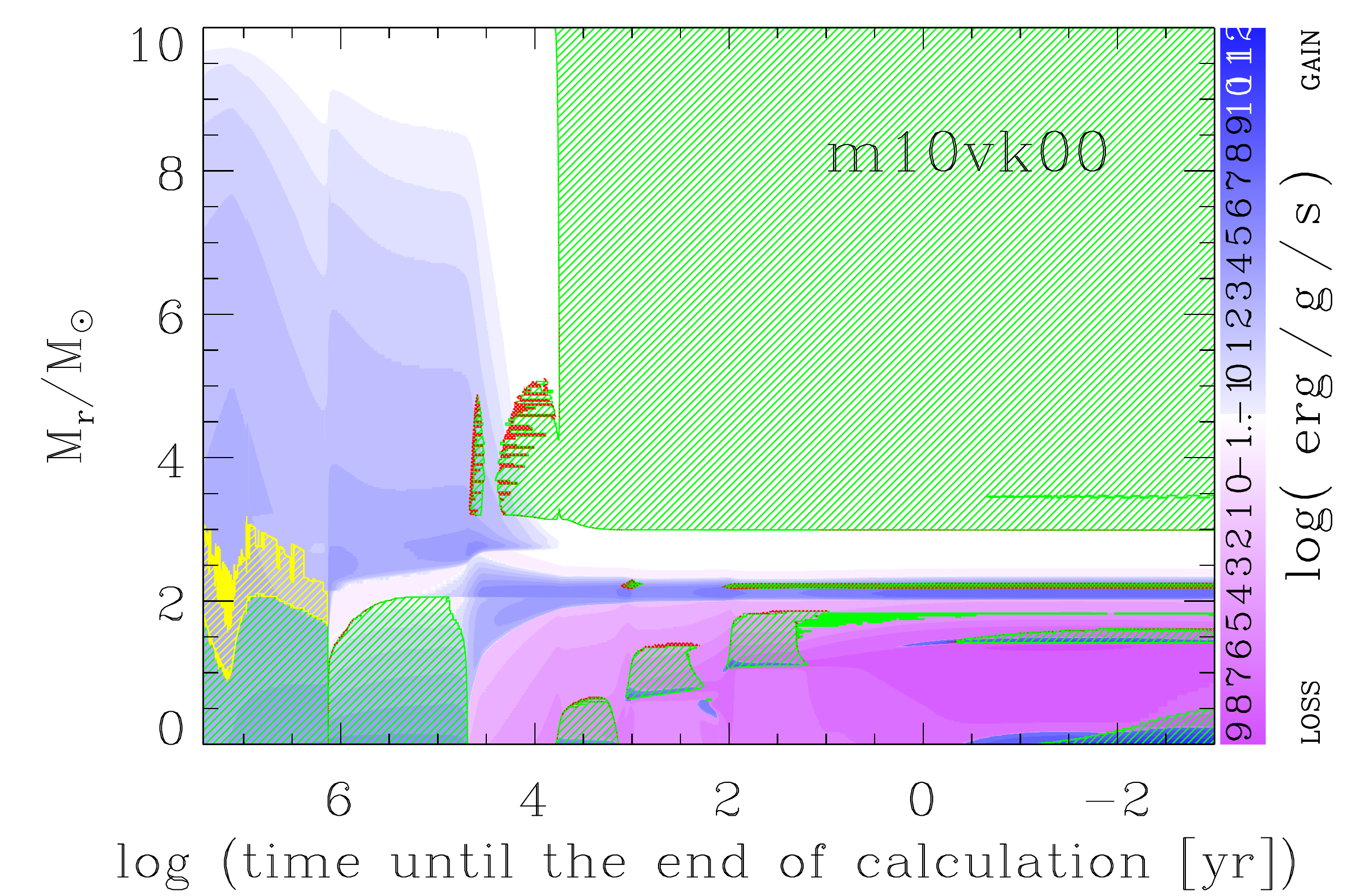}
\includegraphics[width=\columnwidth]{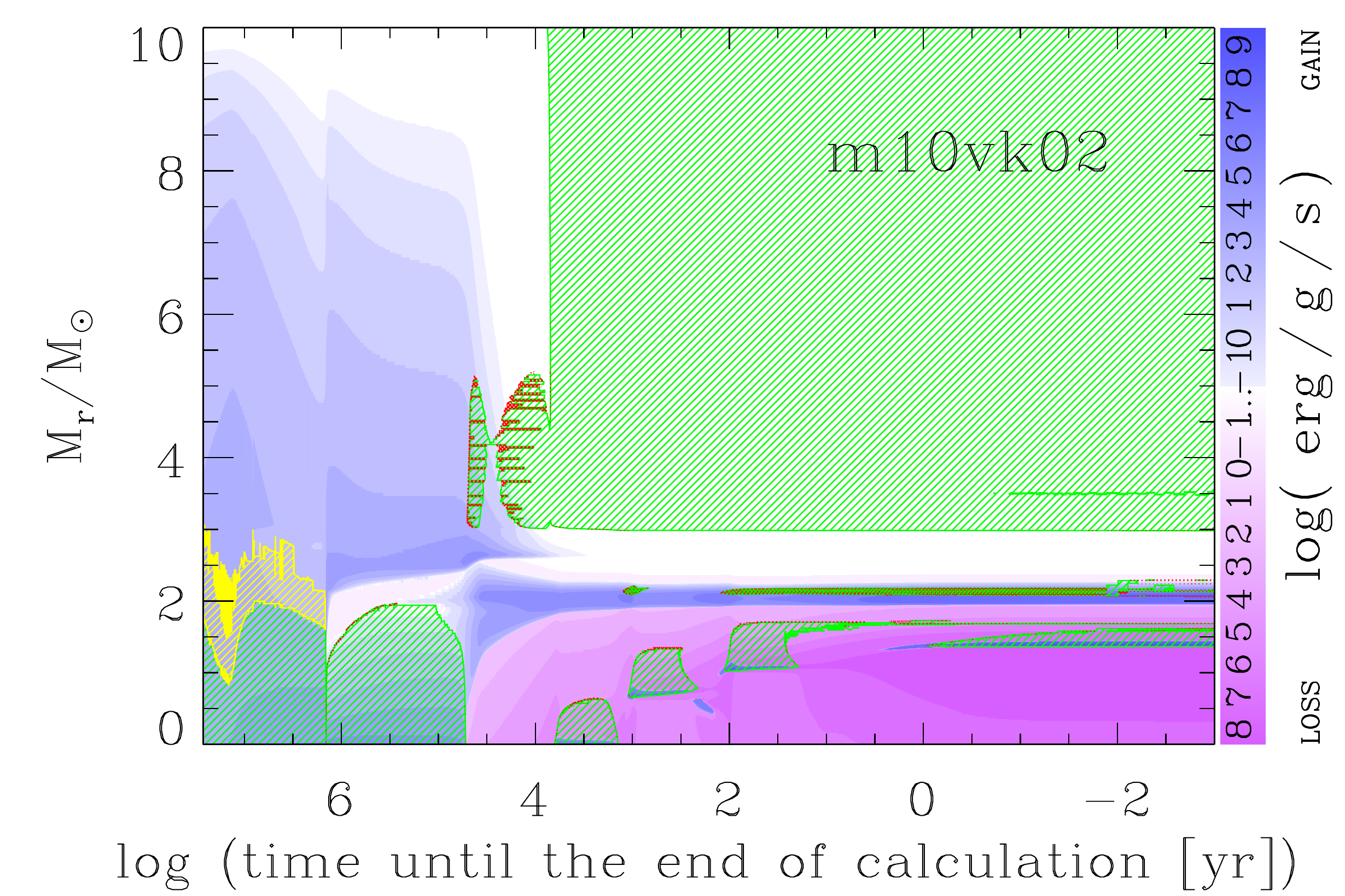}
\includegraphics[width=\columnwidth]{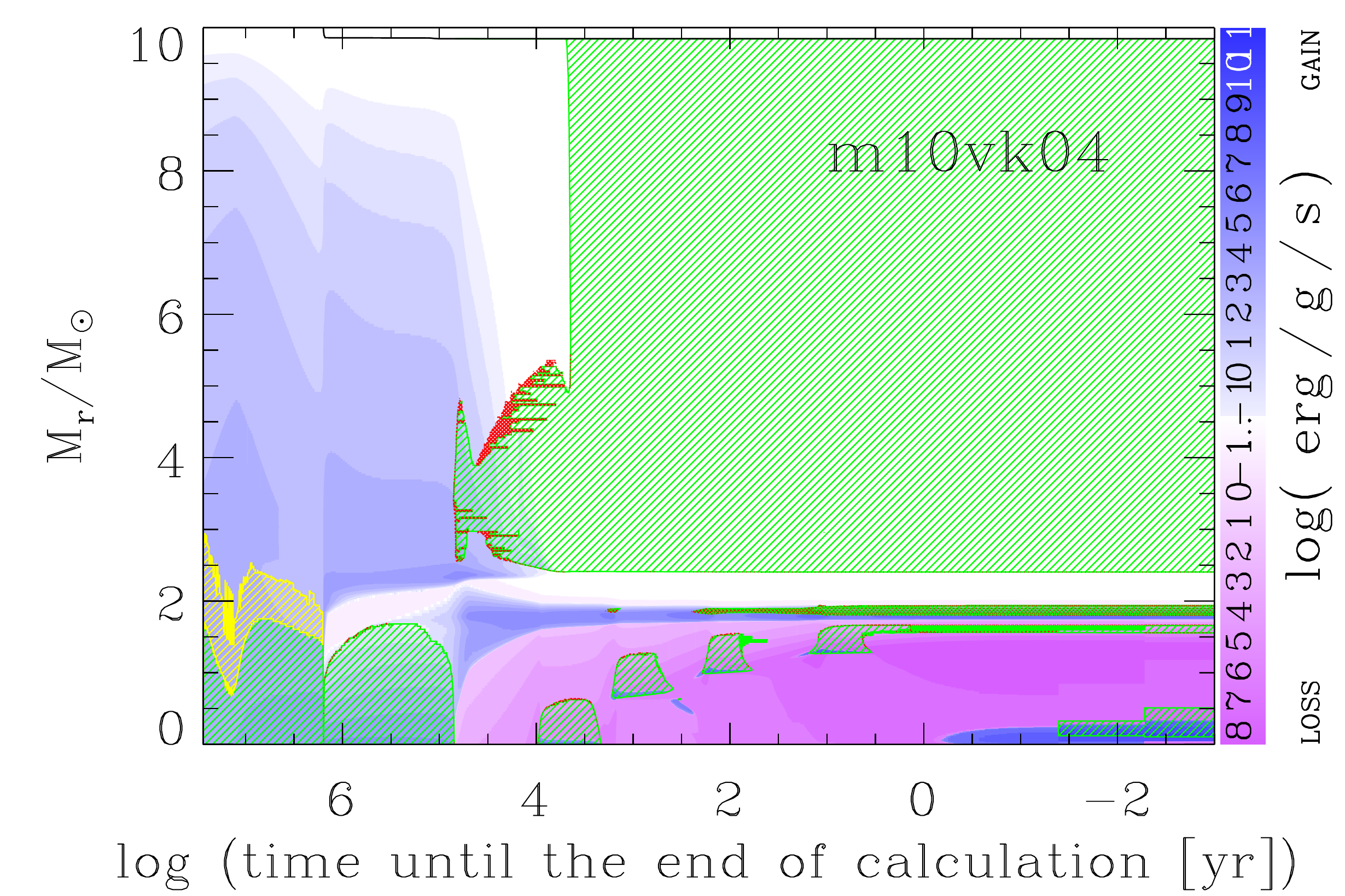}
\includegraphics[width=\columnwidth]{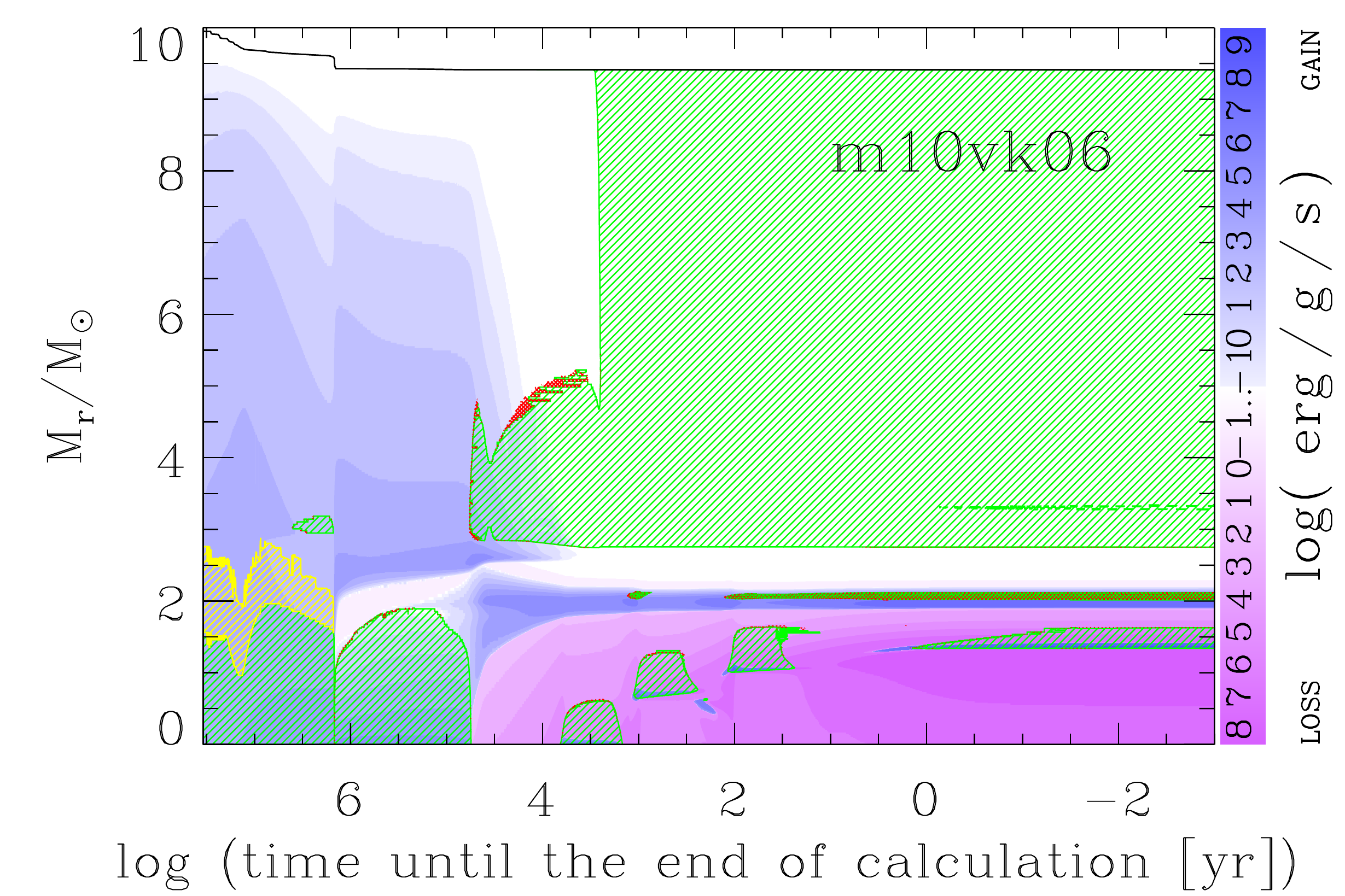}
\includegraphics[width=\columnwidth]{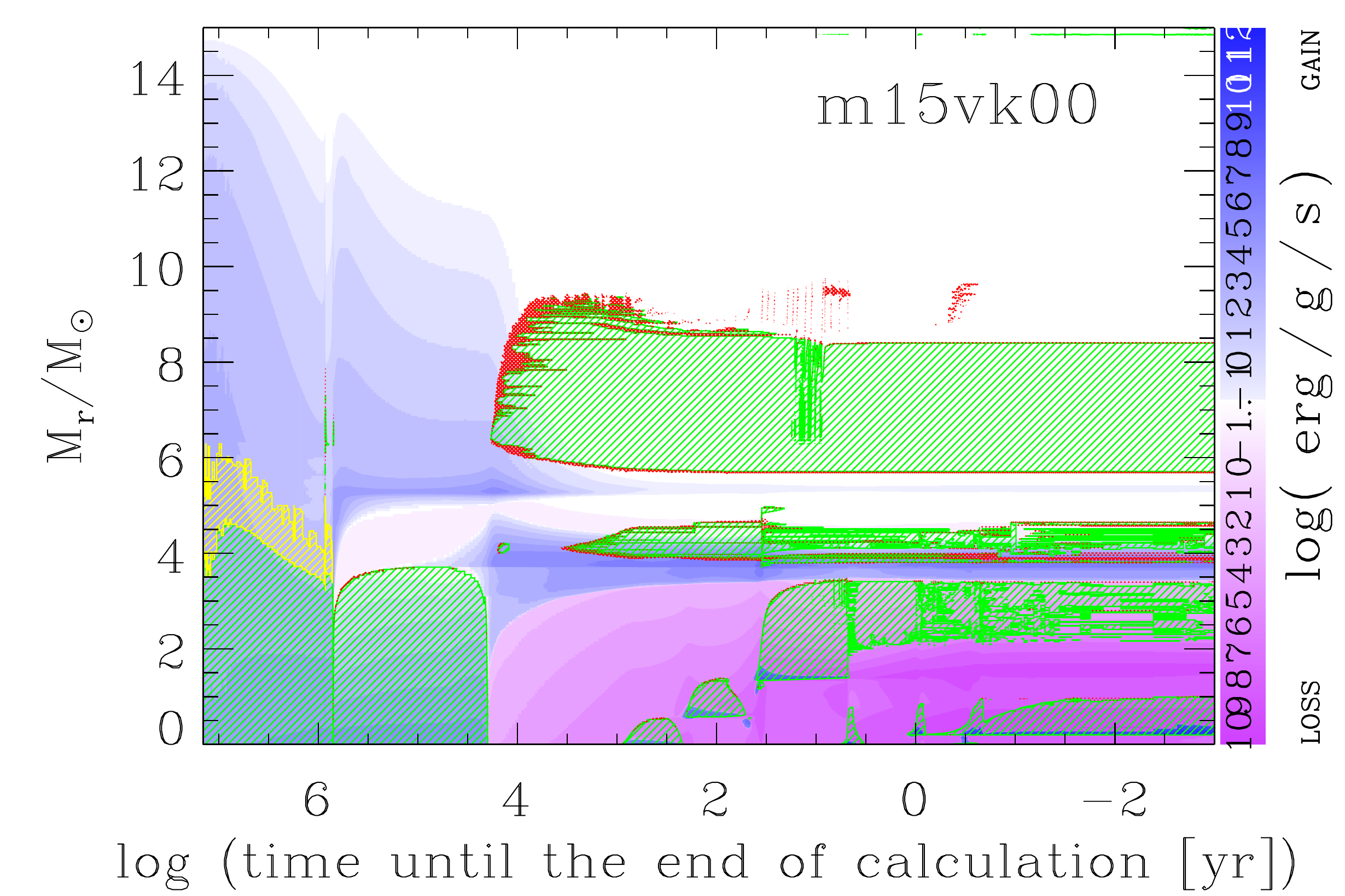}
\includegraphics[width=\columnwidth]{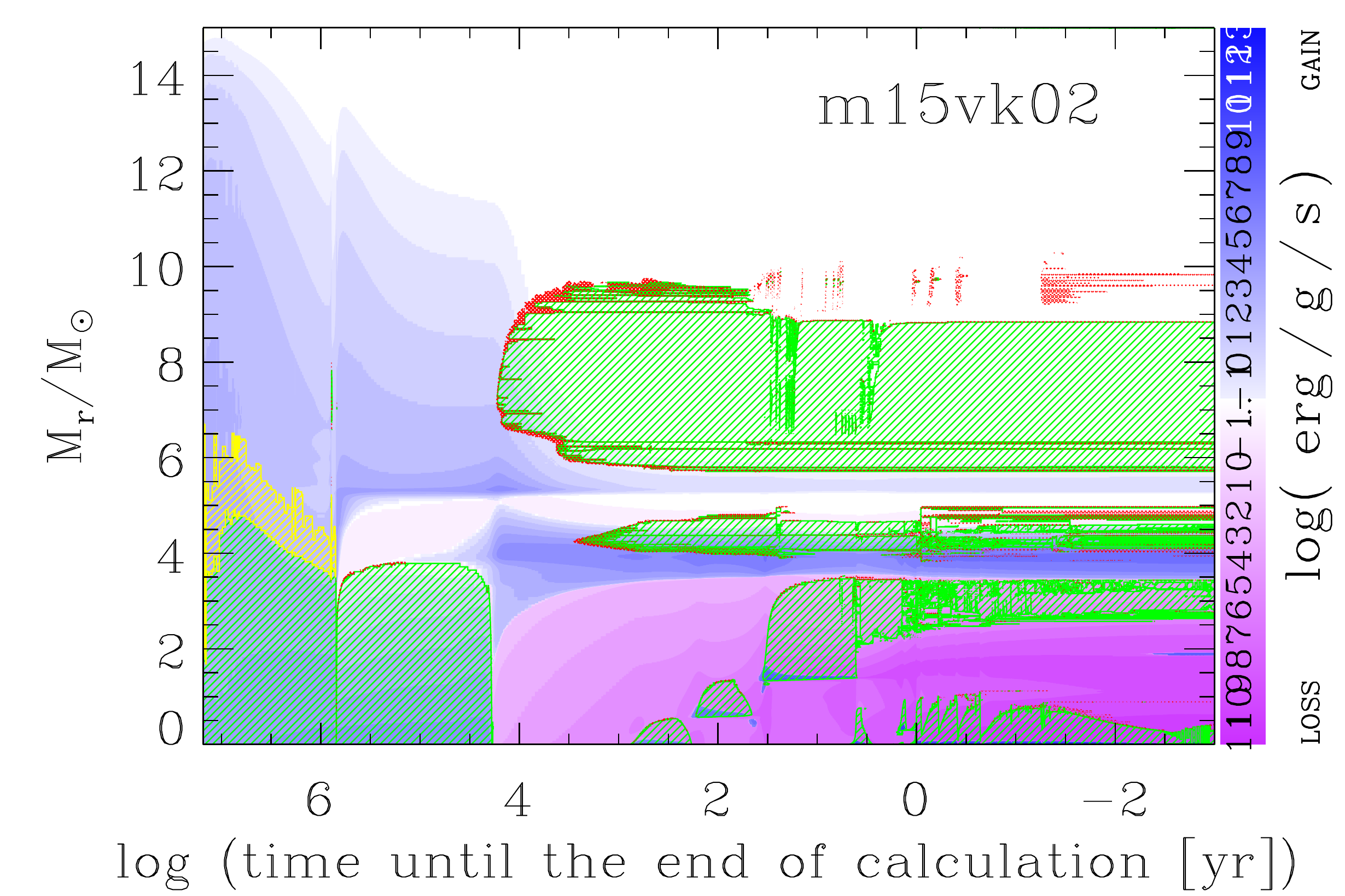}
\includegraphics[width=\columnwidth]{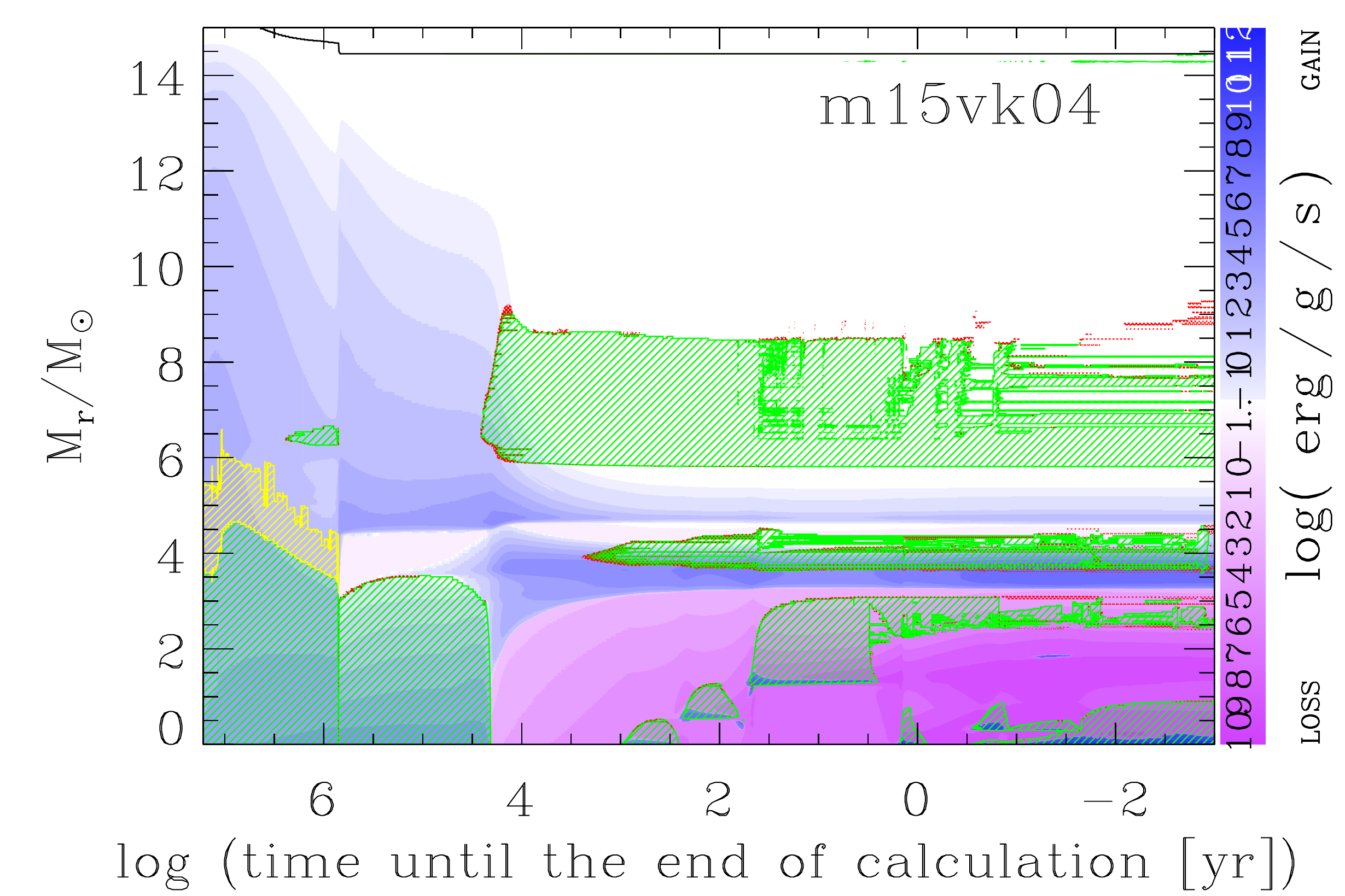}
\includegraphics[width=\columnwidth]{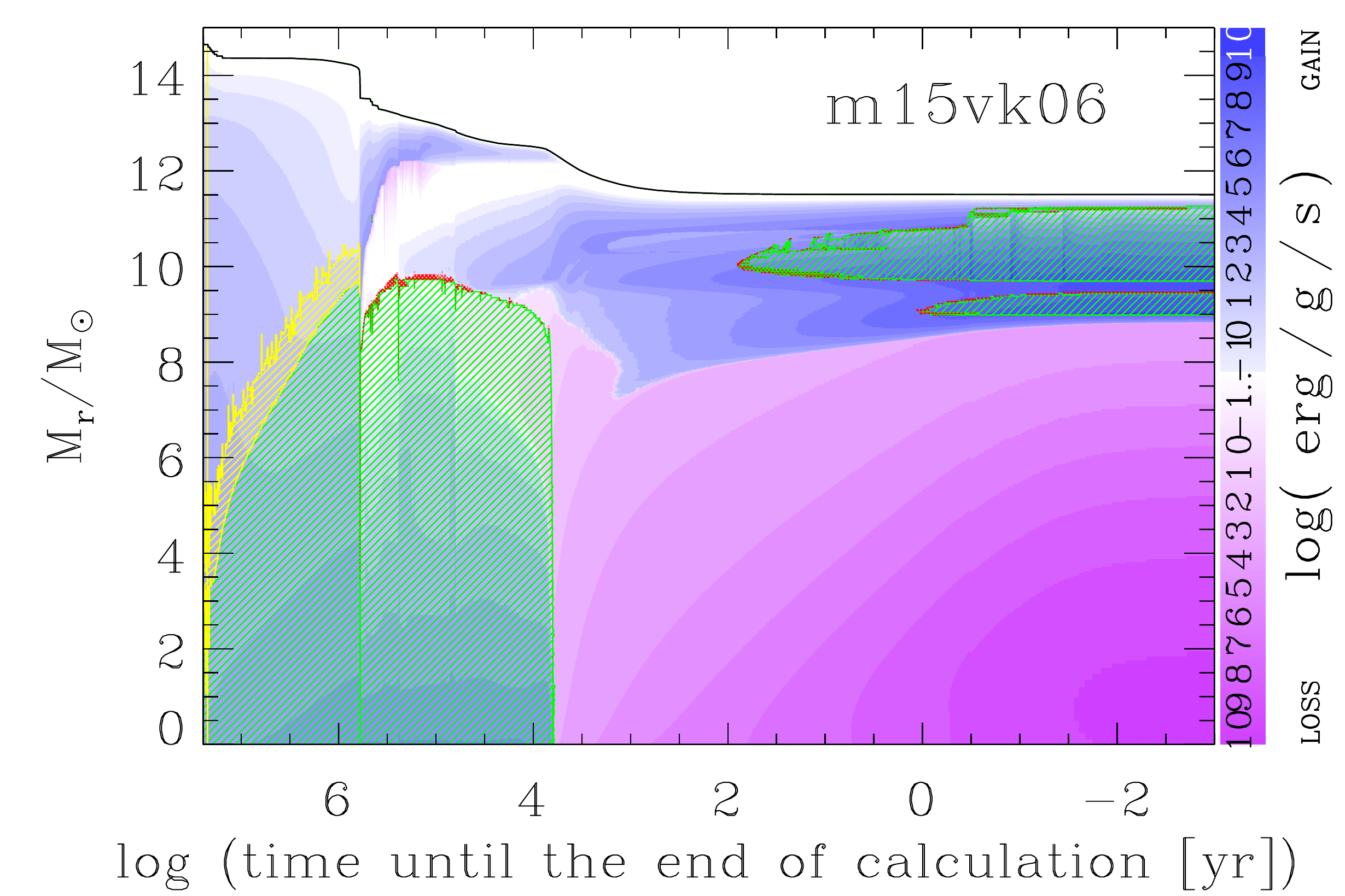}
\caption{Kippenhahn diagram for different model sequences. The sequence number is shown in each diagram. 
Convective and overshooting  layers are marked by green and yellow hatched lines, respectively. 
Semi-convective layers are marked by red dots. The net amount of energy loss and gain is indicated
by color shading. The surface of the star is marked by a black solid line. 
}
\label{fig:kipp1}
\end{center}
\end{figure*}

\begin{figure*}
\begin{center}
\includegraphics[width=\columnwidth]{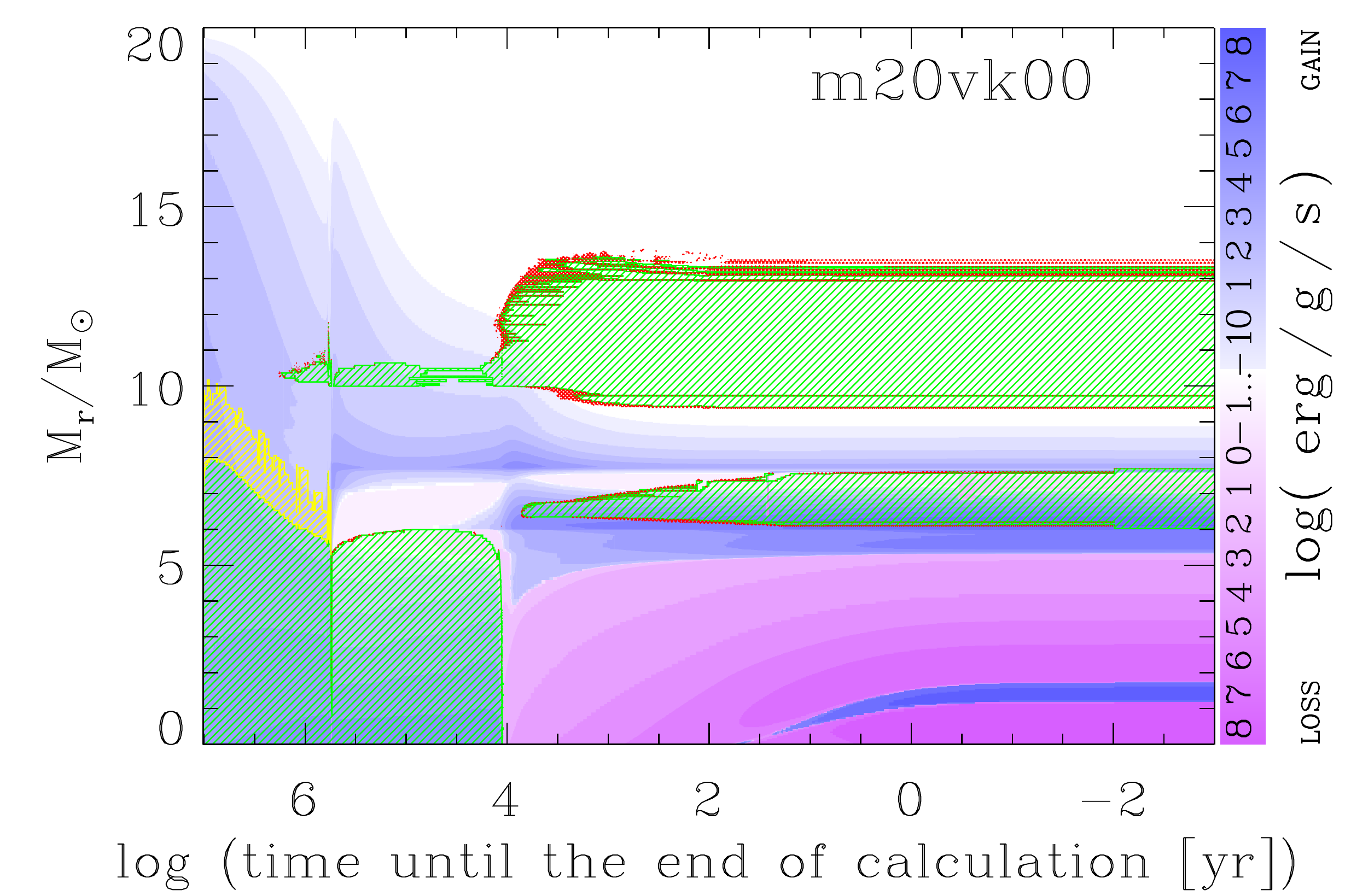}
\includegraphics[width=\columnwidth]{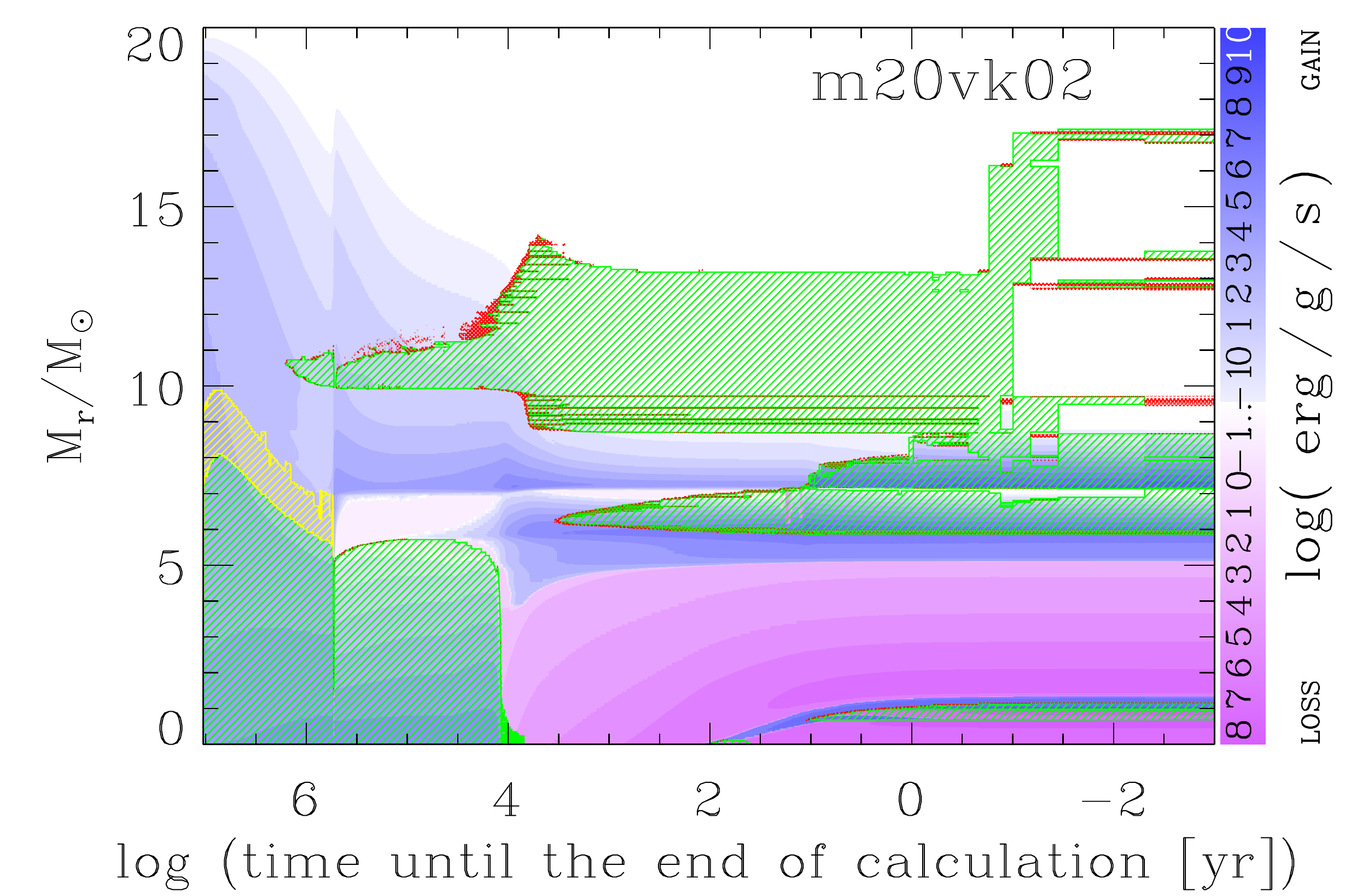}
\includegraphics[width=\columnwidth]{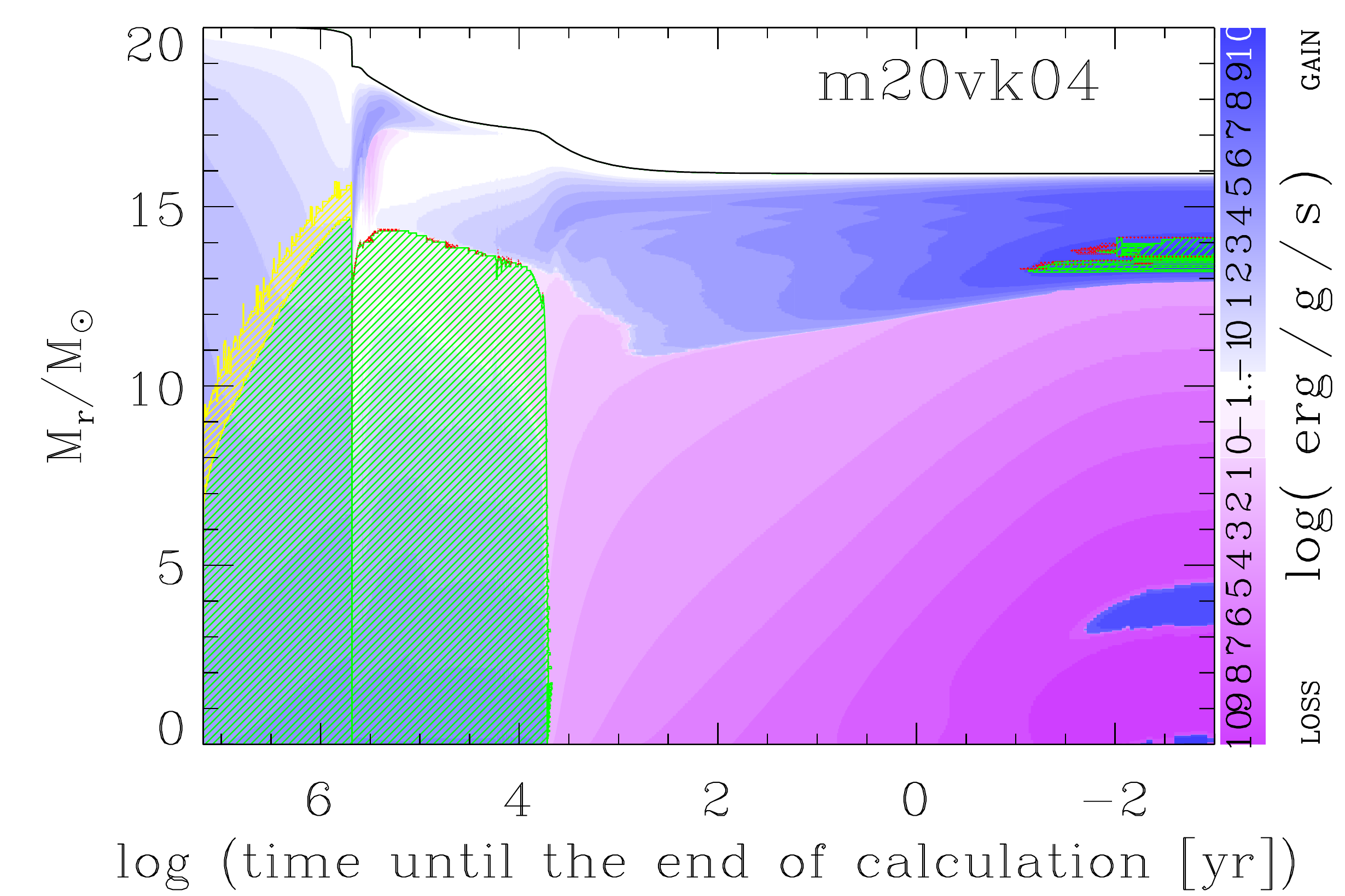}
\includegraphics[width=\columnwidth]{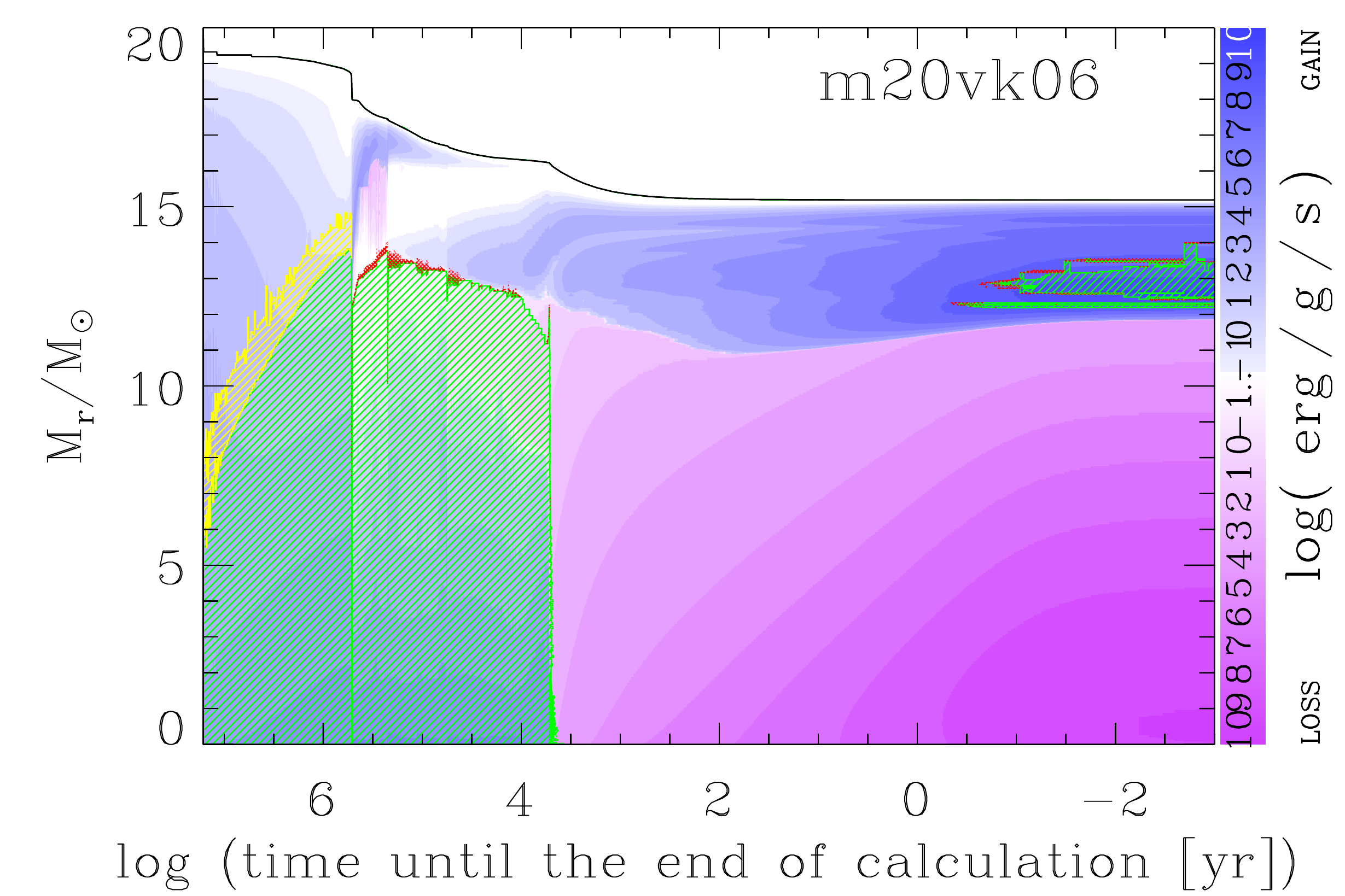}
\includegraphics[width=\columnwidth]{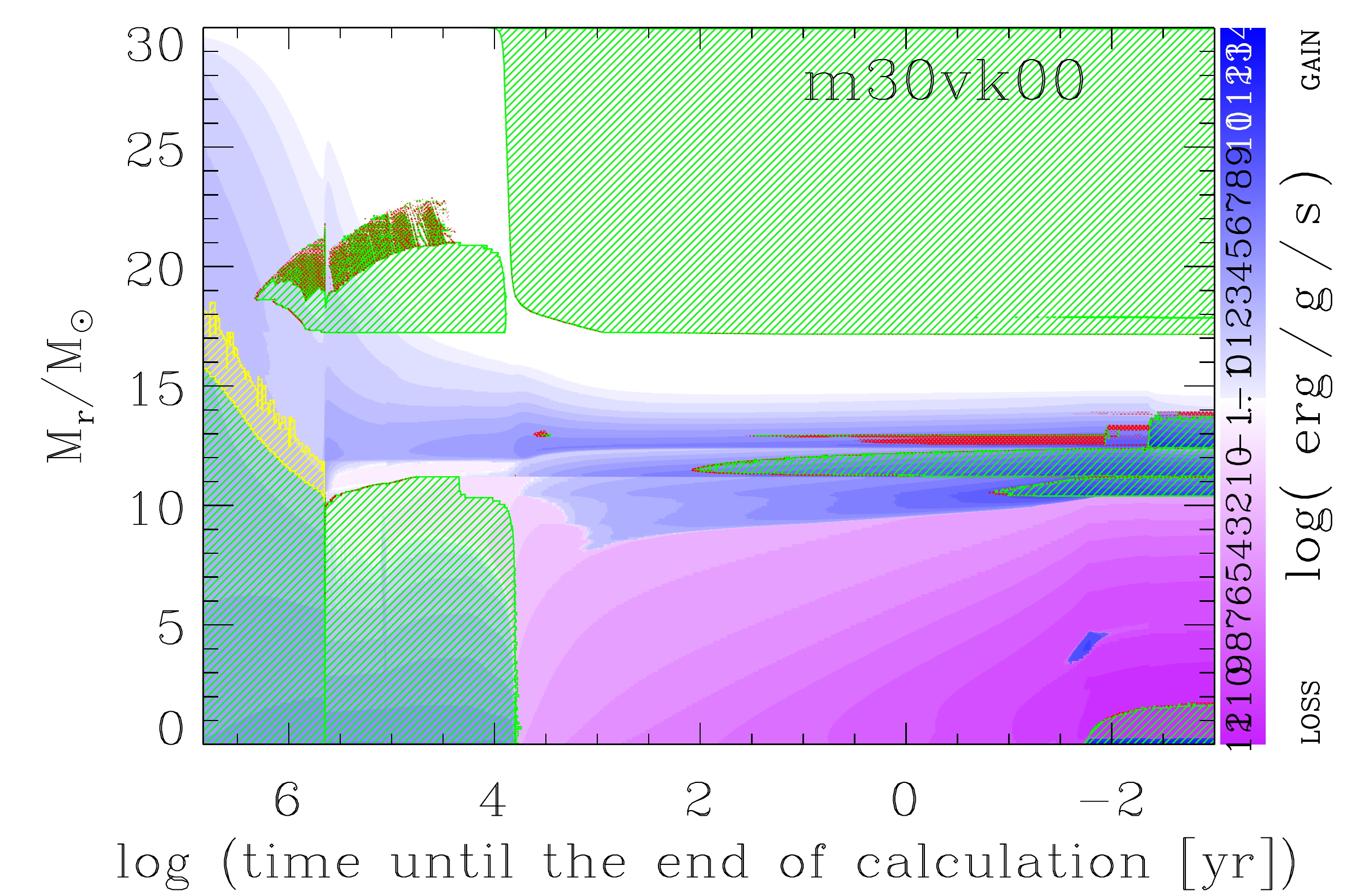}
\includegraphics[width=\columnwidth]{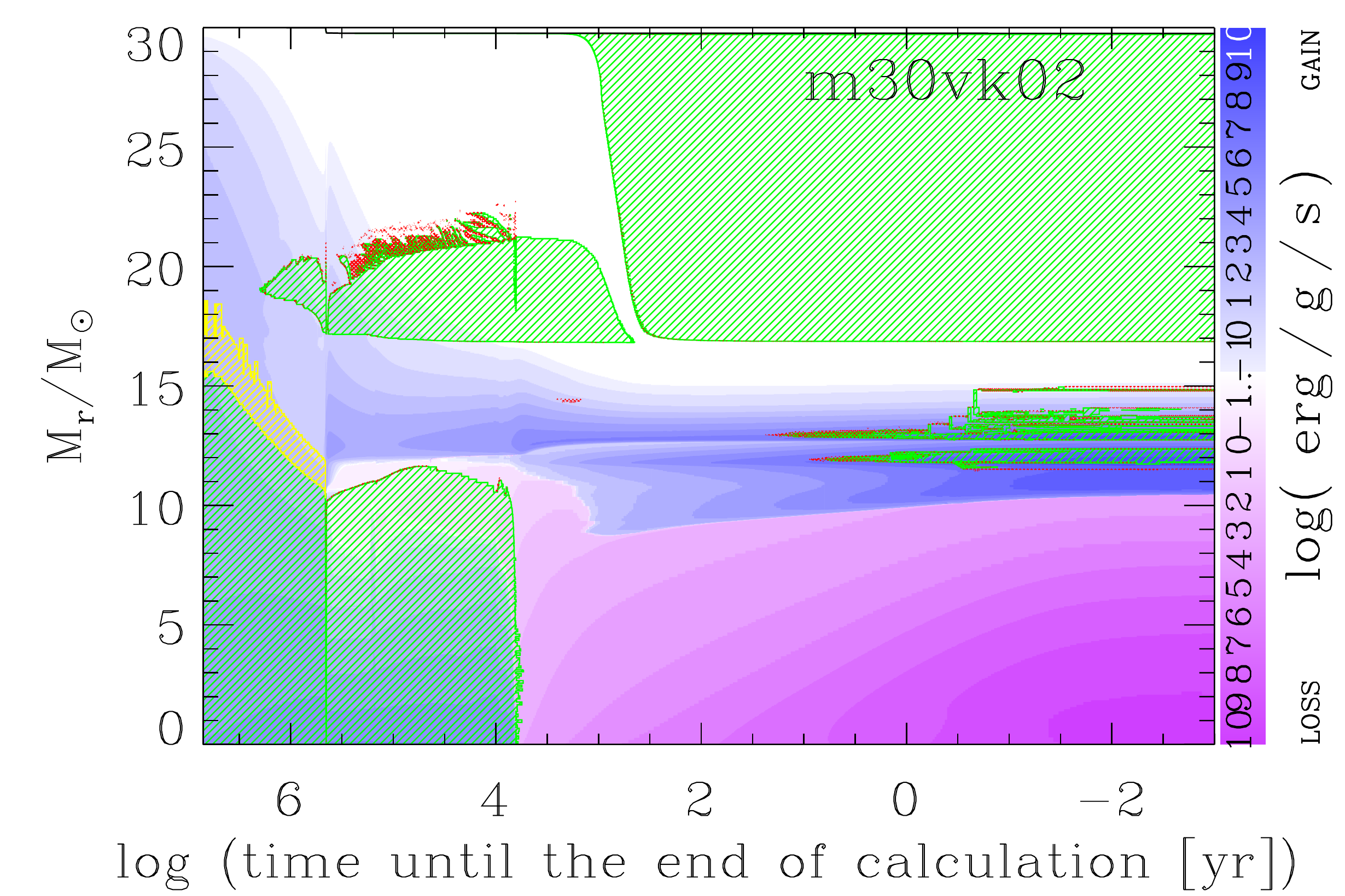}
\includegraphics[width=\columnwidth]{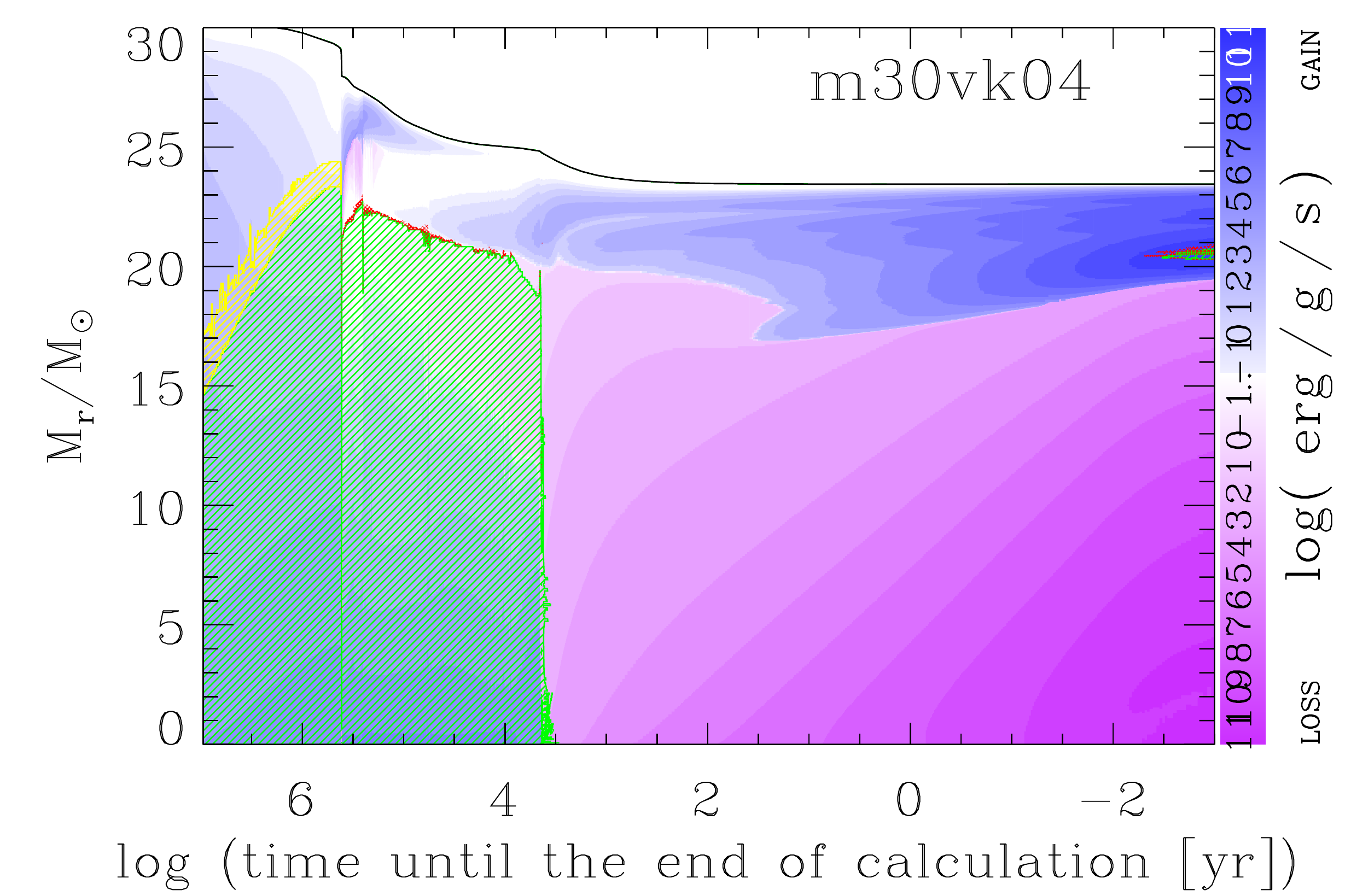}
\includegraphics[width=\columnwidth]{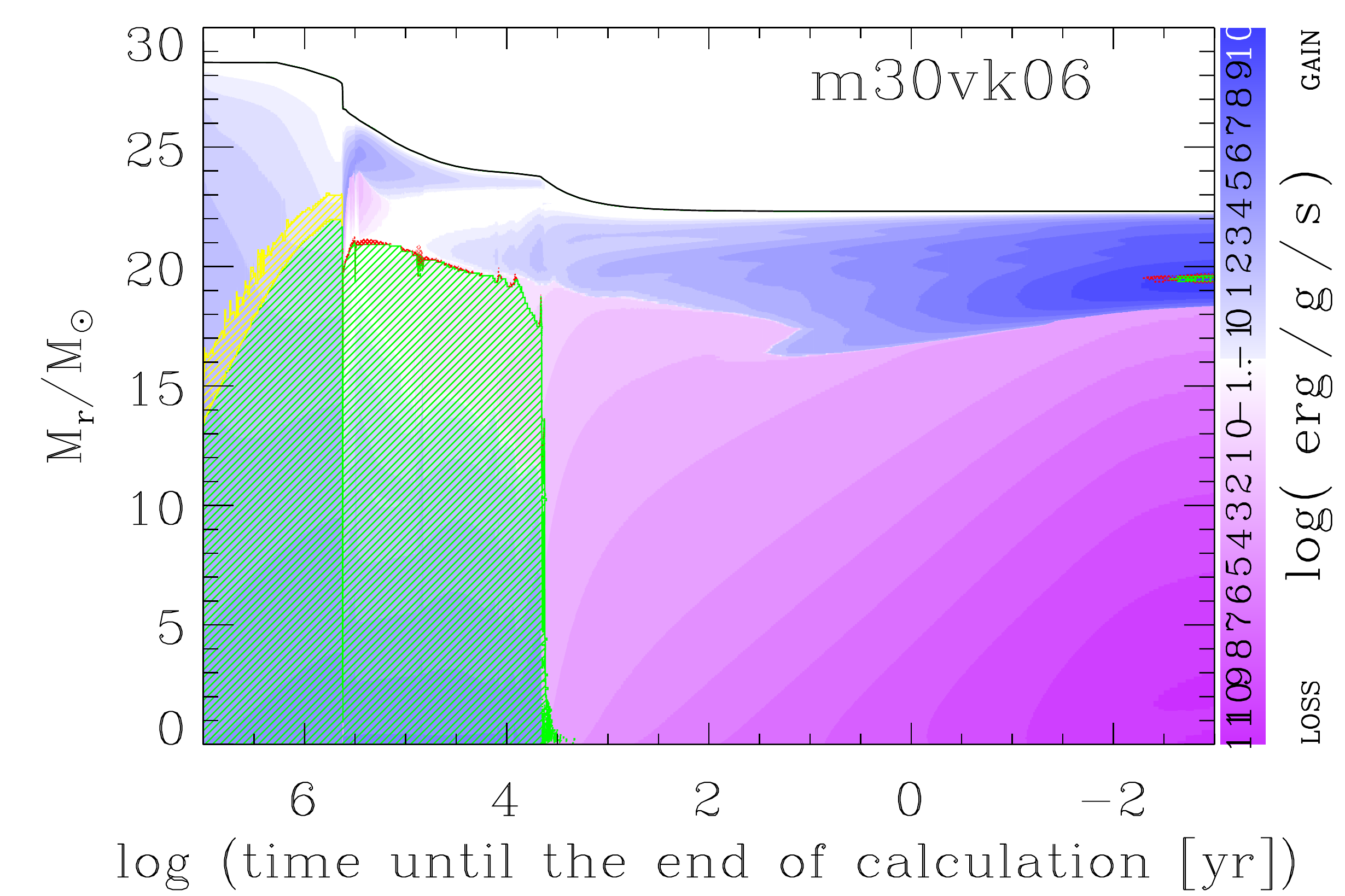}
\end{center}
{\bf Fig.~\ref{fig:kipp1}.}  Continued  
\end{figure*}

\begin{figure*}
\begin{center}
\includegraphics[width=\columnwidth]{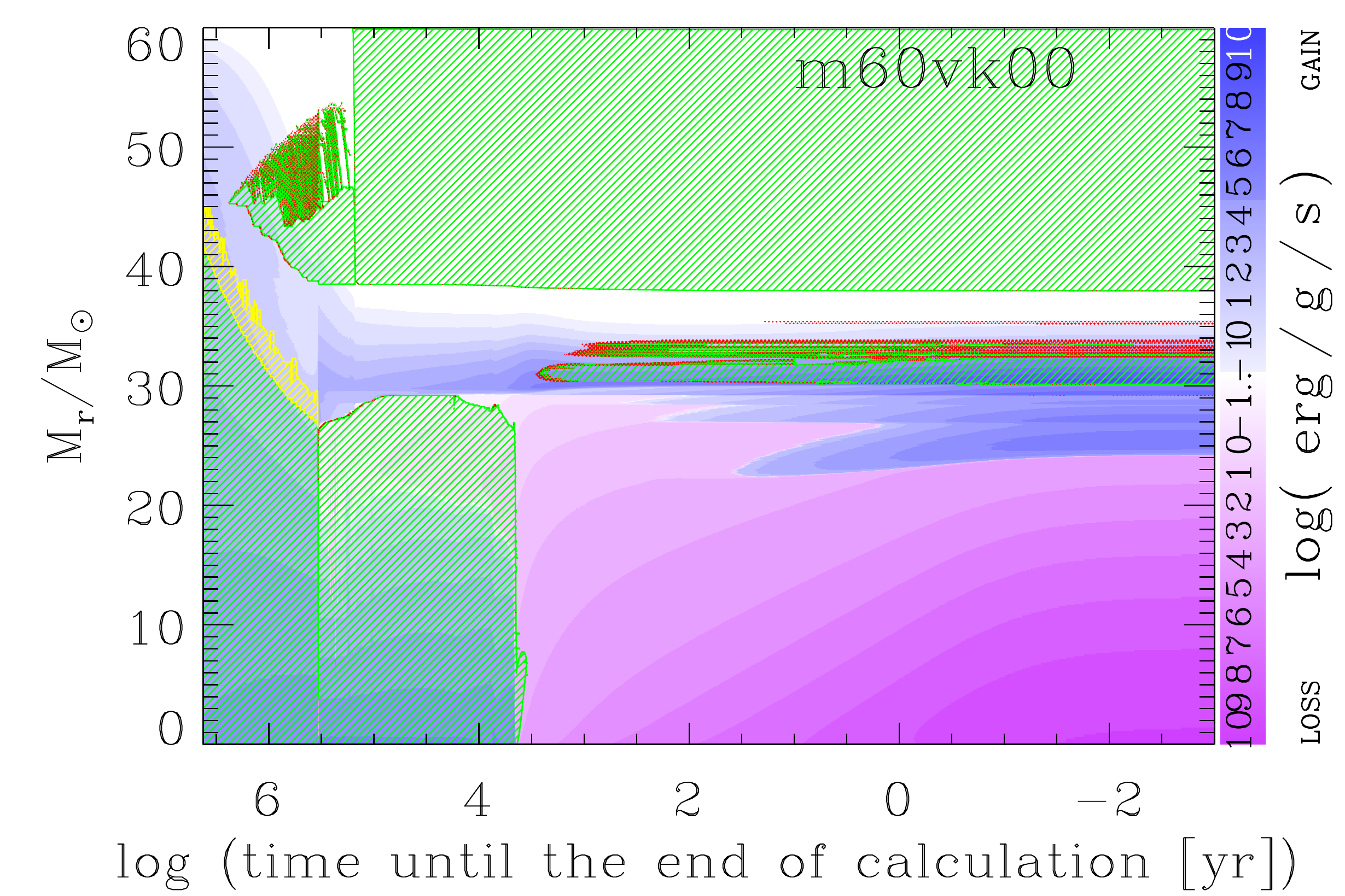}
\includegraphics[width=\columnwidth]{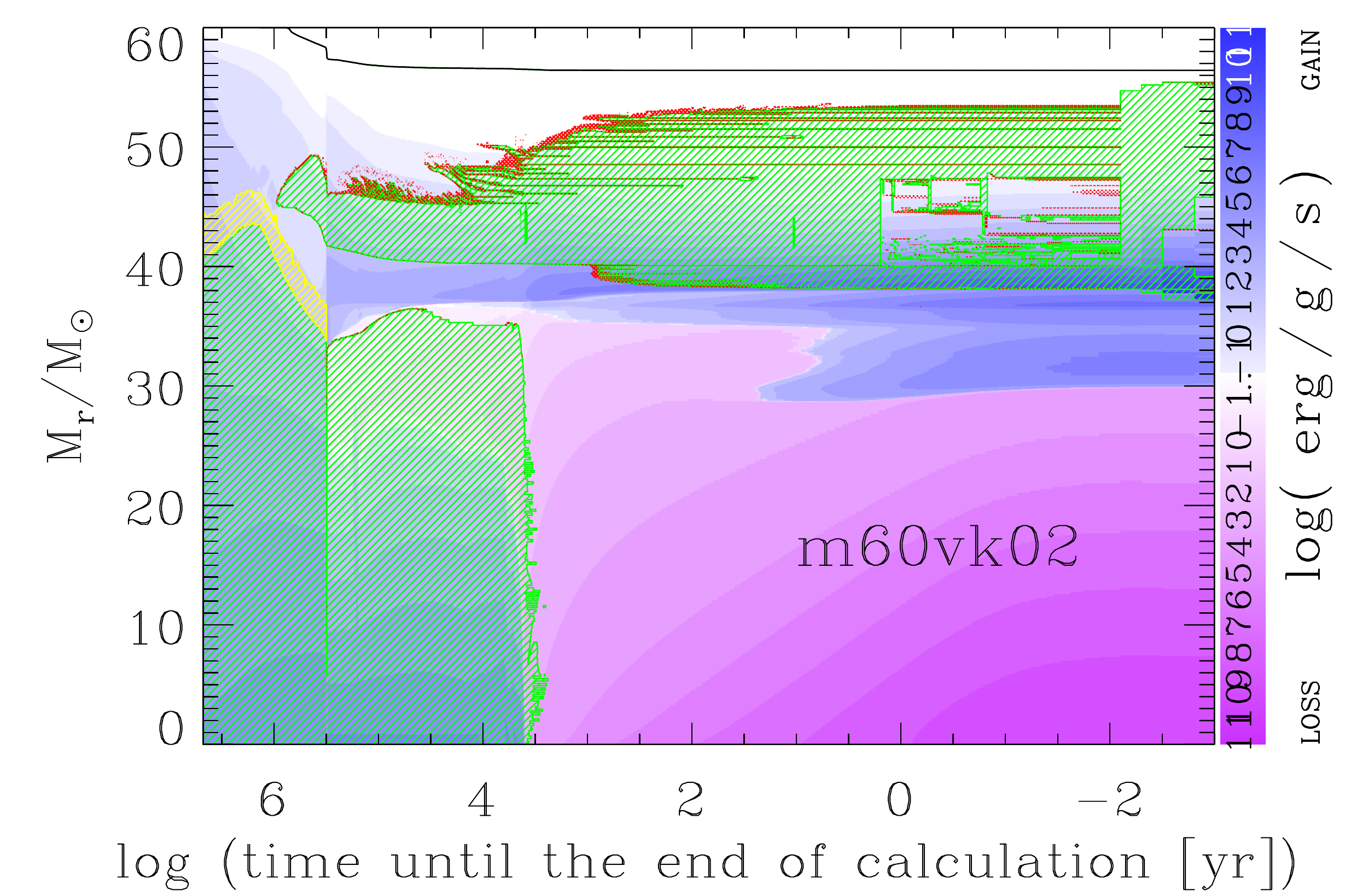}
\includegraphics[width=\columnwidth]{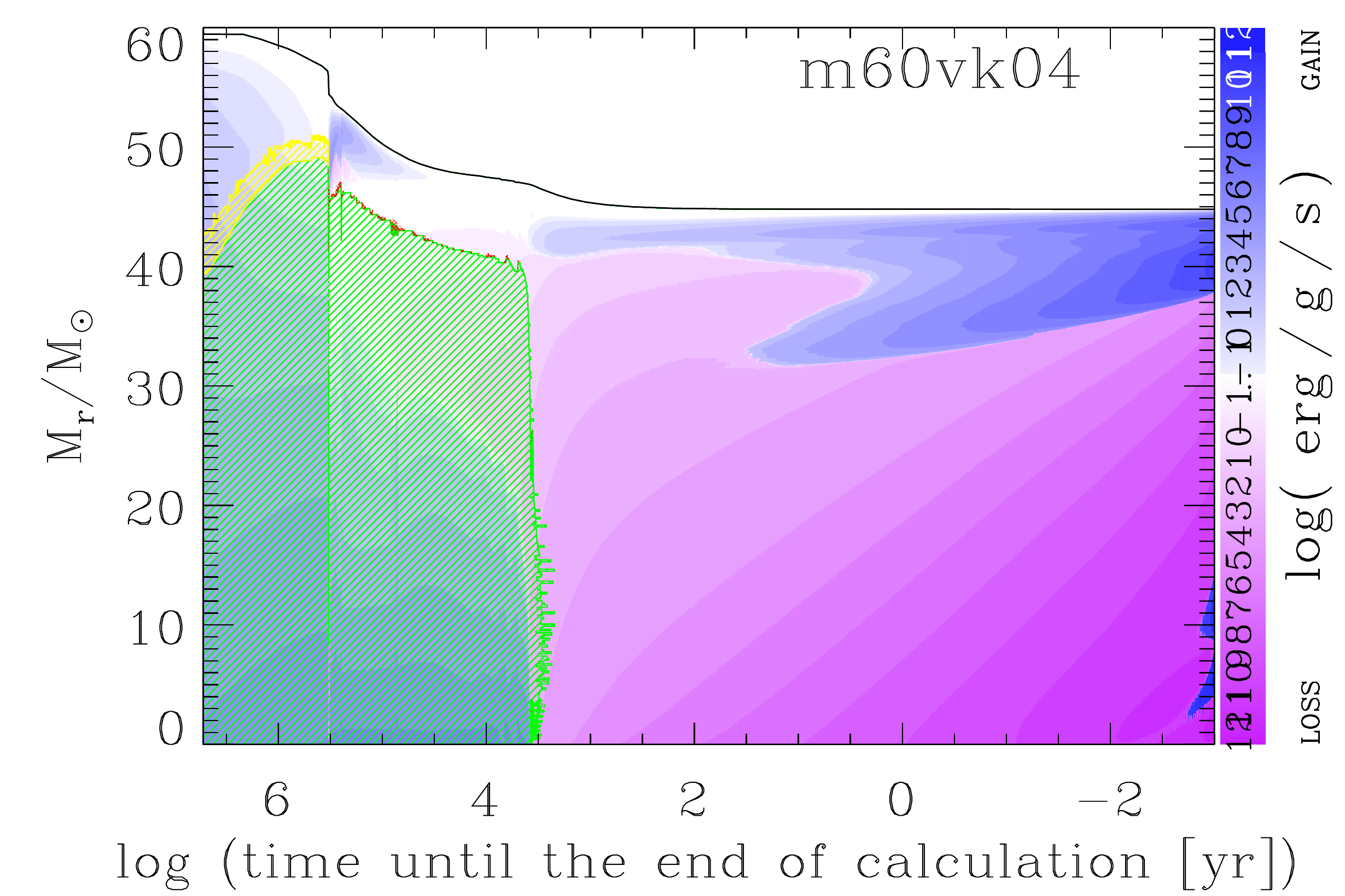}
\includegraphics[width=\columnwidth]{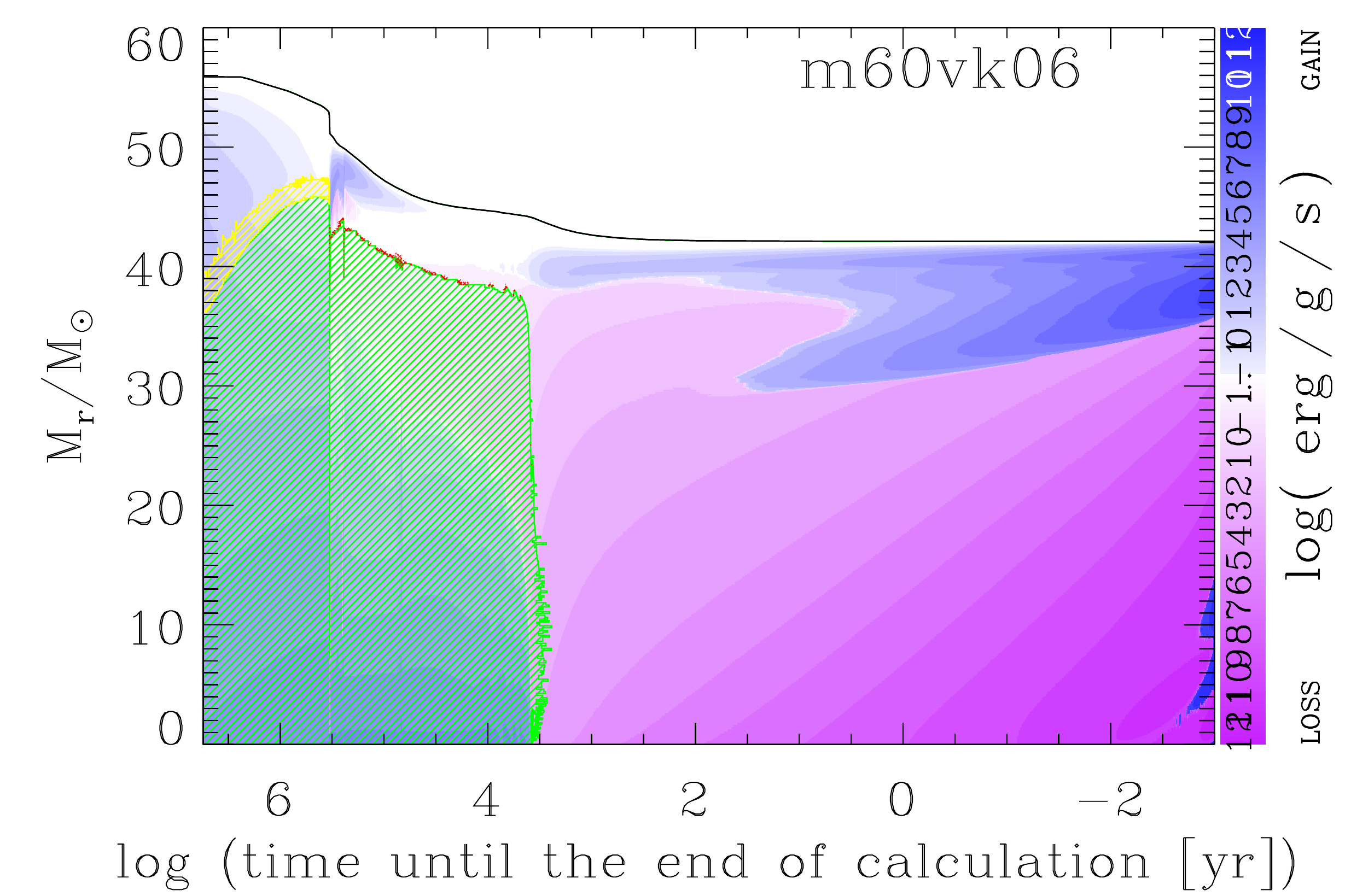}
\includegraphics[width=\columnwidth]{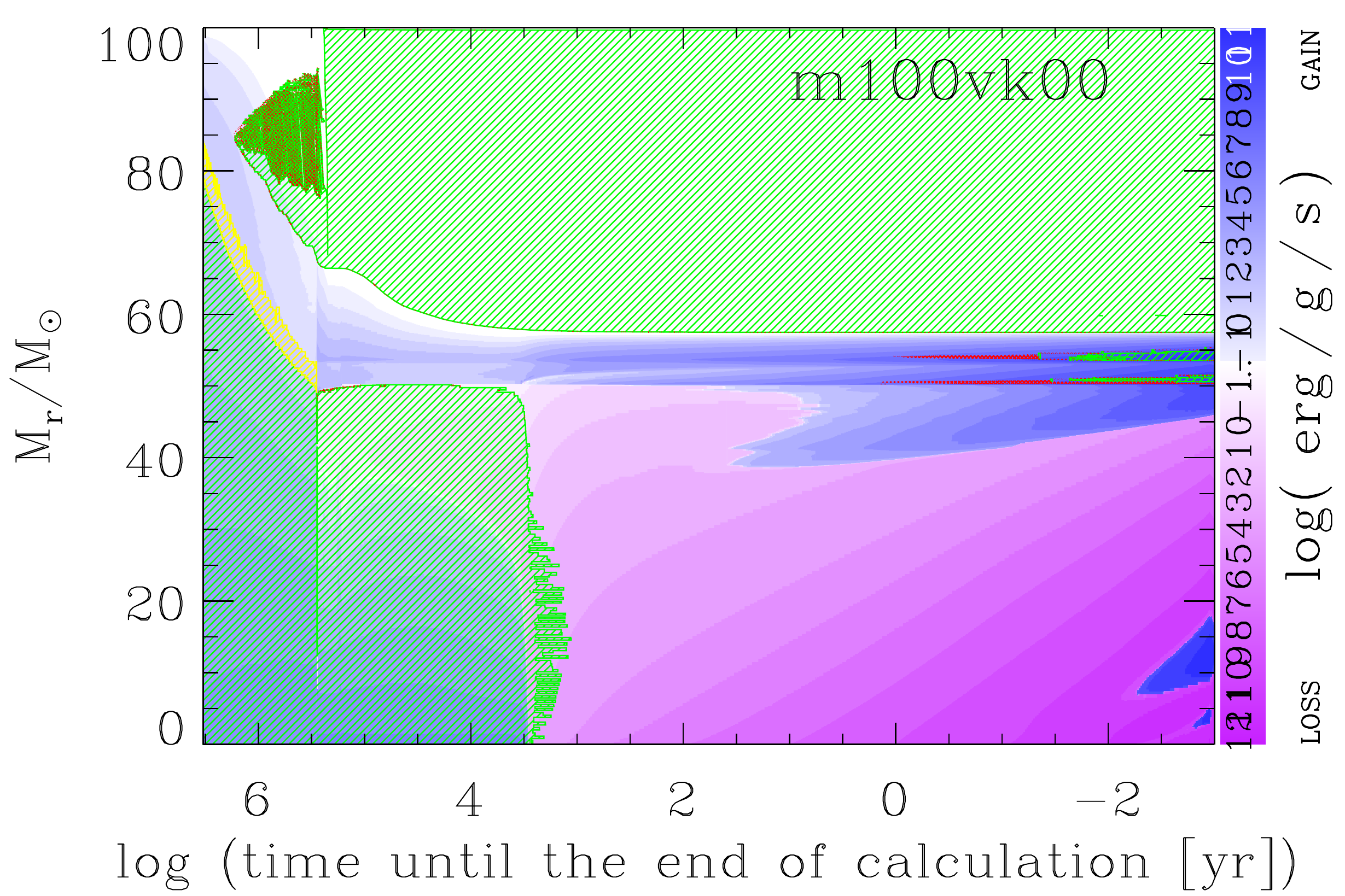}
\includegraphics[width=\columnwidth]{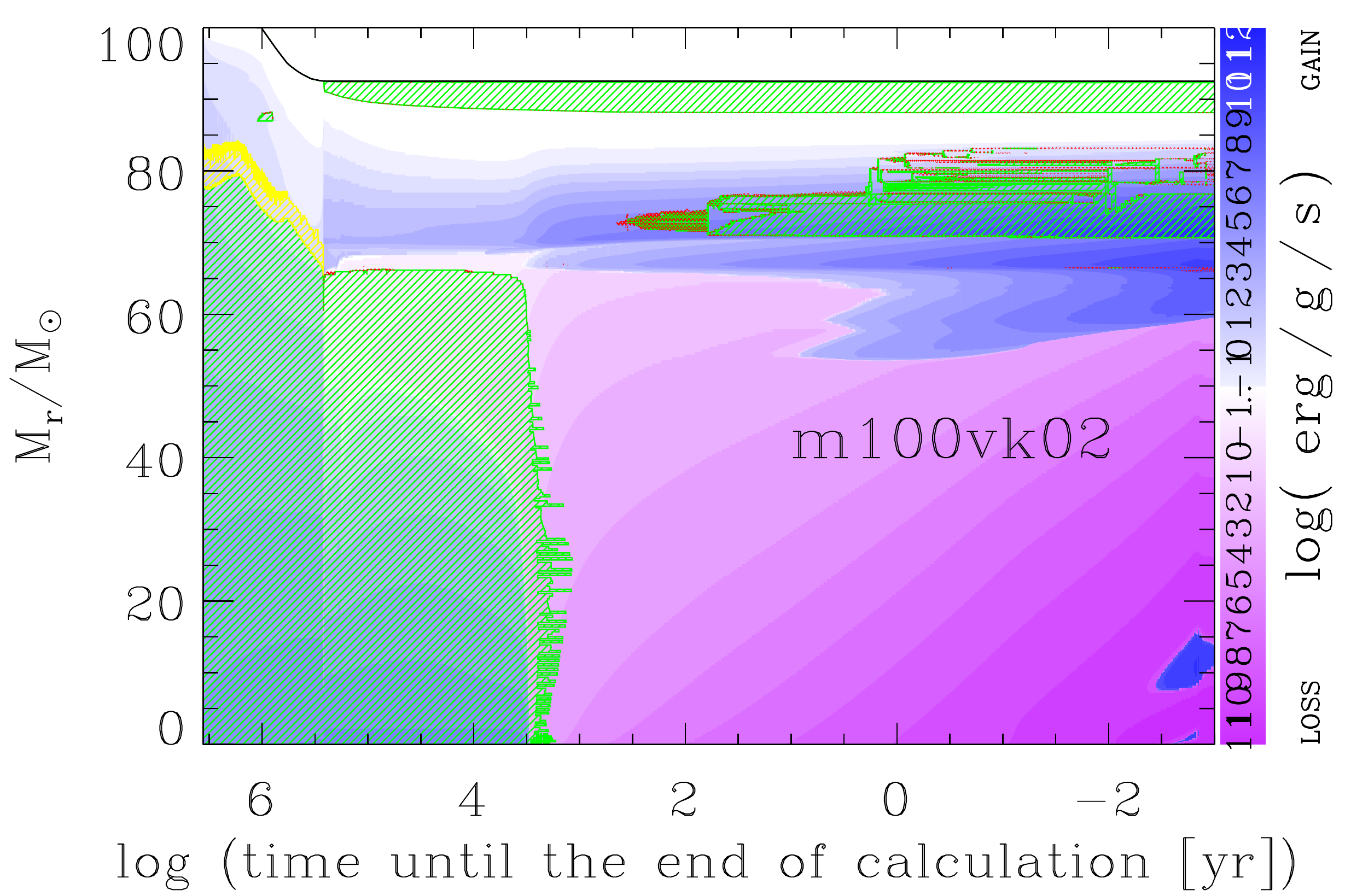}
\includegraphics[width=\columnwidth]{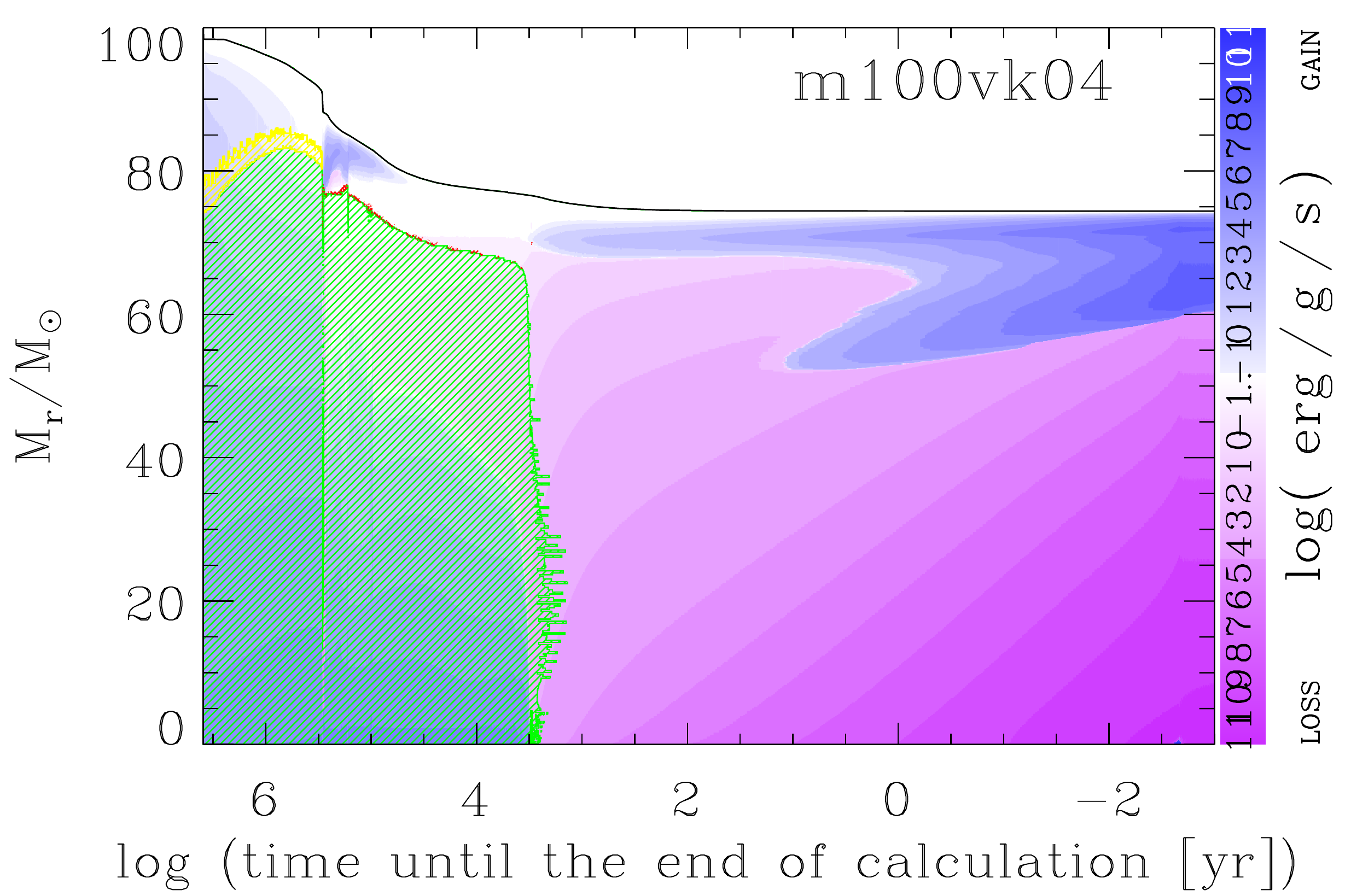}
\includegraphics[width=\columnwidth]{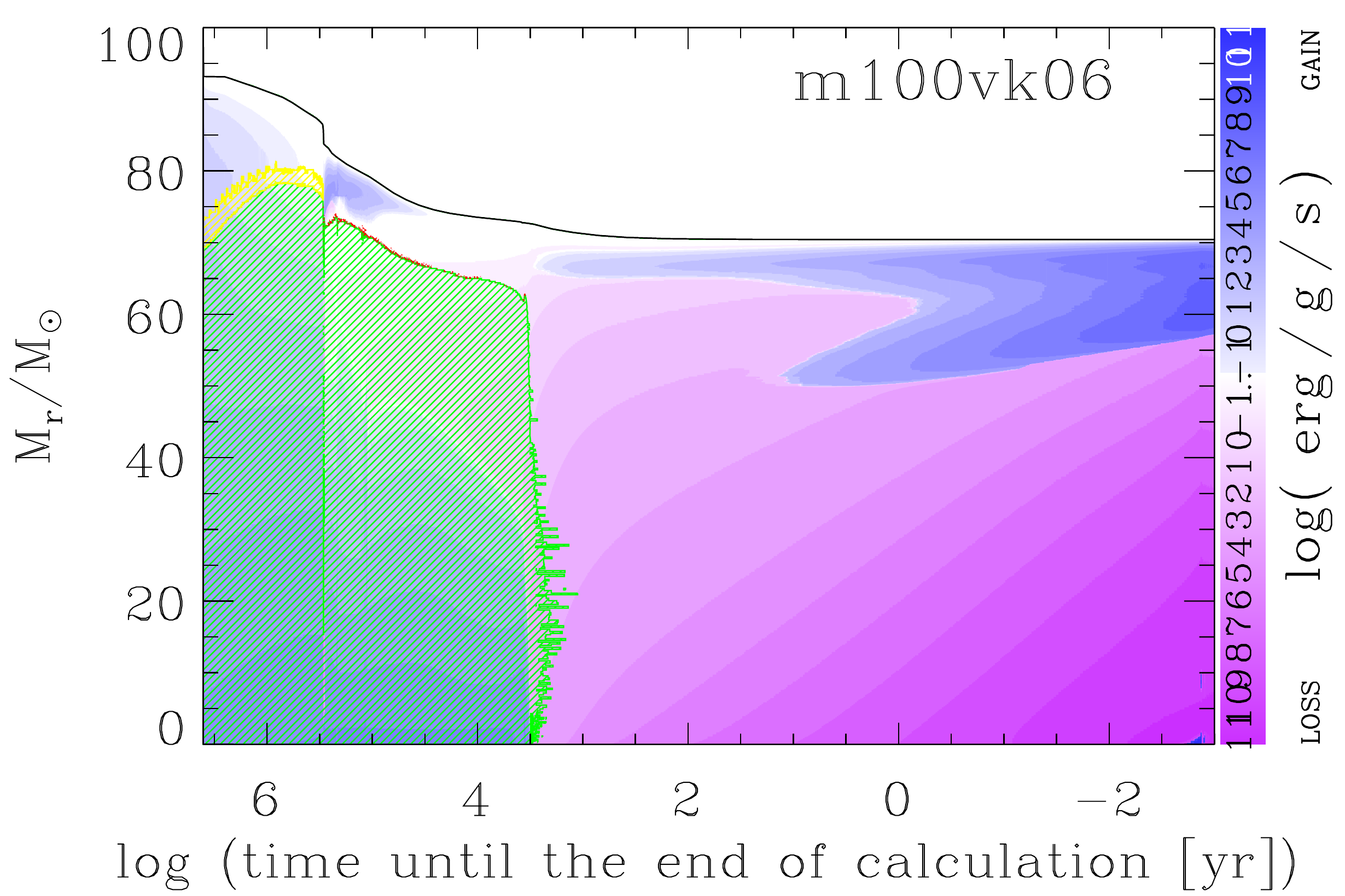}
\end{center}
{\bf Fig.~\ref{fig:kipp1}.}  Continued  
\end{figure*}

\begin{figure*}
\begin{center}
\includegraphics[width=\columnwidth]{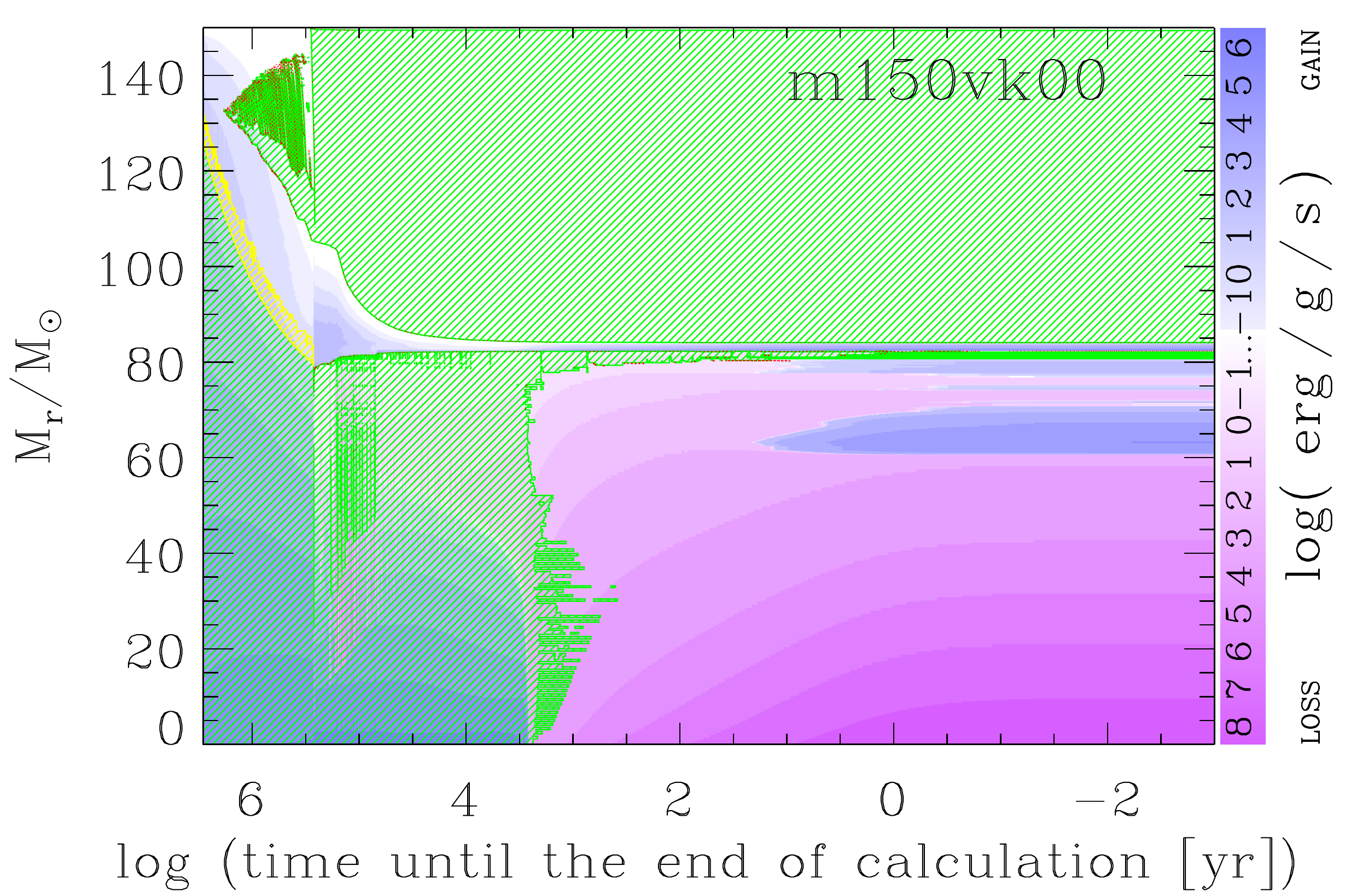}
\includegraphics[width=\columnwidth]{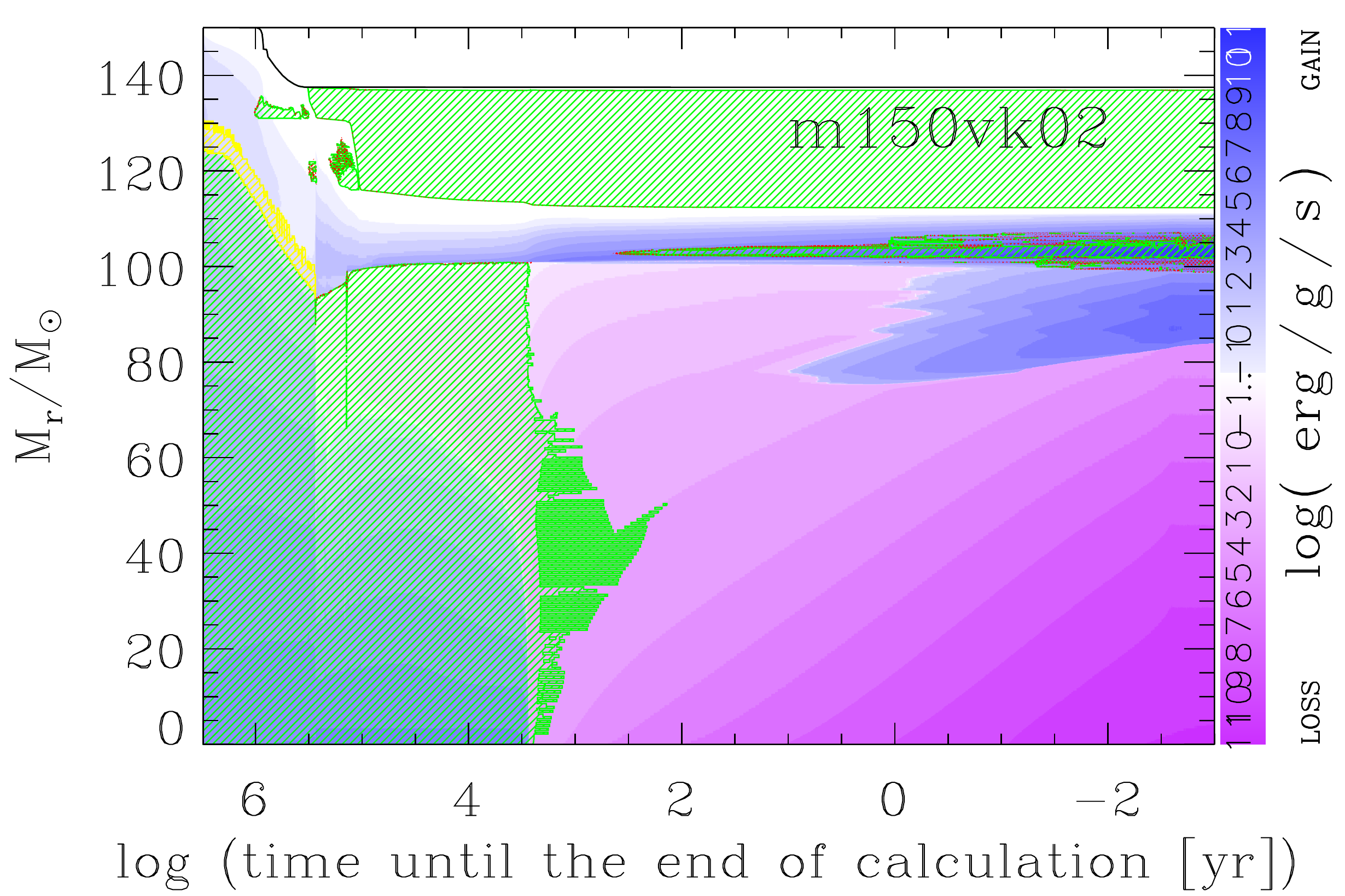}
\includegraphics[width=\columnwidth]{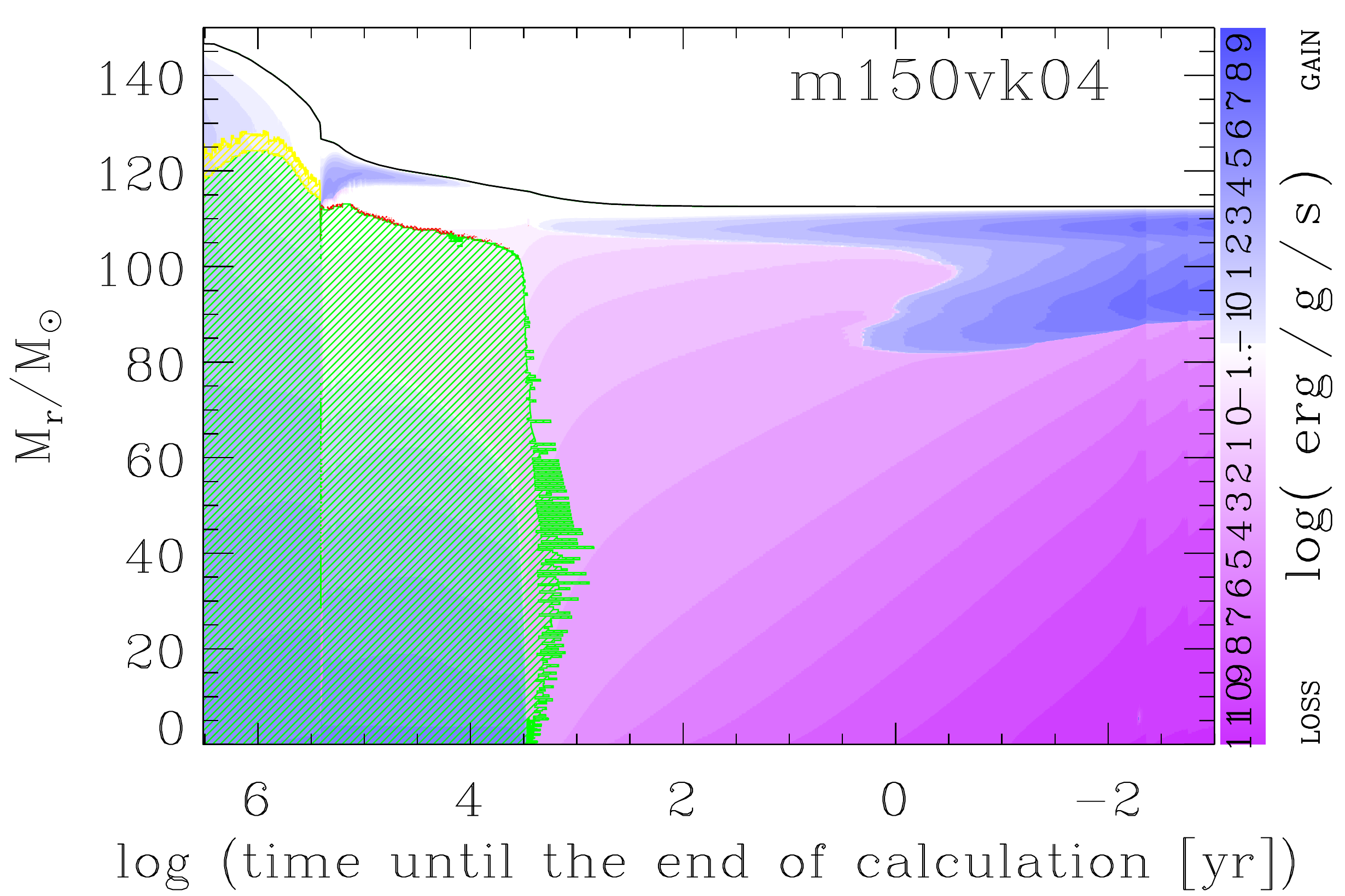}
\includegraphics[width=\columnwidth]{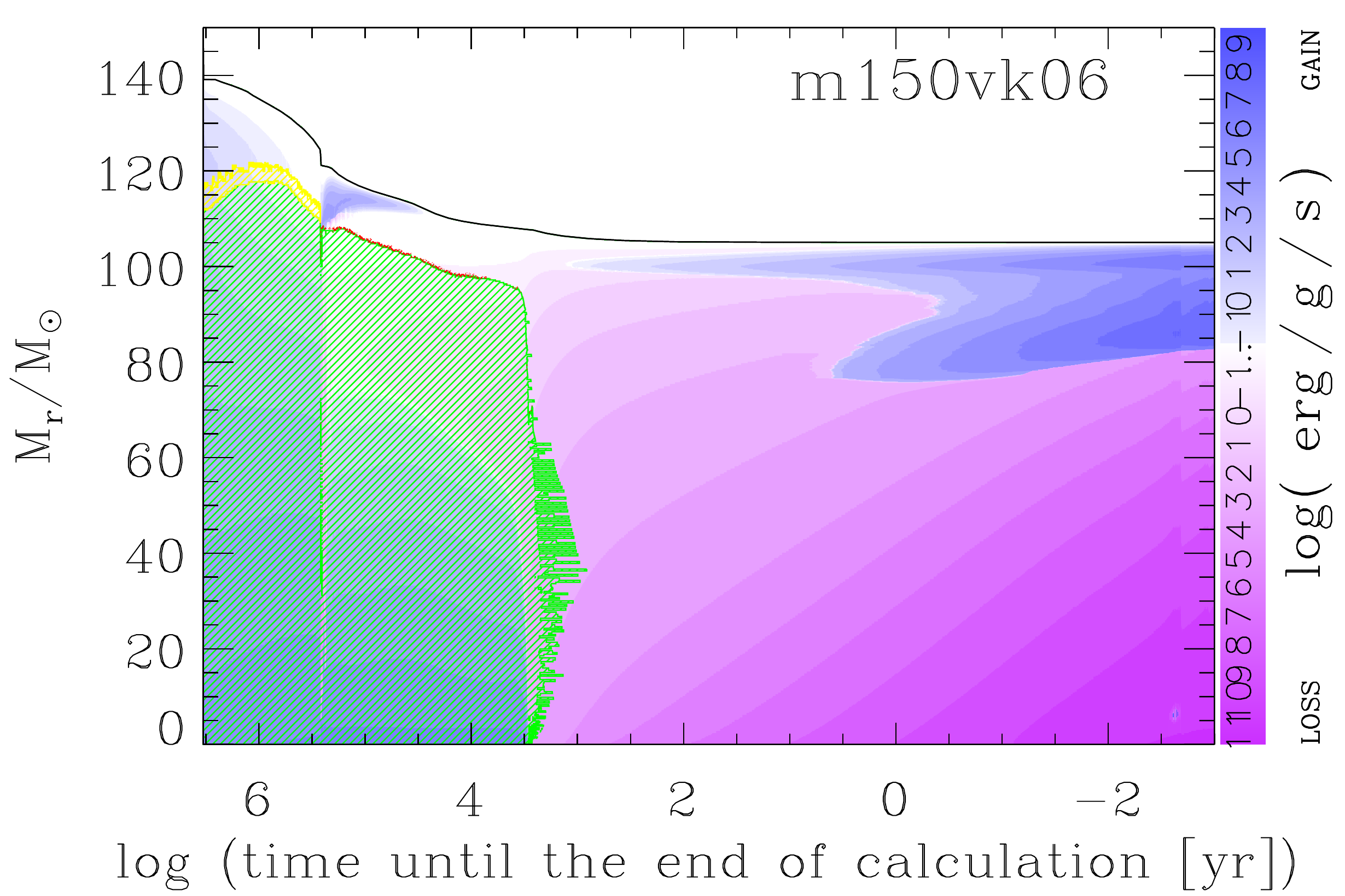}
\includegraphics[width=\columnwidth]{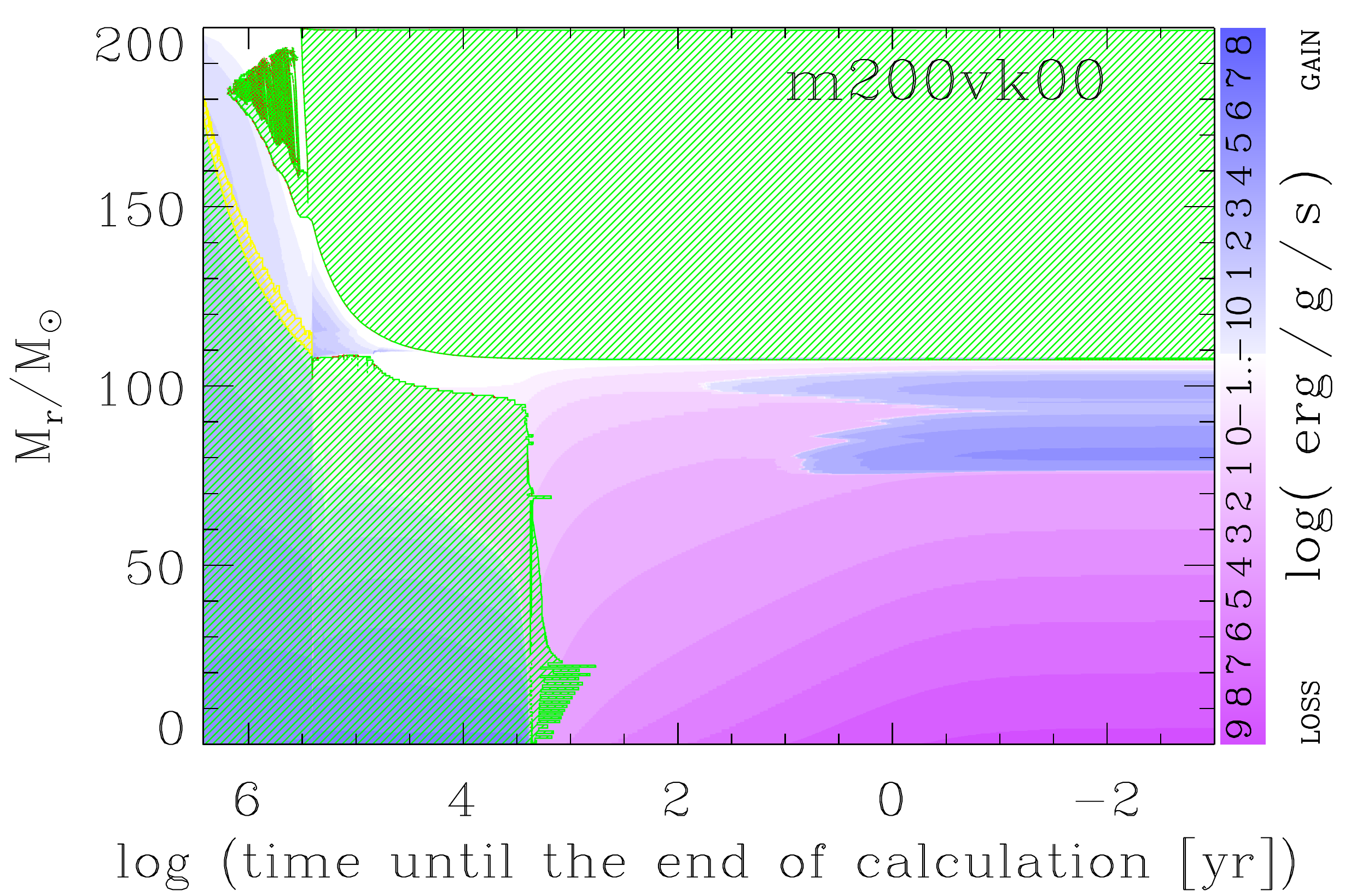}
\includegraphics[width=\columnwidth]{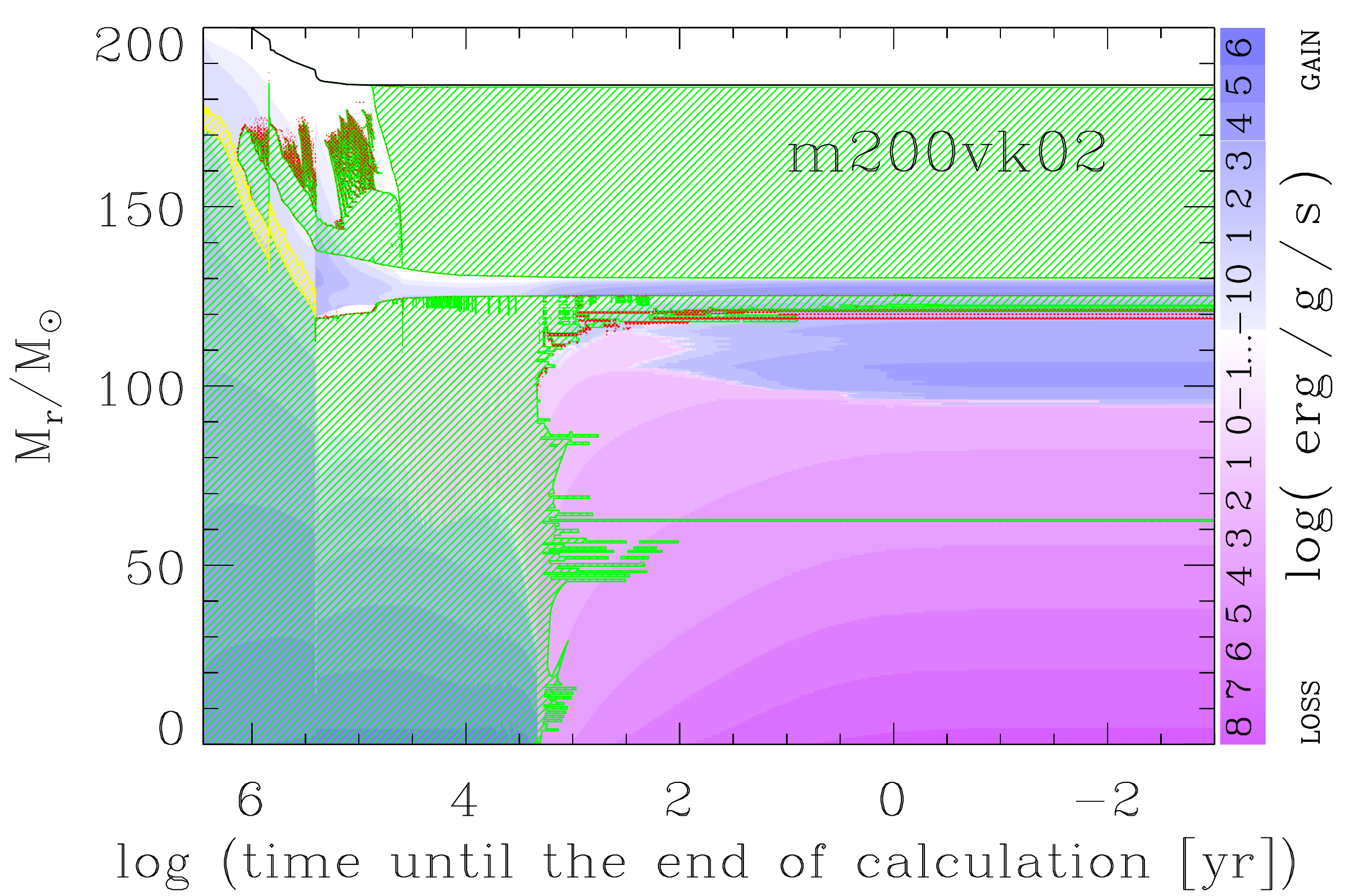}
\includegraphics[width=\columnwidth]{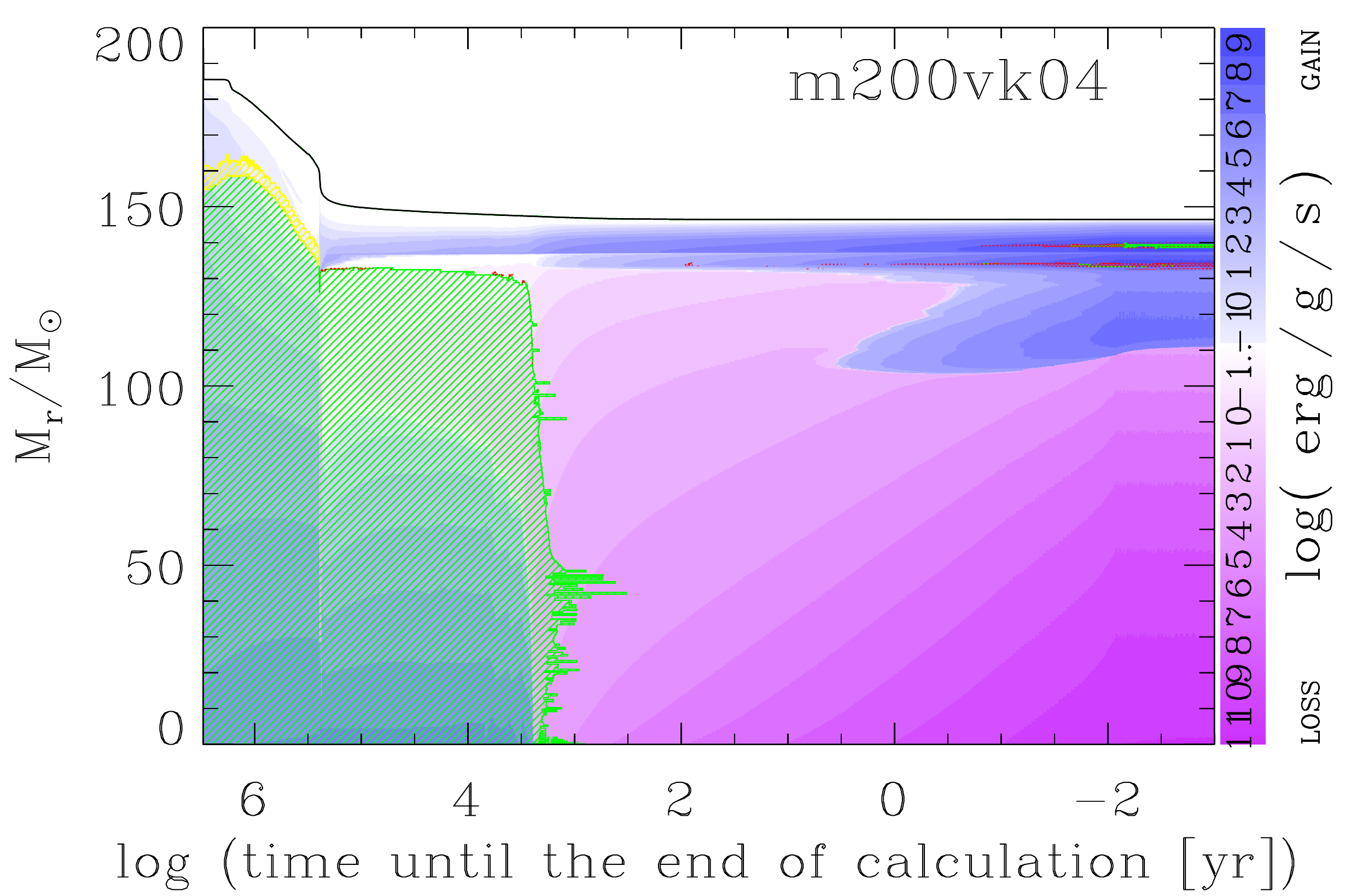}
\includegraphics[width=\columnwidth]{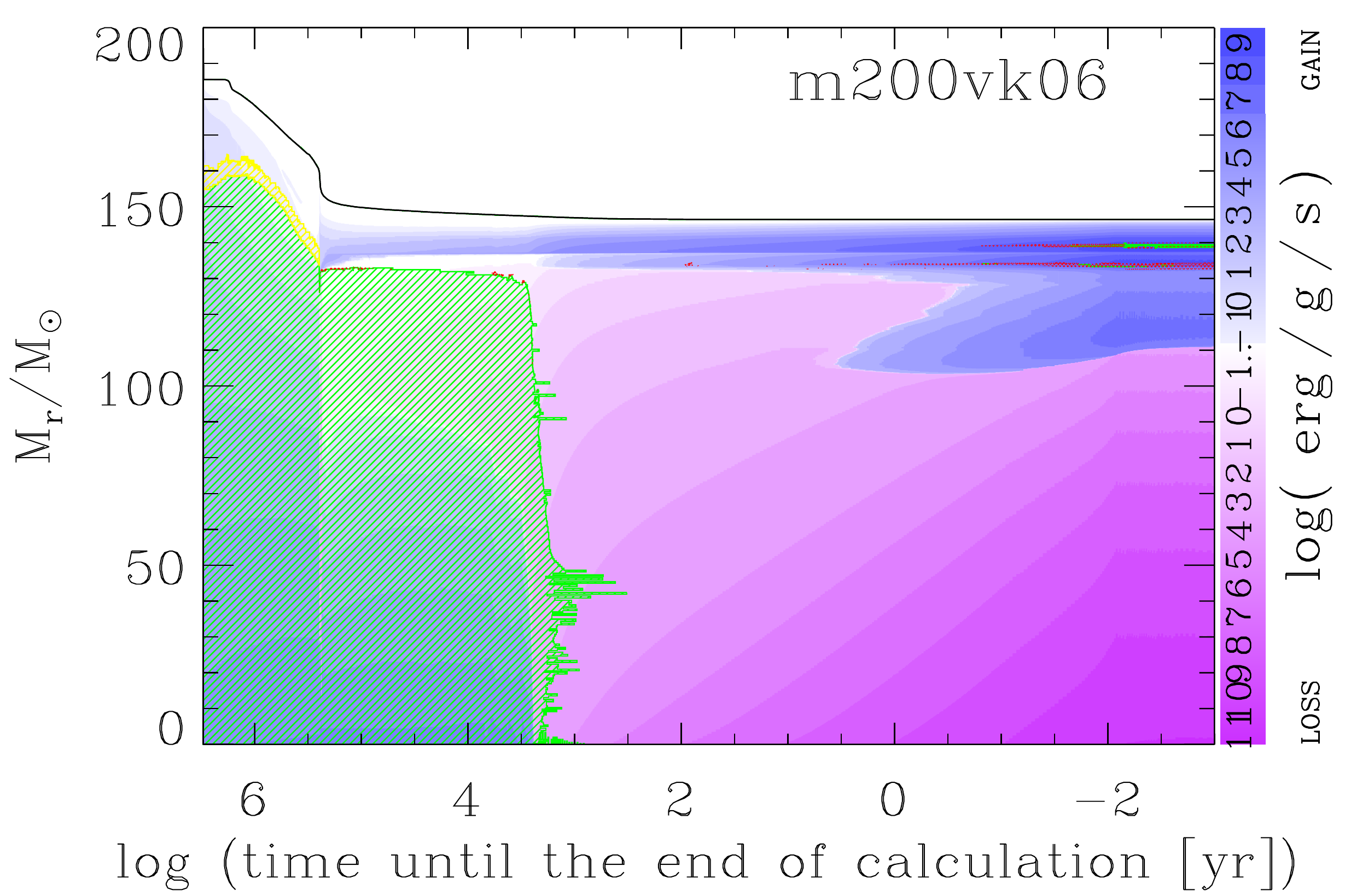}
\end{center}
{\bf Fig.~\ref{fig:kipp1}.}  Continued  
\end{figure*}

\begin{figure*}
\begin{center}
\includegraphics[width=\columnwidth]{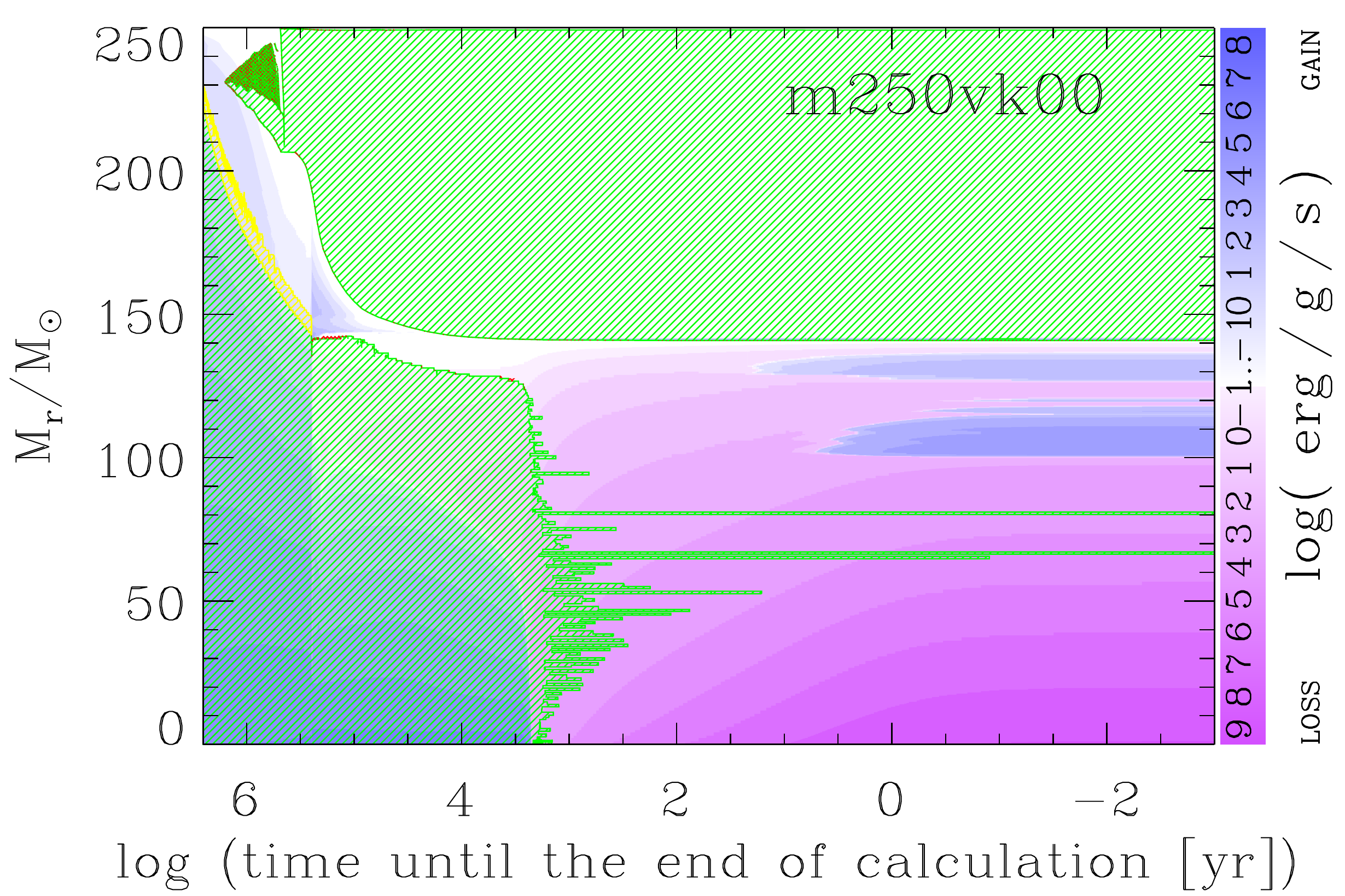}
\includegraphics[width=\columnwidth]{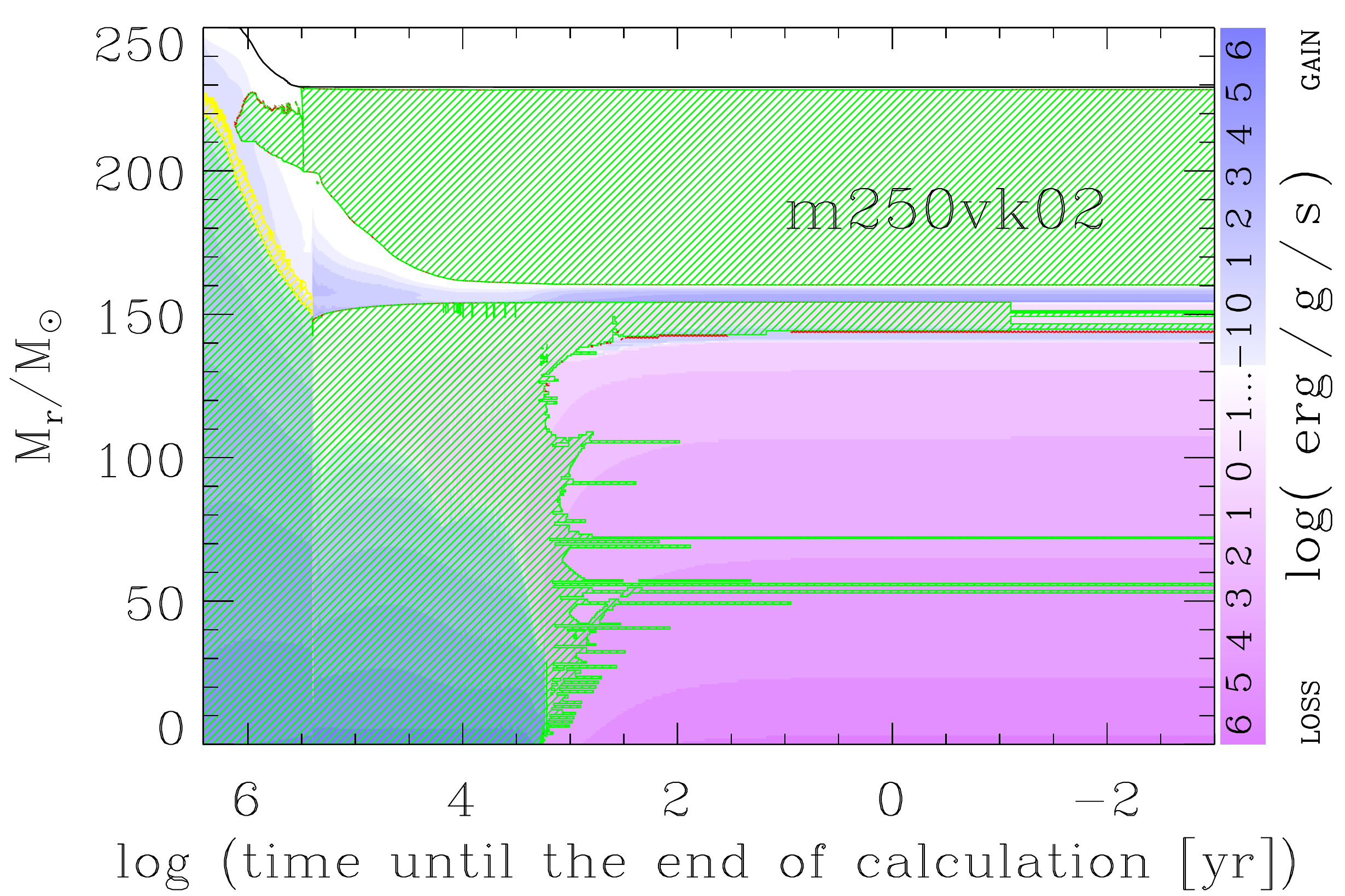}
\includegraphics[width=\columnwidth]{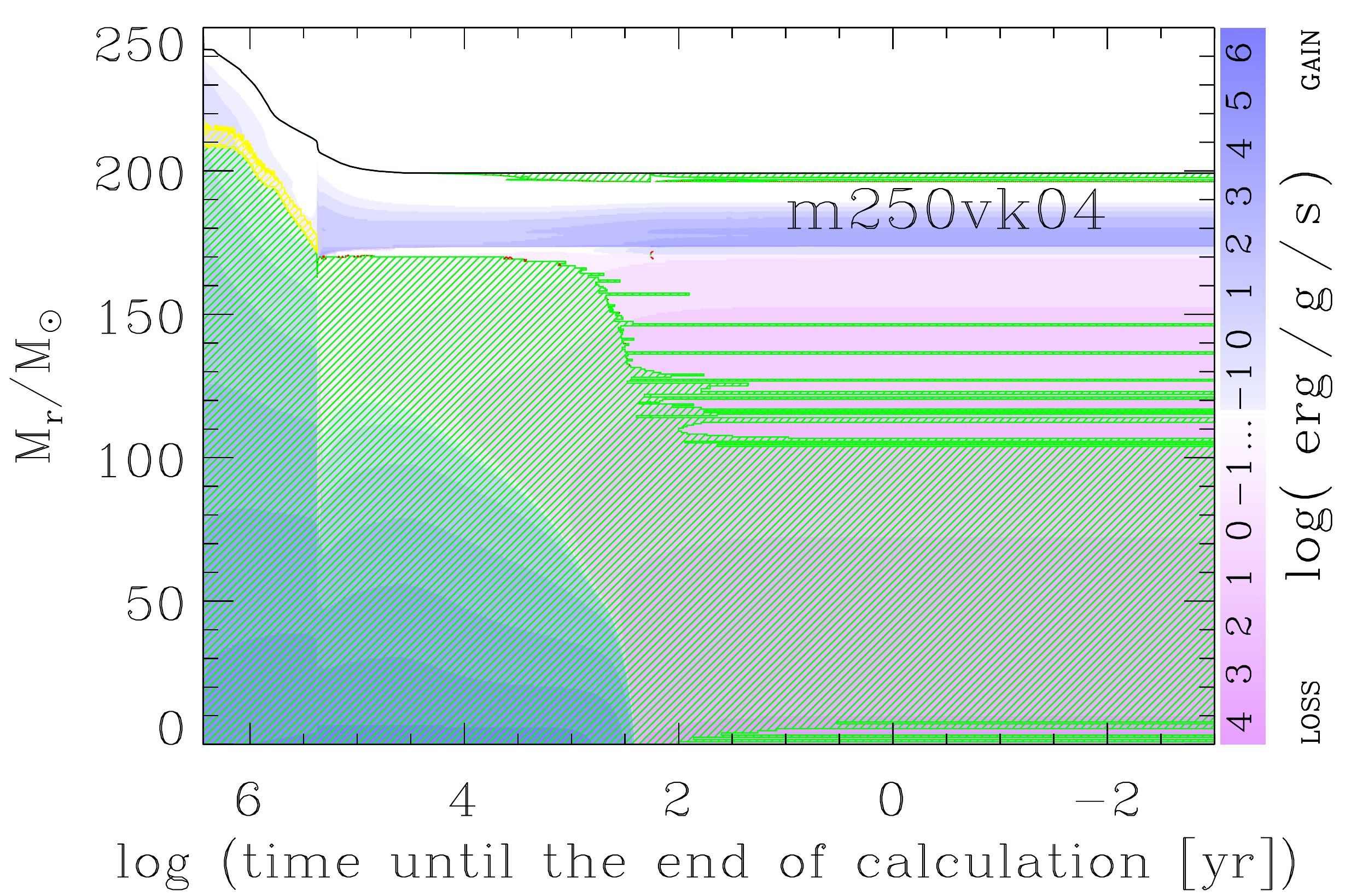}
\includegraphics[width=\columnwidth]{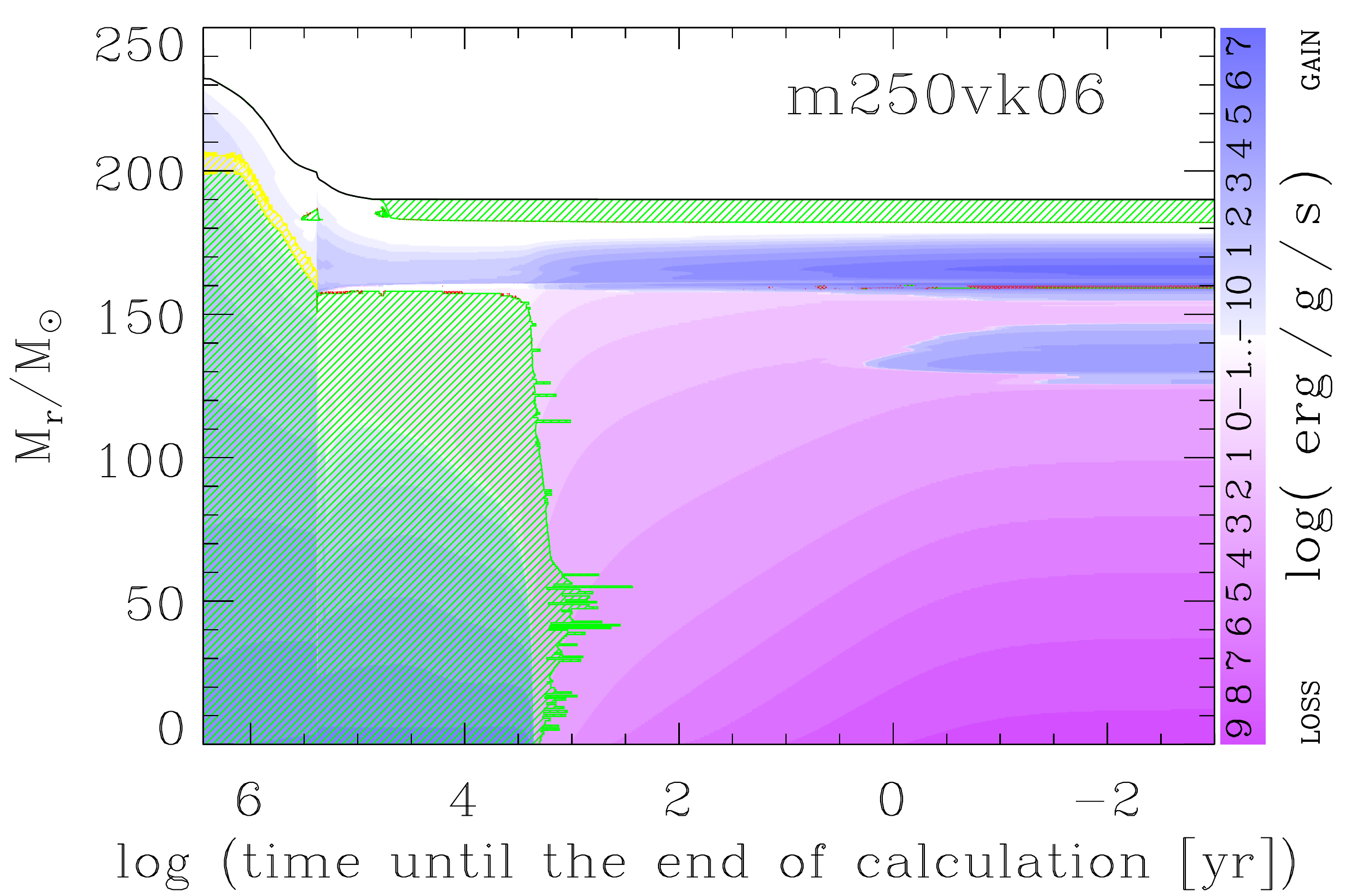}
\includegraphics[width=\columnwidth]{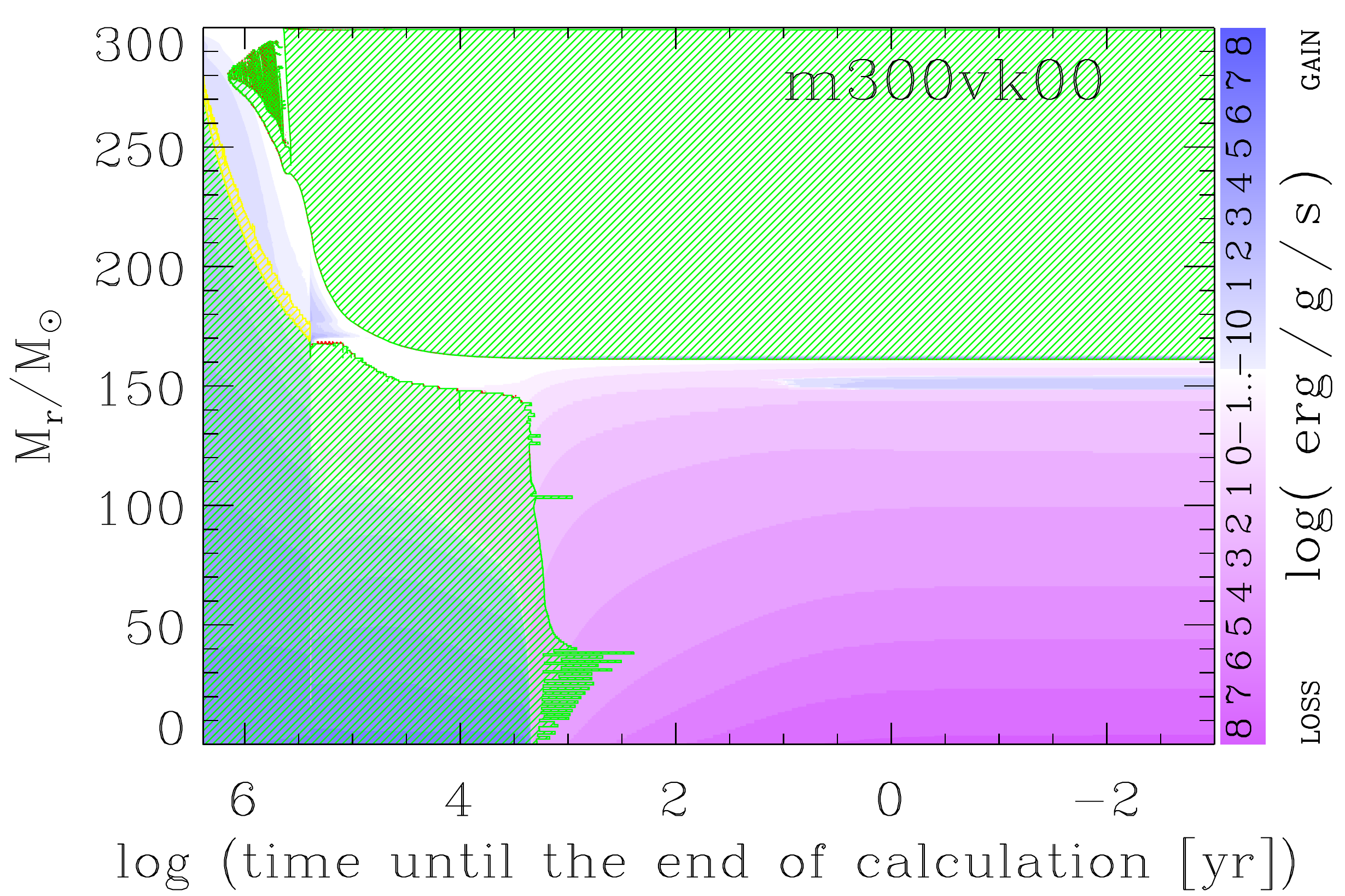}
\includegraphics[width=\columnwidth]{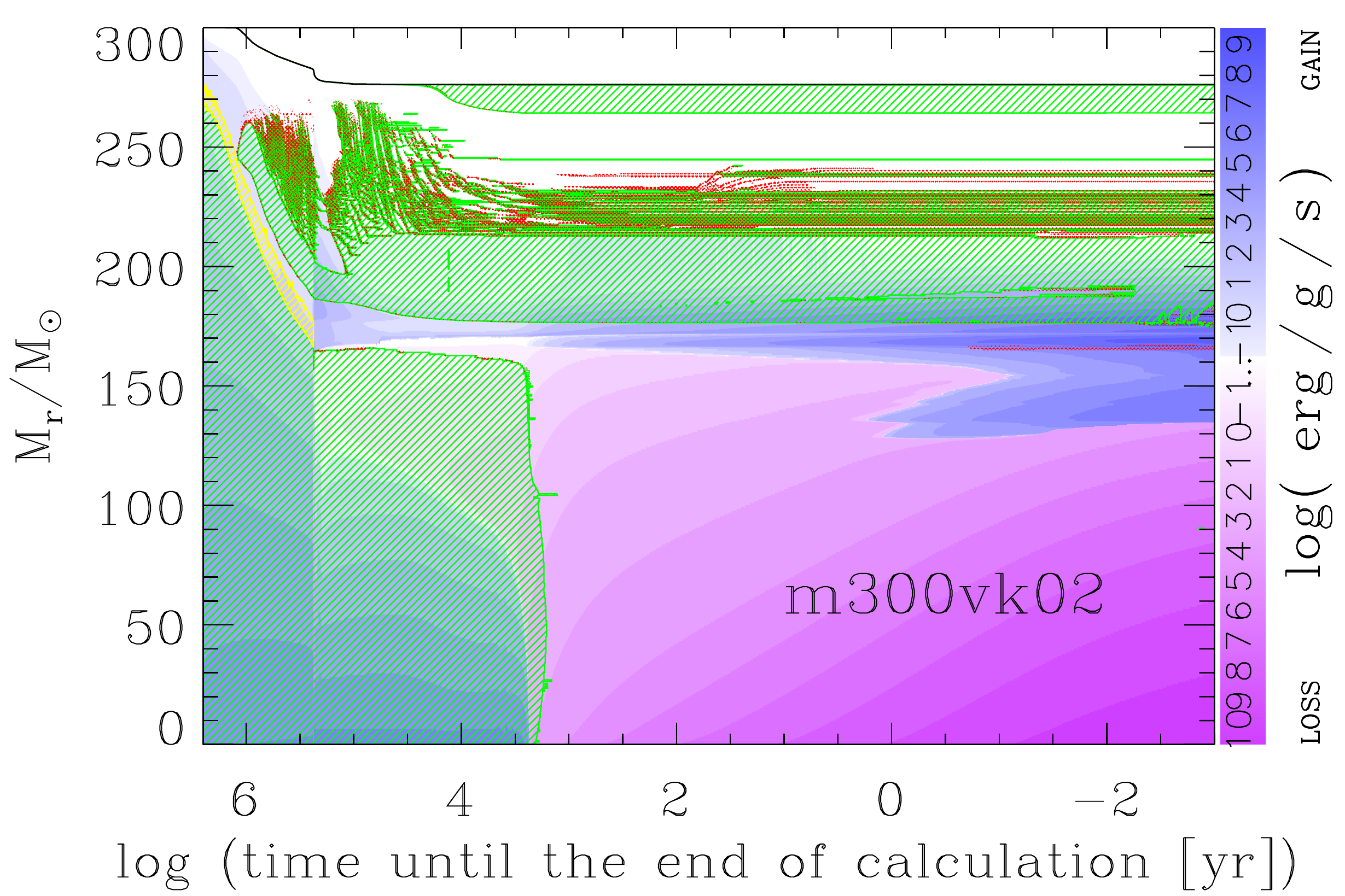}
\includegraphics[width=\columnwidth]{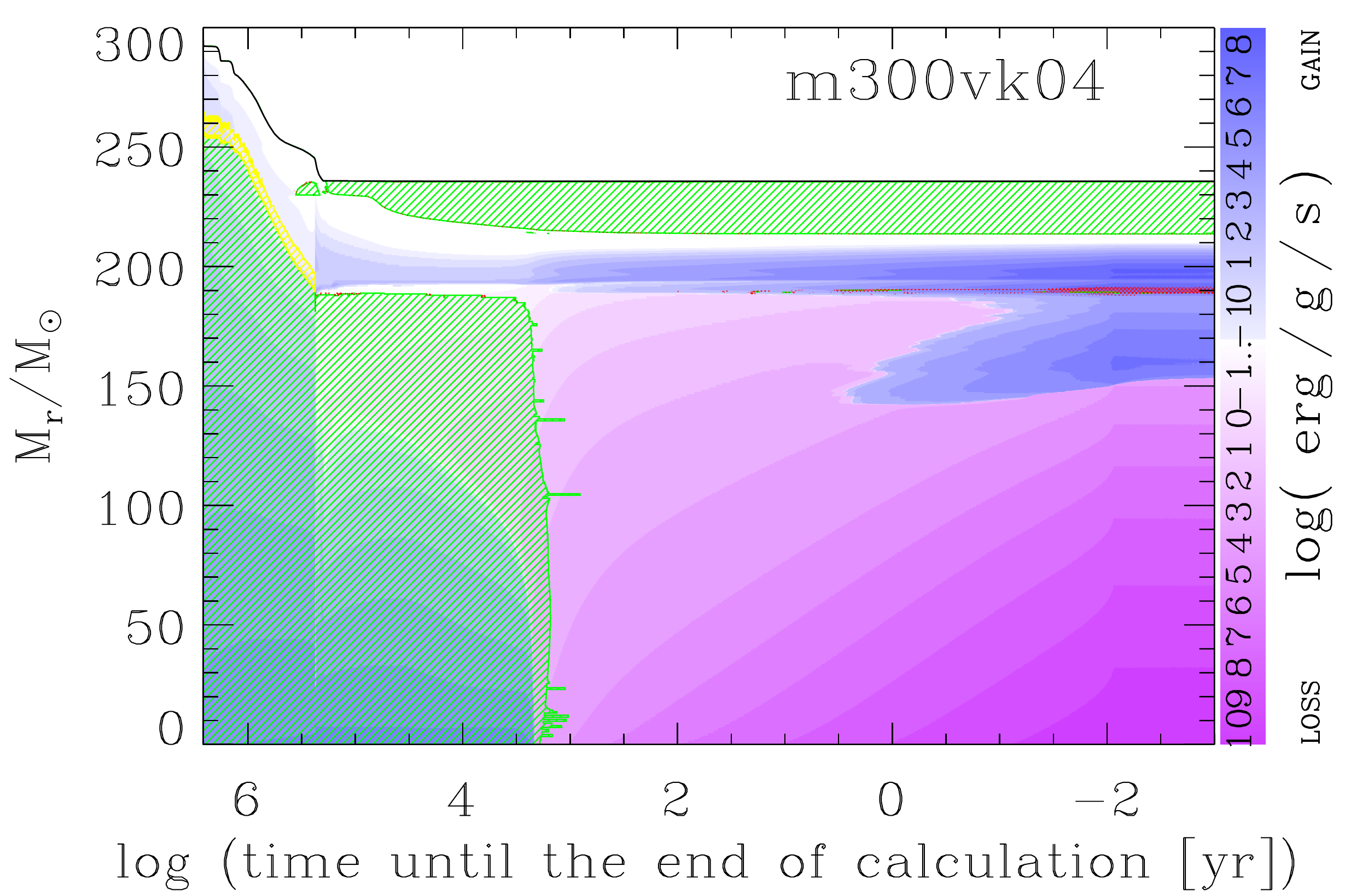}
\end{center}
{\bf Fig.~\ref{fig:kipp1}}  Continued  
\end{figure*}

\begin{figure*}
\begin{center}
\includegraphics[width=\columnwidth]{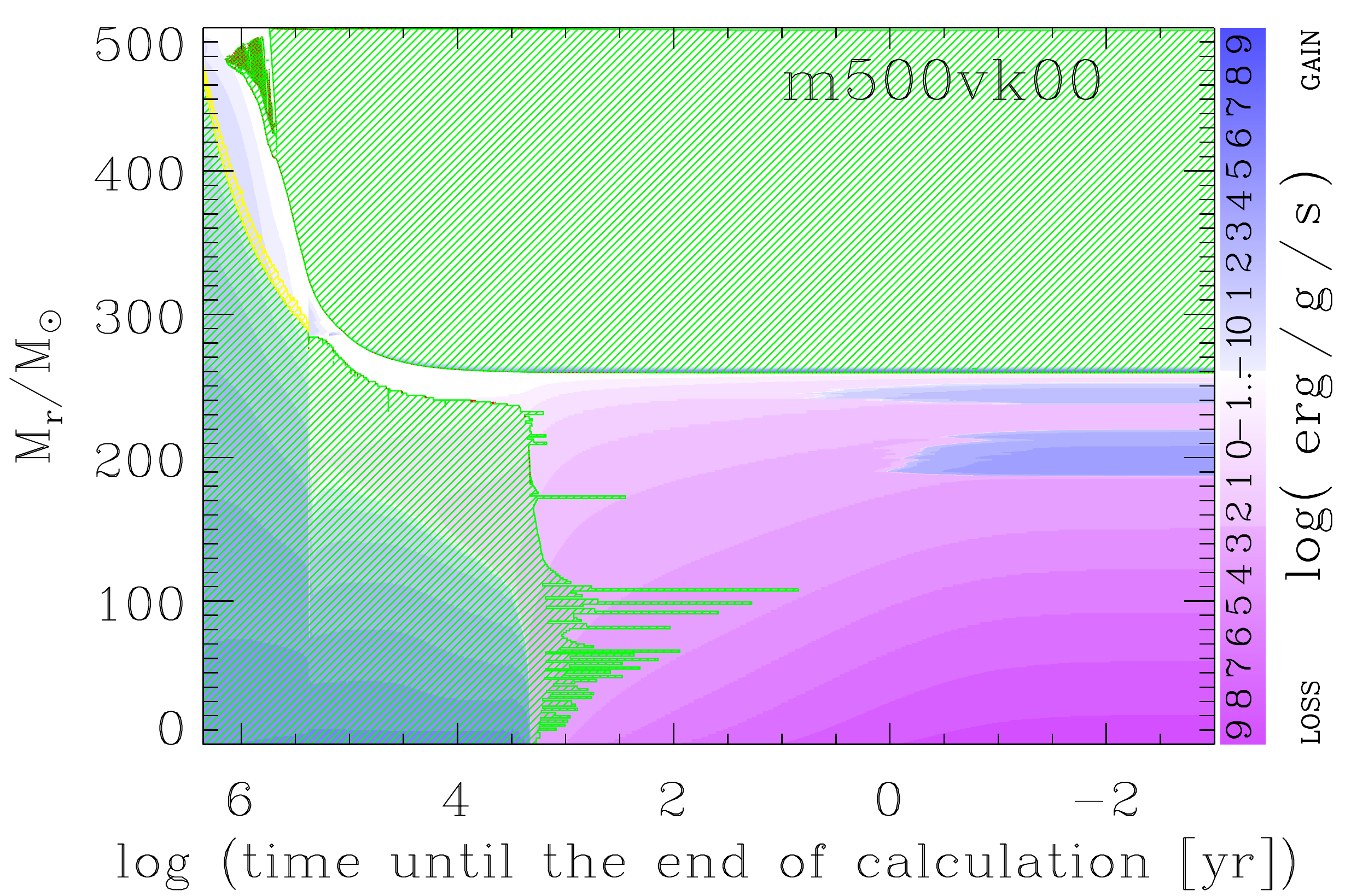}
\includegraphics[width=\columnwidth]{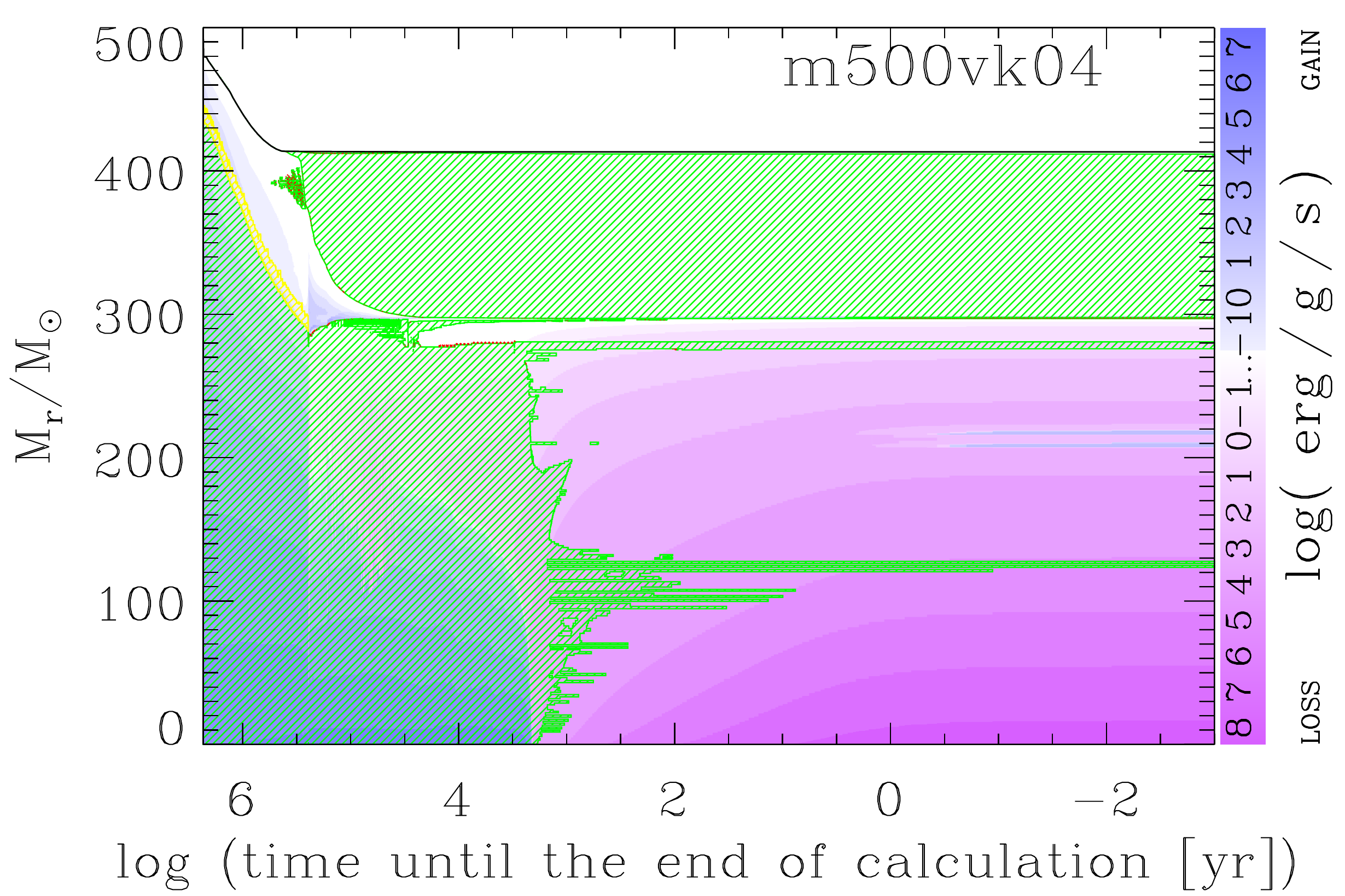}
\end{center}
{\bf Fig.~\ref{fig:kipp1}.}  Continued  
\end{figure*}

\end{appendix}

\end{document}